\documentclass[aps,prb,longbibliography,twocolumn]{revtex4-1}
\usepackage{appendix}
\usepackage{epsfig}
\usepackage{graphicx}
\usepackage{amsfonts}
\usepackage{amsmath}
\usepackage{mathtools}
\usepackage{psfrag}
\usepackage{amssymb,amsmath,amsthm}
\usepackage{verbatim}
\usepackage{soul}
\usepackage[normalem]{ulem}
\usepackage{color}
\newcommand{\beq}{\begin{equation}}
\newcommand{\eeq}{\end{equation}}
\newcommand{\bea}{\begin{eqnarray}}
\newcommand{\eea}{\end{eqnarray}}
\newcommand{\bwt}{\begin{widetext}}
\newcommand{\ewt}{\end{widetext}}
\newcommand{\mM}{{\mathsf M}}

\newcommand{\om}{\omega}
\newcommand{\mb}{\mathbf}
\newcommand{\fl}[1]{\textcolor{red}{#1}}
\newcommand{\flc}[1]{\textcolor{blue}{#1}}
\allowdisplaybreaks
\begin{document}

\title{ {Diffusion of elastic waves in a continuum solid with a random array of pinned dislocations}}

\author{Dmitry Churochkin$^1$ and Fernando Lund$^2$}
\affiliation{$^1$Saratov State University, Saratov, 410012 Russia\\$^2$Departamento de F\'\i sica and CIMAT, Facultad de Ciencias
F\'\i sicas y Matem\'aticas, Universidad de Chile, Santiago, Chile}

\date{{\today}}
\begin{abstract}
 {The propagation of incoherent elastic energy in a three-dimensional solid due to the scattering by many, randomly placed and oriented,  {pinned} dislocation segments, is considered in a continuum mechanics framework. The scattering mechanism is that of an elastic string of length $L$ that re-radiates as a response to an incoming wave. The scatterers are thus not static but have their own dynamics. A Bethe-Salpeter (BS) equation is established, and a  Ward-Takahashi Identity (WTI) is demonstrated.} The BS equation is written as a spectral problem that, using the WTI, is solved in the diffusive limit.  {To leading order a diffusion behavior indeed results, and an explicit formula for the diffusion coefficient is obtained. It can be evaluated in an Independent Scattering Approximation (ISA) in the absence of intrinsic damping. It depends not only on the bare longitudinal and transverse wave velocities but also on the renormalized velocities, as well as attenuation coefficients, of the coherent waves. The influence of the length scale given by $L$, and of the resonant behavior for frequencies near the resonance frequency of the strings, can be explicitly identified. A {Kubo representation} for the diffusion constant can be identified. {P}revious generic results, obtained {with} an energy transfer formalism, {are} recovered when the number of dislocations per unit volume is small. {This includes the equipartition of diffusive energy density which, however, does not hold in general.} The formalism bears a number of similarities with the behavior of electromagnetic waves in a medium with a random distribution of dielectric scatterers;  the elastic interaction, however, is momentum dependent.}
\end{abstract}

\vskip 0.2cm 

\vskip 0.5cm \maketitle
\section{Introduction}

Dislocations have long been known to be a crucial component in the mechanical behavior of metals and alloys. In other areas of condensed-matter physics, however, they have often been considered rather a nuisance. Nevertheless, in recent years  increasing evidence has become available to the effect that dislocations, rather than an obstacle, can become a useful tool to increase the performance of functional materials. For example, dislocations have been shown to drive the amorphization of phase-change materials\cite{Nam2012}; they can contribute to the control of polarization in bulk ferroeelectrics\cite{Hofling2021}, and they considerably alter the distribution of electronic and ionic defects in oxides\cite{Adepalli2017,Porz2021}. Importantly for optoelectronic devices, Massabau et al.\cite{Massabau2017} have reported evidence for carrier localization in the vicinity of dislocations in InGaN. However, progress along these lines has been hampered by a lack of understanding of the basic physics of dislocations, considered as one-dimensional, extended, topological defects in a three-dimensional material.

Additionally, from a condensed matter physics point of view, {surprisingly little appears to have been studied about the influence of dislocations on thermal transport, although experimental evidence of a measurable effect have been reported. Indeed,} Kotchetkov et al.\cite{Kotchetkov2001} showed, using a relaxation time approximation that dislocations have a measurable effect on the thermal conductivity of GaN layers. Kamatagi et al.\cite{Kamatagi2007} and Ma et al.\cite{Ma2013} have studied the effect of point defects and dislocations on bulk wurtzite GaN, and found it to be significant. The same is true for free standing GaN thin films\cite{Kamatagi2009}. A relaxation time approximation was also used by Singh et al.\cite{Singh2006} to study the effect of stacking faults and dislocations on the phonon conductivity of plastically deformed LiF and Ge, with satisfactory results. Recently, the role of dislocations has become the focus of much attention, and there is an increasing quantitative evidence linking a decrease in thermal conductivity with an increase in dislocation density\cite{Shuai2016,Wu2019,You2018,Xin2017,Zhou2018,Yu2018}. Additionally, a numerical experiment\cite{Giaremis2020} has  concluded that decorated dislocation engineering can lead to interesting fabrication strategies for themoelectric devices.

Importantly, lack of a detailed understanding of phonon transport seriously hampers the fabrication of practical thermoelectric materials\cite{Minnich2009}, and there is a significant activity around this issue.  It is worth mentioning here, for example, the calculation of thermal conductivity using first principles atomistic simulations and the Boltzmann transport equation\cite{Lindsay2013,Lee2014}. However, current simulation tools appear to be still insufficient to gauge the impact of defects, particularly extended, resonant defects such as dislocations, on phonon transport\cite{Tian2014}. Molecular dynamics methods have also been used\cite{Jund1999,Muller1997}, but shortcomings have recently been pointed out by  Bedoya-Mart\'\i nez et al.\cite{Bedoya2014} Quite recently, and after decades of the formulation of the traditionally used theoretical models for the phonon-dislocation interaction\cite{Klemens1955,Carruthers1959}, dislocation dynamics such as it is used in the present work has been incorporated into the understanding of thermal transport\cite{Lund2019,Lund2020}.

 {The interaction of acoustic waves---phonons---with dislocations has a long and distinguished history of scholarship\cite{Granato1956a,Granato1956b,Lucke1981,Kneezel1982,Shilo2007}. However, only in recent years it has been possible to make sufficient quantitative progress to have, say, explicit formulae for the scattering cross section of an elastic wave by an oscillating dislocation segment in three dimensions for arbitrary wave polarization, dislocation and Burgers vector orientation\cite{Maurel2005a}. Use of the resulting formalism together with a multiple scattering approach has led to a new way to characterize dislocation densities in metals and alloys through Resonant Ultrasound Spectroscopy (RUS)\cite{Mujica2012}} and in-situ time-of-flight measurements\cite{Barra2015,Salinas2017,Espinoza2018}.

Maurel et al.\cite{Maurel2005b}, working within the framework of the continuum theory of elasticity, have developed a perturbation 
scheme for the  propagation of elastic waves through a random array of pinned vibrating dislocations. 
On the grounds of that model, the problem of  coherent propagation, and attenuation, has been investigated thoroughly 
in the Independent Scattering Approximation (ISA)\cite{Churochkin2016}. The coherent propagation regime carries only part of the information about the transport properties of a given physical system\cite{Sheng2006}. A complete treatment requires the investigation of incoherent behavior. Of special interest is the diffusive range, which is determined by the transfer of energy density {and typically starts at transport distances a bit larger than a few attenuation lengths}. The general approach to this problem is based on the asymptotic solution of the  {Bethe-Salpeter} (BS) equation accompanied with the relevant  {Ward-Takahashi identity} (WTI). In turn, the form of the WTI depends on the specifics of the system under consideration\cite{Sheng2006}.

Diffusion techniques for incoherent waves  were developed to treat the problem of electron localization \cite{Vollhardt1980,Wolfle1984,Bhatt1985}, and  were  later used for the description of the localization of (scalar) acoustic waves moving through a random array of hard scatterers\cite{Kirkpatrick1985}. An  eigenvalue method to solve the BS equation developed by W\"olfle et al.\cite{Wolfle1984,Bhatt1985} was extended to  the problem of light diffusion in a random medium of dielectric scatterers which complies with the generalized WTI by Barabanenkov and Ozrin\cite{Barabanenkov1991,Barabanenkov1995}. In a similar vein, the diffusion of light  in a general anisotropic turbid media was studied by Stark and Lubensky\cite{Stark1997}.

 {The multiple scattering of acoustic and elastic waves has been dealt with in the literature: Kirkpatrick\cite{Kirkpatrick1985} 
 studied the problem of the localization of scalar acoustic waves in a medium with hard scatterers, 
 both in two and three dimensions, using a diagrammatic approach. A diffusion behavior appears in a Boltzmann approximation 
 as a result of the summation of the ladder diagrams. Weaver\cite{Weaver1990} studied the diffusion of ultrasound 
 in a polycrystalline material, introducing disorder through randomly fluctuating elastic constants, 
 and obtained an equation of radiative transfer. Van Tiggelen and collaborators have studied the coherent backscattering 
 of elastic waves in an infinite isotropic medium\cite{Tiggelen2001}, their radiative 
 transfer in a generalized diffusion approximation\cite{Tregoures2002a}, and their multiple scattering within a plate\cite{Tregoures2002b}. 
 The Schr\"odinger-like description used in the last work has been carried over by Trujillo et al.\cite{Trujillo2010} 
 to the description of elastic waves in dry granular media.}  The issue of localization of elastic waves, 
 a phenomenon that may appear when the diffusion constant vanishes because of wave interference, has been addressed experimentally by Cobus et al.\cite{Cobus2018} 
 and Go\"{\i}coechea et al.\cite{Goicoechea2020}. 

On a different perspective, the interaction of sound with the Volterra dislocations that are used in the present paper has been shown to lead to an improved understanding of the acoustic properties of glasses in the THz range\cite{Lund2015,Bianchi2020}. The use of continuum mechanics, without an intrinsic length scale, offers a powerful tool since it applies to all glasses in the appropriate length scale. The same point of view can be helpful to advance our understanding of thermal transport in amorphous solids. Indeed, as emphasized for example by Beltukov et al.\cite{Beltukov2018} through numerical simulations, there is a complex dynamics underlying energy transport by phonons in these materials.

The purpose of this article is to address the above issues from a macroscopic point of view; specifically, to study the diffusion of elastic waves moving through a random array of vibrating dislocations. To this end, we describe the dynamics of a single dislocation following the Granato-L\"ucke vibrating string model\cite{Granato1956a}. It is assumed that we deal with an ensemble of noninteracting dislocations (or, more precisely, that they interact solely through the scattering of elastic waves). On this foundation, we extend the formalism developed by Barabanenkov and Ozrin\cite{Barabanenkov1995} for electromagnetic waves to the case of elastic waves with different polarizations that interact with  scatterers that obey the generalized Granato-L\"ucke string equation\cite{Maurel2005a}.


This paper is organized as follows: Section \ref{sectionpreviousresults}  sets up the formalism for the problem. It is an inhomogeneous wave equation in which the inhomogeneous term describes the interaction between wave and dislocation. This interaction term is dubbed ``the potential term'' by analogy with the case of de Broglie waves describing electrons. We shall use a perturbation approach, in which the potential term is considered a small perturbation. Previous results are briefly recalled. A Bethe-Salpeter equation is derived in Section \ref{sec:BS}.  Following the approach of \cite{Barabanenkov1991,Barabanenkov1995} a Ward-Takahashi identity is obtained in Section \ref{sec:WTI}.
The eigenvalue problem for the BS equation is formulated, and solved, in Section \ref{sec:diffube}. A specific expression for the diffusion constant is obtained. This result is discussed in Section \ref{sec:relation}. It is shown that the diffusion constant can be cast in a Kubo-like expression \cite{Barabanenkov1995}, and that, in the low frequency and low density of scatterers limit, it reduces to the expression obtained in a radiation transfer formalism\cite{Ryzhik1996}. Section \ref{sec:disc} offers a final conclusion and outlook. A number of the more technical calculations are described in six appendices.

\section{Problem set-up and previous results}
\label{sectionpreviousresults}

In the linear theory of elasticity, the dynamics of an isotropic medium with mass density $\rho$ and elastic constants $c_{ijkl}=\lambda \delta_{ij}\delta_{kl}+ \mu ( \delta_{ik}\delta_{jl}+\delta_{il} \delta_{jk} )$ with $(\lambda,\mu)$ the Lam\'e constants, is described by displacements  $\mb u (\mb x,t)$ as a function of an equilibrium position $\mb x$ at time $t$. Velocity $\mb v$ is the time derivative, $\mb v = \partial \mb u / \partial t$. The speed of sound is $c_L \equiv \sqrt{\lambda + 2\mu/\rho}$, the speed of shear waves is $c_T \equiv \sqrt{\mu/\rho}$ and we shall denote their ratio by $\gamma \equiv c_L/c_T$. The vibration of edge dislocations of length $L$ that are pinned at the ends, and characterized by the Burgers vector $\mb b$ with a local tangent oriented along $\hat{\mb{\tau}}$ and situated in the equilibrium state at the point $\mb X_0$ perturbs the medium in such a way that the whole system is governed by the wave equation with a source  \cite{Maurel2005b,Churochkin2016}:
\begin{equation}\label{waveeq}
\rho \frac{\partial^2}{\partial t^2}
v_i(\mb x,t)-c_{ijkl}\frac{\partial^2}{\partial x_j\partial x_l} v_k
(\mb x,t)= V_{ik} v_k (\mb x,t)
\end{equation}
where the perturbation potential is defined as
\begin{equation}
\left. V_{ik}=  {\cal A} \; {\mathsf  M}_{ij}
 \frac{\partial}{\partial x_j}  \delta ( \mb x-\mb X_0 )\;
{\mathsf  M}_{lk}{\frac{\partial}{\partial x_l}} \right|_{\mb x=\mb X_0} \, ,
\label{potential}
\end{equation}
with
\begin{equation}\label{mtensor}
 {\cal A}  \equiv  \frac{8}{\pi^2}\frac{(\mu b)^2L}{m} g(\om) 
 \, ,
\end{equation}
$g(\omega)\equiv [\omega^{2} {+} i \omega (B/m) -\omega^2_{F}]^{-1}$, $\hat{\mb n} \equiv \hat{\mb \tau} \wedge \hat{\mb t}$,  $\hat{\mb t} \equiv \mb b/|\mb b|$ is the unit Burgers vector that indicates the direction of glide, and ${\mathsf  M}_{ij} \equiv t_i n_j + t_j n_i$,
with
\beq
\omega_F \equiv \frac{\pi}{L}\sqrt{\frac{\Gamma}{m}}
\label{eigenom}
\eeq
the fundamental frequency  of a vibrating string characterized by effective mass per unit length $m$, line tension $\Gamma$, and damping $B$, which represent the dislocation dynamics. Only glide motion, that is, along $\hat t$, is allowed, a fact that translates into $\tau_{i}V_{ik}\equiv 0$. Dislocation climb implies mass transport and is not allowed.\cite{Lund1988} {The medium is considered linear everywhere outside the dislocations core. Consequently, when more than one dislocation is present, their effect is obtained simply by addition of the individual terms.} {Note that the potential (\ref{potential}) involves two gradients, a feature that will lead, in momentum space, to a dependence on the square of the momentum. Care will have to be exercised then at short wavelengths.}

An important quantity for the analysis is the Green's tensor, or impulse response function, for Eqn. (\ref{waveeq}).  Its average properties provide information about both coherent and incoherent wave behavior.  {In the frequency domain,} it obeys the equation \cite{Maurel2005b,Churochkin2016}
\begin{multline}
 \rho \omega^2 G_{im}(\mb x, \mb x',\omega)+c_{ijkl}\frac{\partial^2}{\partial x_j\partial x_l} G_{km}  (\mb x,\mb x',\omega)=  \\
 \hspace{1em} - \sum_{\text{disloc. lines}} V_{ik}G_{km}(\mb x,\mb x', \omega) -
  \delta_{im} \delta(\mb x-\mb x')  \,\, .
\label{greeneq}
  \end{multline}
Eqn. (\ref{greeneq}) carries information about the asymptotic behavior of outgoing waves at large distances from the source. For  convenience, we have not written explicitly the second argument in the Green tensor: $G_{im}(\mb x,\omega)$ must be understood as $G_{im}(\mb x,\mb x^{\prime},\omega)$ with $\mb x$ the detection point and $\mb x^{\prime}$ the source point. The poles of the {Fourier transformed} averaged Green tensor yield the modified spectrum of $T$ ({transversal}) and $L$ ({longitudinal}) modes present in the medium. A solution of Eqn. (\ref{greeneq})  can be found perturbatively. Applying  the ISA approach {(i.e., that the random variables associated with each one of the dislocation segments are statistically independent of each other)} we have found  the averaged Green's tensor for outgoing waves $\langle \mathbf{G} \rangle^{+}({\bf k},\omega)$  as\cite{Churochkin2016}
\beq
\label{green}
\langle \mb G \rangle^{+}({\bf k},\omega)=G_{T}\left(\textbf{I}-P_{\textbf{\^{k}}}\right)+
G_{L}P_{\textbf{\^{k}}}
\eeq
 {with}
\beq
G_{T,L}=\frac{1}{\rho \omega^{2}\left\{\frac{k^2}{K_{T,L}^{2}}-1\right\}}\nonumber
\eeq
as well as the self-enegy tensor $\mathbf{\Sigma}^{+}({\bf k},\omega)$  {defined through the Dyson equation\cite{Maurel2005b}
\beq
\label{Dyson}
\langle \mb G \rangle^{-1} = ({\mb G}^0)^{-1} -\mb \Sigma
\eeq with ${\mb G}^0$ the Green's tensor for free space and}
\beq
\label{self-energy}
\mathbf{\Sigma}^{+}({\bf k},\omega)=\Sigma_{T}\left(\textbf{I}-P_{\textbf{\^{k}}}\right)+\Sigma_{L}P_{\textbf{\^{k}}}
\eeq
with
\beq
\Sigma_{T,L}=\rho\left(c_{T,L}^{2}-\frac{\omega^{2}}{K_{T,L}^{2}}\right)k^{2}\nonumber \, ,
\eeq
\begin{eqnarray}\label{poles}
K_{T} & = & \frac{\omega}{c_{T}} \left[1+\frac{n{\cal A}}{5\rho c_{T}^{2}\left(1+i {\cal A}I\right)}\right]^{-1/2} \, , \\
K_{L} & =& \frac{\omega}{c_{L}} \left[1+\frac{4n{\cal A}}{15\rho c_{L}^{2}\left(1+i {\cal A}I\right)}\right]^{-1/2} \, , \nonumber
\end{eqnarray}
and
\beq\label{parameters}
I=\frac{1}{30\pi^{}}\left[\frac{3\gamma^{5}+2}{\gamma^{5}}\right]\frac{\omega^3}{\rho c_{T}^{5}}
\eeq
 {where} $P_{\textbf{\^{k}}}=\textbf{\^{k}}^{t}\textbf{\^{k}}$ and $\textbf{\^{k}}^{t}$ is the transposed unit vector along $\textbf{k}$.
The incoming waves, related to $<\mathbf{G}>^{-}({\bf k},\omega)$  {and} $\mathbf{\Sigma}^{-}({\bf k},\omega)$, are described by the complex conjugate form of  Eqns. (\ref{green}) and (\ref{self-energy}).

{The average $\langle \cdot \rangle$ is over dislocation position, orientation, and Burgers vector. It has been described in detail by Maurel et al.\cite{Maurel2005b} On average, the medium is homogeneous and isotropic.} The effective wave numbers $K_{T,L}$ define an effective phase velocity for wave propagation
\beq
v_{T,L} \equiv \frac{\om}{Re[K_{T,L}]}
\label{eq:effvel}
\eeq
and attenuation length
\beq
l_{T,L} \equiv \frac{1}{2Im[K_{T,L}]} \, .
\label{eq:attenlength}
\eeq
These quantities will appear explicitly in the diffusion constant that will be discussed in Section \ref{sec:diffube}.

\section{ {Bethe-Salpeter equation for an elastic medium with many vibrating dislocation segments}}
\label{sec:BS}
{We have tested the methods of this paper in a simplified setting: that of the incoherent behavior of elastic waves in a two dimensional continuum with a random distribution of screw dislocations\cite{Churochkin2017} {and of edge dislocations\cite{Churochkin2021}}. {The screw case} is a scalar problem that keeps the whole basic physics of the diffusion behavior of elastic waves when propagating incoherently among a maze of dislocations. Being scalar the algebra is much simpler. {The edge case keeps the full vector nature of the three-dimensional problem, but the algebra is still simpler in two dimensions, particularly since dislocations are points and not lines}. The physics of the present problem is much richer because the dislocations have a finite length, a precise orientation and Burgers vector, and the elastic waves have two polarizations that travel at different speeds. The algebra, however, is quite close to that of \cite{Churochkin2021}}, and we shall refer to this reference for the details of the computation.

To track the wave transport after the phase coherence is lost we have to focus on the evolution of the corresponding configurationally averaged intensity which is qualitatively represented in momentum space as  {the two-point correlation of the Green's tensor}\cite{Sheng2006}
\begin{multline}\label{intensity}
\mathbf{\Phi}({\bf k},{\bf k}^{\prime};{\bf q},\Omega) \equiv \Phi_{kl,mn}({\bf k},{\bf k}^{\prime};{\bf q},\Omega)  \\ \equiv <G^{+}_{km}(\mathbf{k}^{+},\mathbf{k}^{\prime +},\omega^{+})G^{-}_{nl}(\mathbf{k}^{\prime-},\mathbf{k}^{-},\omega^{-})>
\end{multline}
 {with}
\beq
\mathbf{k}^{\pm}=\mathbf{k}\pm\frac{\mathbf{q}}{2},\quad\omega^{\pm}=\omega\pm\frac{\Omega}{2} \, .
\eeq
The reciprocity of the Green's tensor, $G_{im}(\mb x,\mb x^{\prime},\omega)=G_{mi}(\mb x^{\prime},\mb x,\omega)$,  {implies}
\beq
\Phi_{kl,mn}({\bf k},{\bf k}^{\prime};{\bf q},\Omega)=\Phi_{mn,kl}({\bf k}^{\prime},{\bf k};{\bf q},\Omega)\nonumber \, .
\eeq
In this approach, ``diffusive behavior'' means that the the two-point correlation tensor (\ref{intensity}) has a specific pole structure in terms of the diffusive variables $\mathbf{q}$ and $\Omega$. Just as the Dyson equation yields the pole structure for the averaged Green's tensor, the BS equation yields the pole structure for the intensity\cite{Sheng2006}. Using the standard formalism\cite{Sheng2006,Vollhardt1980,Kirkpatrick1985,Churochkin2021}, the BS equation for the elastic wave diffusion in the medium with dislocations is found to be ({See Appendix \ref{BSE}})
\begin{widetext}
\beq
\label{BS}
\left[\imath\omega\Omega\mathbf{E}+\mathbf{P}({\bf k};{\bf q})\right]:\mathbf{\Phi}({\bf k},{\bf k}^{\prime};{\bf q},\Omega)+
\int\limits_{\bf{k}^{\prime\prime}}\mathbf{U}({\bf k},{\bf k}^{\prime\prime};{\bf q},\Omega):\mathbf{\Phi}({\bf k}^{\prime\prime},{\bf k}^{\prime};{\bf q},\Omega)
=\delta_{{\bf k},{\bf k}^{\prime}}\mathbf{\Delta G}({\bf k};{\bf q},\Omega)
\eeq
Where
\bea
\label{potBS}
\mathbf{U}({\bf k},{\bf k}^{\prime};{\bf q},\Omega) &\equiv& U_{ij,kl}({\bf k},{\bf k}^{\prime};{\bf q},\Omega)  \\
& \equiv & \Delta\Sigma_{ij,kl}({\bf k};{\bf q},\Omega)\delta_{{\bf k},{\bf k}^{\prime}}- \Delta G_{ij,mn}({\bf k};{\bf q},\Omega)K_{mn,kl}({\bf k},{\bf k}^{\prime};{\bf q},\Omega)  \\
\label{delta}
\mathbf{\Delta G}({\bf k};{\bf q},\Omega) &\equiv& \Delta G_{ij,mn}({\bf k};{\bf q},\Omega)  \\
&\equiv &
\frac{1}{2\imath\rho}\left(\delta_{im}<G>^{-}_{nj}(\mathbf{k}^{-},\omega^{-})-<G>^{+}_{im}(\mathbf{k}^{+},\omega^{+})\delta_{nj}\right) \, ,    \\
\label{deltasigma}
\mathbf{\Delta\Sigma}({\bf k};{\bf q},\Omega) &\equiv & \Delta\Sigma_{ij,mn}({\bf k};{\bf q},\Omega) \\
& \equiv &
\frac{1}{2\imath\rho}\left(\delta_{im}\Sigma^{-}_{nj}(\mathbf{k}^{-},\omega^{-})-\Sigma^{+}_{im}(\mathbf{k}^{+},\omega^{+})\delta_{nj}\right)
\eea
\end{widetext}
 {and}
\bea
\label{potBS2}
\mathbf{P}({\bf k};{\bf q}) &\equiv& P_{ij,kl}({\bf k};{\bf q})  \\
& \equiv &
\frac{1}{2\imath\rho}\left(\delta_{ik}L_{lj}(\mathbf{k}^{-})-L_{ik}(\mathbf{k}^{+})\delta_{lj}\right)   \\
\mathbf{E} &=& E_{ij,kl}=\delta_{ik}\delta_{lj} \\
 L_{ij}(\mathbf{k}^{\pm})&=&-c_{ikjl}k^{\pm}_{k}k^{\pm}_{l} .
\eea
 {Here,} $K_{mn,kl}({\bf k},{\bf k}^{\prime};{\bf q},\Omega)$ is the irreducible vertex, {explicitly presented in Appendix  \ref{BSE}}. We denote
 \beq
  \int\limits_{\bf{p}^{\prime\prime}}=(2\pi)^{-3}\int d\bf{p}^{\prime\prime},
\eeq
and $:$ is the inner tensor product {defined in components for arbitrary fourth rank tensors as  $\mathcal{E}:\mathcal{F}\equiv \mathcal{E}_{ij,kl}\mathcal{F}_{kl,mn}$}.

\section{ {Ward-Takahashi identity}}
\label{sec:WTI}
 Energy conservation, formulated in the form of a WTI, underlie the theoretical description of incoherent transport of classical waves\cite{Nieh1998}. For specific forms of the perturbation potential the WTI has been obtained on the basis of Lagrangian\cite{Nieh1998} as well as pre-WTI methods\cite{Barabanenkov1991,Barabanenkov1995}, an issue that was the object of some debate\cite{Barabanenkov2001,Nieh2001}.
In this paper we shall use the pre-WTI method\cite{Barabanenkov1991,Barabanenkov1995} that deals directly with the equations of motion.

\subsection{Pre-WTI}
\label{sec:pre-WTI}
We establish, as a preliminary step, a relation between the average Green's function and its two-point correlation that does not explicitly involve the interaction $V_{ik}$. To this end we start with Eqn. (\ref{greeneq}) written for Green's tensors at two different sets of variables: $G_{i_1m_1}(\mb x_{1},\mb x^{\prime}_{1},\omega_{1})$ and $G_{i_2m_2}(\mb x_{2},\mb x^{\prime}_{2},\omega_{2})$. 
and we take the two-sided Fourier transform of these relations with the definitions
\begin{eqnarray}\label{ift}
\mathbf{F}(\mb k,\mb k^{\prime};\omega)&=&\int\int d\mb x d\mb x^{\prime} e^{-\imath \mb k\mb x}\mathbf{F}(\mb x,\mb x^{\prime};\omega) e^{\imath \mb k^{\prime}\mb x^{\prime}}   \\
\label{dft}
\mathbf{G}(\mb x,\mb x^{\prime};\omega)&=&\int\limits_{\mb k} \int\limits_{\mb k^{\prime}}e^{\imath \mb k\mb x} \mathbf{G}(\mb k,\mb k^{\prime};\omega)e^{-\imath \mb k^{\prime}\mb x^{\prime}} \, .
\end{eqnarray}
{
We now act on the first and second equations of the system from the right by $g^{*}(\omega_{2})(G)^{-1}_{m1n1}(\mb k^{\prime}_{1},\mb k^{\prime\prime}_{1};\omega_{1})$ and $g(\omega_{1})(G^{*})^{-1}_{m2n2}(k^{\prime}_{2},\mb k^{\prime\prime}_{2};\omega_{2})$, respectively.
{The next step consists in subtraction  of the second equation from the first, and {evaluating at} $i_1=n_2=i$, $n_1=i_2=n$;
$\mb k_{2}\rightarrow\mb k^{\prime\prime}_{1}$, $\mb k^{\prime\prime}_{2}\rightarrow\mb k_{1}$.  Otherwise, it is not possible to eliminate the remaining parts of the potentials in both equations that are subject to the substraction from each other, since the parts imply not only summation over defects but also contain components of the second rank tensor, i. e. to achieve identity of those parts between each other, the components must be also identical. Noting the explicit expression of the bare Green's function
\beq
(G^{0})^{-1}_{ik}(\mb k,\omega)=-\left(\rho\omega^{2}\delta_{ik}-c_{ijlk}k_{j}k_{l}\right)
 \label{g0ft}
\eeq
{We obtain
\begin{multline}\label{preWTIft}
\lim\limits_{{\substack{\mb k_{2}\rightarrow\mb k^{\prime\prime}_{1}\\ \mb k^{\prime\prime}_{2}\rightarrow \mb k_{1}}}}
\left(-(G^{0})^{-1}_{in}(\mb k_{1},\omega_{1})\delta_{\mb k_{1},\mb k^{\prime\prime}_{1}}g^{*}(\omega_{2}) \right.  \\
+g^{*}(\omega_{2})(G)^{-1}_{in}(\mb k_{1},\mb k^{\prime\prime}_{1};\omega_{1}) +(G^{*0})^{-1}_{ni}(\mb k_{2},\omega_{2})\delta_{\mb k_{2},\mb k^{\prime\prime}_{2}}g(\omega_{1})   \\ \left. -g(\omega_{1})(G^{*})^{-1}_{ni}(\mb k_{2},\mb k^{\prime\prime}_{2};\omega_{2})\right)\equiv 0 \, .
\end{multline}
Now, multiplying this identity on the right by
\beq
\lim\limits_{{\substack{\mb k_{2}\rightarrow\mb k^{\prime\prime}_{1}\\ \mb k^{\prime\prime}_{2}\rightarrow \mb k_{1}}}} G(\mb k^{\prime\prime}_{1},\mb k^{\prime\prime\prime}_{1} ;\omega_{1})_{nl}G^{*}(\mb k^{\prime\prime}_{2},\mb k^{\prime\prime\prime}_{2};\omega_{2})_{ij} \, ,
\eeq 
 averaging, and using the following notation:
\begin{align}\label{notat}
\mb k_{1}=&\mb k^{+} & \mb k^{\prime\prime}_{2}=&\mb k^{-} &\mb k^{\prime\prime}_{1}=&\mb k^{\prime\prime+} & \mb k_{2}=&\mb k^{\prime\prime-} \nonumber \\
\mb k^{\prime\prime\prime}_{1}=&\mb k^{\prime\prime\prime+} & \mb k^{\prime\prime\prime}_{2}=&\mb k^{\prime\prime\prime-}  & \omega_{1}=&\omega_{+} & \omega_{2}=&\omega_{-} \\ G=&G^{+} &  G^{*}=&G^{-} & G^{0}=&G^{0+} & G^{0*}=&G^{0-}\nonumber
\end{align}
the following pre-WTI is obtained
\begin{widetext}
\begin{multline}\label{preWTI}
{\int\limits_{\mathbf{k}}}\left((G^{0-})^{-1}_{ni}({\bf k};{\bf q},\Omega)g(\omega_{+})-(G^{0+})^{-1}_{ni}({\bf k};{\bf q},\Omega)g^{\ast}(\omega_{-})\right)\Phi_{ni,lj}({\bf k},{\bf k}^{\prime\prime\prime};{\bf q},\Omega)\\ +g^{\ast}(\omega_{-})<G>^{-}(\mb k^{\prime\prime\prime};{\bf q},\Omega)_{lj}-g(\omega_{+})<G>^{+}(\mb k^{\prime\prime\prime};{\bf q},\Omega)_{lj}\equiv 0
\end{multline}
\end{widetext}
If we use Eq.(\ref{intensity}) and recall that
\begin{eqnarray}\label{intensityft}
(G^{0\pm})^{-1}_{in}(\mb k^{\pm},\omega_{\pm})&=&(G^{0\pm})^{-1}_{in}({\bf k};{\bf q},\Omega) \\
<G^{\pm}(\mb k^{\prime\prime\prime\pm},\mb k^{\prime\prime\prime\pm} ;\omega_{\pm})_{jl}>&=&<G>^{\pm}(\mb k^{\prime\prime\prime};{\bf q},\Omega)_{jl} \, ,\nonumber
\end{eqnarray}
{we see that the pre-WTI relates, in Fourier space, the averaged Green's function, with its two-point correlations without the explicit appearance of the interaction $V_{ij}$.}}
%

\subsection{WTI}
{{The relation between averages obtained at the end of the last subsection is now turned into a relation between their ``irreducible" parts,}  {the irreducible vertex} $\mathbf{K}$ and {the mass operator} $\mathbf{\Sigma}$. Multiplying Eq. (\ref{preWTI}) on the right by $\Phi^{-1}_{lj,mt}({\bf k}^{\prime\prime\prime},{\bf k}^{\prime\prime\prime\prime};{\bf q},\Omega)$, 
 using (\ref{ApBSIrred}) and (\ref{Dyson}), the following WTI {is obtained}
\begin{multline}\label{WTIevent}
\left(\Sigma^{-}_{mt}({\bf k}^{\prime\prime\prime\prime};{\bf q},\Omega)g(\omega_{+})-\Sigma^{+}_{mt}({\bf k}^{\prime\prime\prime\prime};{\bf q},\Omega)g^{\ast}(\omega_{-})\right) \\ \equiv
  {\int\limits_{\mathbf{k}^{\prime\prime\prime}}}\left(g^{\ast}(\omega_{-})
<G>^{-}(\mb k^{\prime\prime\prime};{\bf q},\Omega)_{lj} \right. \\ \left. -g(\omega_{+})
<G>^{+}(\mb k^{\prime\prime\prime};{\bf q},\Omega)_{lj}\right) K_{lj,mt}({\bf k}^{\prime\prime\prime},{\bf k}^{\prime\prime\prime\prime};{\bf q},\Omega)
\end{multline}

In terms of the general, i. e. symbolical, representation of the WTI there are two differences compared to a well-known tensorial version of the WTI for {electromagnetic} waves \cite{Barabanenkov1995}:
first, $g$ is a complex valued resonance like function; second, the tensor rank of the WTI is two rather than four as in the case of {electromagnetic} waves \cite{Barabanenkov1995}. {In our case, this is all we need to solve the problem at hand.}

The WTI can be written in the more compact form
\beq\label{WLWTI}
\int\limits_{\mathbf{k}^{\prime\prime}}\mathbf{\overline{U}}(\mathbf{k}^{\prime\prime},\mathbf{k}^{\prime};\mathbf{q},\Omega)= \frac{i}{2}\mathbf{\overline{A}}(\mathbf{k}^{\prime};\mathbf{q},\Omega)\left(g(\omega_{+})-g^{\ast}(\omega_{-})\right)
\eeq
with the following notation:
\begin{align}\label{WLUtensor}
\mathbf{\overline{U}}(\mathbf{k}^{\prime\prime},\mathbf{k}^{\prime};\mathbf{q},\Omega)=&U_{ii,mt}({\bf k},{\bf k}^{\prime};{\bf q},\Omega)\\
\mathbf{\overline{A}}(\mathbf{k}^{\prime};\mathbf{q},\Omega)=&A_{nn,mt}(\mathbf{k}^{\prime};\mathbf{q},\Omega)\nonumber\\
 \mathbf{A}(\mathbf{k}^{\prime};\mathbf{q},\Omega)=&\frac{2}{g(\omega_{+})+g^{\ast}(\omega_{-})}
\left({\cal R}\mathbf{\Sigma}(\mathbf{k}^{\prime};\mathbf{q},\Omega) \right. \\
& \left.+\int\limits_{\mathbf{k}^{\prime\prime}}{\cal R}\mathbf{G}(\mathbf{k}^{\prime\prime};\mathbf{q},\Omega):\mathbf{K}({\bf k^{\prime\prime}},{\bf k^{\prime}};{\bf q},\Omega)\right) \, , \nonumber\\
{\cal R}\mathbf{\Sigma}(\mathbf{k}^{\prime};\mathbf{q},\Omega)=&
\frac{1}{2\rho}\left(\mathbf{I}\otimes\mathbf{\Sigma}^{-}(\mathbf{k}^{\prime-},\omega^{-}) \right. \\& \left. +\mathbf{\Sigma}^{+}(\mathbf{k}^{\prime+},\omega^{+})\otimes\mathbf{I} \right) \, , \nonumber
\end{align}
The tensor $\mb U$ is given by (\ref{potBS}). The operation ${\cal R}$ is defined here for the self-energy tensor $\mathbf{\Sigma}$, it is similarly defined for  the Green's tensor $\mathbf{G}$.

\subsection{Low $\Omega$, low ${\bf q}$ behavior}
%
The diffusion behavior appears in the limit $\Omega \, ,{\bf q} \to 0.$ In this case,
the following relations for the self energy and for the Green's function will prove useful:
\begin{align}\label{SEDiff}
\mathbf{\Delta\Sigma}({\bf k};{\bf 0},0)
  = & \mathbf{\Delta\Sigma({\bf k})}=\Delta\Sigma_{im,tk}({\bf k})  \nonumber \\
 =&\frac{1}{2\imath\rho}\left(\delta_{it}\Sigma^{*}_{km}(\mathbf{k})-\Sigma_{it}(\mathbf{k})\delta_{km}\right)
\end{align}
and its trace over two indices is given by
\bea
\Delta\Sigma_{ii,tk}({\bf k})&=&
 \frac{-1}{\rho}\left((\delta_{kt}-\hat k_k \hat k_t)Im[\Sigma_{T}(\mathbf{k})]  \right. \nonumber  \\
 &&\left.\hspace{2em} +\hat k_k \hat k_t Im[\Sigma_{L}(\mathbf{k})]\right) \, .
 \eea
Similarly
\begin{align}
\label{GFDiff}
\mathbf{\Delta G}({\bf k};{\bf 0},0)
  = & \mathbf{\Delta G({\bf k})}=\Delta G_{im,tk}({\bf k}) \\
  = &\frac{1}{2\imath\rho}\left(\delta_{it}G^{*}_{km}(\mathbf{k})-G_{it}(\mathbf{k})\delta_{km}\right)\\
\end{align}
so that its trace is
\begin{align}
\Delta G_{ii,tk}({\bf k})
  \approx & \frac{-\pi(\delta_{kt}-\hat k_k \hat k_t)k^{2}}{\rho^{2}\omega^{2}}\delta\left(k^{2}-Re[K_{T}^{2}]\right) \\
  &+\frac{-\pi\hat k_k \hat k_t k^{2}}{\rho^{2}\omega^{2}}\delta\left(k^{2}-Re[K_{L}^{2}]\right),
 \label{deltaGdelta}
\end{align}
The last approximation holds in the limit $| Im[K_{T,L}^{2}] | \ll | k^{2}-Re[K_{T,L}^{2}] |$. {(The meaning of this inequality in terms of the dislocation parameters is explored in  Section \ref{discone}).} Also, an abbreviated notation has been introduced: $<G^{+}>_{km}({\bf k},\omega) = G_{km}({\bf k})$, $<G^{-}>_{km}({\bf k},\omega) = G^{*}_{km}({\bf k}))$ and similarly for $\Sigma^{\pm}_{km}({\bf k},\omega)$.

\subsection{Lossless case, $B=0$, and independent scattering approximation (ISA)}
{When $B=0$, i.e. when $g$ is real,  ${\bf q}, \Omega$ tend to zero, and the standard ISA expressions for  ${\bf \Sigma}$ and ${\bf K}$ tensors, {Eq. (\ref{WTIisaexpressions}) below}, are taken (See Appendix \ref{OT}), the optical theorem is obtained. Explicitly, the WTI reads in this case
\begin{multline}
\label{WTIeventzero}
\left(\Sigma^{\ast}_{mt}({\bf k}^{\prime\prime\prime\prime})-\Sigma_{mt}({\bf k}^{\prime\prime\prime\prime})\right)\equiv \\{\int\limits_{\mathbf{k}^{\prime\prime\prime}}}\left(G^{0\ast}(\mb k^{\prime\prime\prime})_{lj}-G^{0}(\mb k^{\prime\prime\prime})_{lj}\right)
K_{lj,mt}({\bf k}^{\prime\prime\prime},{\bf k}^{\prime\prime\prime\prime})
\end{multline}
with the following expressions, valid to leading order in $n$, the density of scatterers:
\bea
\label{WTIisaexpressions}
\Sigma_{mt}({\bf k}^{\prime\prime\prime\prime})&=&\Sigma_{mt}({\bf k}^{\prime\prime\prime\prime};{\bf 0},0) \nonumber \\
&\approx & n<t>_{mt}({\bf k}^{\prime\prime\prime\prime})  \\
K_{lj,mt}({\bf k}^{\prime\prime\prime},{\bf k}^{\prime\prime\prime\prime};{\bf 0},0)&= & K_{lj,mt}({\bf k}^{\prime\prime\prime},{\bf k}^{\prime\prime\prime\prime}) \\
&\approx & n<t_{lm}({\bf k}^{\prime\prime\prime},{\bf k}^{\prime\prime\prime\prime})t^{*}_{tj}({\bf k}^{\prime\prime\prime\prime},{\bf k}^{\prime\prime\prime})> \nonumber \\
<G>(\mb k^{\prime\prime\prime};{\bf 0},0)_{jl}&= & <G>(\mb k^{\prime\prime\prime})_{jl} \nonumber \\
&\approx & G^{0}(\mb k^{\prime\prime\prime})_{jl}
\eea

\section{Diffusion behavior}
\label{sec:diffube}
The similarity that has been established between the WTI for elastic and electromagnetic waves motivates us to employ the well-developed formalism\cite{Barabanenkov1991,Barabanenkov1995,Stark1997,Berman2000} in the treatment of the diffusion problem. In that approach, we deal with the BS equation through the exploration of the eigenvalue problem for the operator with the kernel
\begin{eqnarray}\label{Hoperator}
\mathbf{H}=\left[\imath\omega\Omega\mathbf{E}+\mathbf{P}({\bf k};{\bf q})\right]\delta_{\mathbf{k}\mathbf{k}^{\prime\prime}}+\mathbf{U}({\bf k},{\bf k}^{\prime\prime};{\bf q},\Omega) \, .
\end{eqnarray}
 In terms of $\mathbf{H}$, the BS equation (\ref{BS}) can be written as
\beq\label{BSH}
\int\limits_{\bf{k}^{\prime\prime}}\mathbf{H}({\bf k},{\bf k}^{\prime\prime};{\bf q},\Omega):\mathbf{\Phi}({\bf k}^{\prime\prime},{\bf k}^{\prime};{\bf q},\Omega)=\mathbf{\Delta G}({\bf k};{\bf q},\Omega)\delta_{{\bf k},{\bf k}^{\prime}} \, .
\eeq
Moreover, the definition of the kernel $\mathbf{H}$ ensures that it obeys the symmetry property
 \begin{multline}\label{symmetry}
H_{ij,kl}({\bf k},{\bf k}^{\prime\prime};{\bf q},\Omega)\Delta G_{kl,mn}({\bf k}^{\prime\prime};{\bf q},\Omega)\\= H_{mn,kl}({\bf k}^{\prime\prime},{\bf k};{\bf q},\Omega)\Delta G_{kl,ij}({\bf k};{\bf q},\Omega)
\end{multline}
To see this, the explicit form of $\mathbf{U}$, and the reciprocity of the tensor $\mathbf{K}$, must be used.

In accordance with the general formalism\cite{Barabanenkov1991,Barabanenkov1995,Stark1997,Berman2000} the solution of Eqn. (\ref{BSH}) should be found through the consideration of the spectral problem for the corresponding homogeneous equation with $\mathbf{f}^{\mathbf{r}n}_{kl}({\bf k}^{\prime\prime};{\bf q},\Omega)$  (resp. $\mathbf{f}^{\mathbf{l}n}_{kl}({\bf k}^{\prime\prime};{\bf q},\Omega)$) as right (resp. left) eigentensors and $\lambda_{n}\left(\mathbf{q},\Omega\right)$ as eigenvalue:
\beq
\label{Hom}
\int\limits_{\bf{k}^{\prime\prime}}H_{ij,kl}({\bf k},{\bf k}^{\prime\prime};{\bf q},\Omega)\mathbf{f}^{\mathbf{r}n}_{kl}({\bf k}^{\prime\prime};{\bf q},\Omega)=
\lambda_{n}\left(\mathbf{q},\Omega\right)\mathbf{f}^{\mathbf{r}n}_{ij}({\bf k};{\bf q},\Omega)
\eeq
Following\cite{Barabanenkov1991,Barabanenkov1995,Stark1997,Berman2000} we assume the}
  eigentensors in Eqn. (\ref{Hom}) to obey  completeness and orthogonality conditions,
\begin{eqnarray}\label{basis}
\int\limits_{\mathbf{k}}\mathbf{f}_{ij}^{\mathbf{r}m}({\bf k};{\bf q},\Omega)\mathbf{f}_{ij}^{\mathbf{l}n}({\bf k};{\bf q},\Omega)&=& \delta_{mn}\, ,\\
\sum\limits_{n}\mathbf{f}_{ij}^{\mathbf{r}n}({\bf k};{\bf q},\Omega) \mathbf{f}_{kl}^{\mathbf{l}n}({\bf k}^{\prime};{\bf q},\Omega)&=&\delta_{{\bf k}{\bf k}^{\prime}}{\delta_{ik}\delta_{lj}}\nonumber \, .
\end{eqnarray}
 The left and right eigentensors are related, as a consequence of the symmetry properties (\ref{symmetry}) of the operator $\mathbf{H}$, as follows:
\beq
\label{lr}
\mathbf{f}^{\mathbf{r}n}_{mn}({\bf k};{\bf q},\Omega)=\Delta G_{mn,kl}({\bf k};{\bf q},\Omega)
\mathbf{f}^{\mathbf{l}n}_{kl}({\bf k};{\bf q},\Omega) \, .
\eeq
A set of properties for the eigentensors reflected in Eqns. (\ref{basis},\ref{lr})
enable us to form the basis  for the representation of the solution $\mathbf{\Phi}$ as a series over the states $n$:\cite{Barabanenkov1991,Barabanenkov1995,Stark1997,Berman2000}
\begin{eqnarray}\label{solution}
\Phi_{ij,kl}=\sum\limits_{n}\frac{\mathbf{f}^{\mathbf{r}n}_{ij}({\bf k};{\bf q},\Omega)\mathbf{f}^{\mathbf{r}n}_{kl}({\bf k}^{\prime};{\bf q},\Omega)}{\lambda_{n}\left(\mathbf{q},\Omega\right)}
\end{eqnarray}
The concept of diffusion assumes that in the limit $\mathbf{q}\rightarrow0$, $\Omega\rightarrow0$ the function $\mathbf{\Phi}$ has a pole structure, dictating the lowest eigenvalue asymptotics $\lambda_{0}\left(\mathbf{q}\rightarrow0,\Omega\rightarrow0\right)\rightarrow0$, and  being separated from a regular part\cite{Barabanenkov1991,Barabanenkov1995,Berman2000}. Therefore, the whole problem is reduced to the determination of coefficients of perturbative expansion for $\lambda_{0}\left(\mathbf{q},\Omega\right)$ with regard to $\mathbf{q}$ and $\Omega$ up to the second and the first order respectively, taken around the point $\mathbf{q}=0$, $\Omega=0$. To do this, Eqn. (\ref{Hom}) has to be treated perturbatively, with the condition that Eqns. (\ref{WLWTI},\ref{symmetry}) hold at every order of the perturbation in $\mathbf{q}$, and $\Omega$\cite{Barabanenkov1991,Barabanenkov1995,Berman2000}.

\subsection{ {Perturbation approach to the eigenvalue problem}}
 {The solution to Eq. (\ref{Hom}) is developed in a successive approximation scheme, for small $\Omega$ and  small $\mathbf{q}$:}
\begin{eqnarray}\label{Pseries}
\mathbf{H}({\bf k},{\bf k}^{\prime\prime};{\bf q},\Omega) & = &
 \mathbf{H}({\bf k},{\bf k}^{\prime\prime};{\bf 0},0)+\mathbf{H}^{1\Omega}({\bf k},{\bf k}^{\prime\prime};{\bf 0},\Omega) \nonumber \\
& & \hspace{-2em} +\mathbf{H}^{1\mathbf{q}}({\bf k},{\bf k}^{\prime\prime};{\bf q},0)+\mathbf{H}^{2\mathbf{q}}({\bf k},{\bf k}^{\prime\prime};{\bf q},0) + \dots \, , \nonumber \\
& & \nonumber \\
\label{Pseries2}
\mathbf{f}^{\mathbf{r}0}({\bf k}^{\prime\prime};{\bf q},\Omega) & = &
\mathbf{f}({\bf k}^{\prime\prime};{\bf 0},0)+\mathbf{f}^{1\Omega}({\bf k}^{\prime\prime};{\bf 0},\Omega) \\
& &\hspace{-2em} +\mathbf{f}^{1\mathbf{q}}({\bf k}^{\prime\prime};{\bf q},0)+\mathbf{f}^{2\mathbf{q}}({\bf k}^{\prime\prime};{\bf q},0) + \dots \nonumber \\
& & \nonumber \\
\lambda_{0}\left(\mathbf{q},\Omega\right) & = &
\lambda^{1\Omega}\left(\mathbf{0},\Omega\right)+\lambda^{1\mathbf{q}}\left(\mathbf{q},0\right)+\lambda^{2\mathbf{q}}\left(\mathbf{q},0\right) + \dots \nonumber
\label{lambda}
\end{eqnarray}
 {and, b}y deploying the perturbative scheme in detail ({See Appendix \ref{pert}}) the following set of coupled integral equations is obtained
\begin{eqnarray}\label{system0}
\int\limits_{\bf{k}^{\prime\prime}}H_{ij,kl}({\bf k},{\bf k}^{\prime\prime})\mathbf{f}_{kl}({\bf k}^{\prime\prime})& = & 0 \\
\label{Omega}
\int\limits_{\bf{k}^{\prime\prime}}\left(H_{ij,kl}({\bf k},{\bf k}^{\prime\prime})\mathbf{f}^{1\Omega}_{kl}({\bf k}^{\prime\prime}) \right. \hspace{4em} & & \nonumber \\
+ \left. H_{ij,kl}^{1\Omega}({\bf k},{\bf k}^{\prime\prime})\mathbf{f}_{kl}({\bf k}^{\prime\prime})\right) & = &
\lambda^{1\Omega}\mathbf{f}_{ij}({\bf k}) 
\label{system1q}
\end{eqnarray}
\begin{eqnarray}
\int\limits_{\bf{k}^{\prime\prime}}\left(H_{ij,kl}({\bf k},{\bf k}^{\prime\prime})\mathbf{f}^{1\mathbf{q}}_{kl}({\bf k}^{\prime\prime}) \right. \hspace{4em} & &\nonumber \\
\left. +H^{1\mathbf{q}}_{ij,kl}({\bf k},{\bf k}^{\prime\prime})\mathbf{f}_{kl}({\bf k}^{\prime\prime})\right) & = & 0 \\
\label{system2q}
\int\limits_{\bf{k}}{\cal B} P_{ii,kl}({\bf k};{\bf q})\mathbf{f}^{1\mathbf{q}}_{kl}({\bf k}) & = &
\lambda^{2\mathbf{q}}
\end{eqnarray}
where  the arguments $\mathbf{q}$ and $\Omega$ have been omitted.
As shown in Appendix \ref{pert}, the first-order-in-wavenumber contribution to the eigenvalue vanishes:
\beq
\lambda^{1\mathbf{q}} = 0.
\eeq
This result ensures the existence of a diffusion regime for the problem at hand.

Using Eqs. (\ref{WLWTI}) and (\ref{system0}), the eigentensor $\mathbf{f}^{\mathbf{r}0}$ at $\mathbf{q}=0$, $\Omega=0$ is found to be
\beq\label{zeroorder}
f_{ij}({\bf k}^{\prime\prime})={\cal{B}}{\Delta G_{ij,kk}({\bf k}^{\prime\prime})}
\eeq
with
\beq
{\cal{B}}^{-2}=  \int\limits_{\bf{v}}{\Delta G_{jj,kk}({\bf v})}  \, .
\label{eq:B}
\eeq
Integrating Eq. (\ref{Omega}) over ${\bf k}$ and using  the WTI, Eq. (\ref{WLWTI}), at the corresponding order,  the eigenvalue $\lambda^{1\Omega}$ is obtained:
\beq
\label{firstomega}
\lambda^{1\Omega} = i\omega\Omega\left(1+a\right)
\eeq
with
\begin{multline}
a=\frac{1}{\int\limits_{\bf{k}}f_{ss}({\bf k})} \\ 
\times \int\limits_{\bf{k}^{\prime\prime}} \frac{\left(A_{ii,kl}(\mathbf{k}^{\prime\prime};\mathbf{0},\Omega)\left(g(\omega_{+})-g^{\ast}(\omega_{-})\right)\right)^{1\Omega}}{2\omega\Omega}f_{kl}({\bf k}^{\prime\prime}) \, .
\label{aparameter}
\end{multline}

A similar parameter  appears in the diffusion of light and, since it is positive, it renormalizes the phase velocity to a  value that is smaller than the transport velocity}
\cite{Barabanenkov1991,Barabanenkov1995,Tiggelen1993,Livdan1996}. To see that our $a$ is indeed positive, replace  Eqs. (\ref{WLUtensor}) and (\ref{zeroorder}) into Eq. (\ref{aparameter}) to obtain
\bea
\label{aparaprox}
a  &=&
\frac{-\int\limits_{\bf{k}}
Im[{\Sigma_{mn}(\mathbf{k})G_{mn}}({\bf k})]}{\rho^{2}\left(\omega_{r1}^{2}-\omega^{2}\right)\int\limits_{\bf{v}}{\Delta G_{ii,jj}({\bf v})}}\\
&\approx& \frac{2 R_{2T}^{3/2} \left( c_T^2 R_{2T} - \om^2\right) + R_{2L}^{3/2} \left( c_L^2 R_{2L} - \om^2\right) }{ \left( \omega_{F}^{2} -\om^2\right) \left(2R_{2T}^{3/2} +R_{2L}^{3/2} \right) } 
\label{defaaa}
\eea
where $R_{2L,T} \equiv Re [K_{L,T}^2]$ and $I_{2L,T} \equiv Im[K_{L,T}^2]$. 
The last approximation is obtained in the limit of small $Im[K_{T,L}^{2}]$, as explained in Appendix \ref{esti}. {Clearly $a>0$ for wave frequencies $\omega$ smaller that the first fundamental mode of the vibrating string-like dislocation $\omega<\omega_{F}$.}

\subsection{Diffusion constant}
From Eqs. (\ref{solution}-\ref{system2q}), the following leading order expression  for the singular part of the intensity, $\mathbf{\Phi}^{sing}$ is obtained:
\bea\label{sing}
\Phi^{sing}_{ij,kl}&=&\frac{f^{\mathbf{r0}}_{ij}({\bf k};{\bf q},\Omega)f^{\mathbf{r0}}_{kl}({\bf k}^{\prime};{\bf q},\Omega)}{\lambda^{1\Omega}+\lambda^{2\mathbf{q}}} \nonumber \\
&=&\frac{f^{\mathbf{r0}}_{ij}({\bf k};{\bf q},\Omega)f^{\mathbf{r0}}_{kl}({\bf k}^{\prime};{\bf q},\Omega)}{\frac{\lambda^{1\Omega}}{-i\Omega}\left(-i\Omega+\frac{-i\Omega\lambda^{2\mathbf{q}}}{\lambda^{1\Omega}q^{2}}q^{2}\right)} \, .
\eea
Then, using Eqs. (\ref{firstomega},\ref{sing}) the diffusion constant can be simply read off. It is
\begin{eqnarray}\label{Dconstant}
D &\equiv & -\frac{i\Omega\lambda^{2\mathbf{q}}}{q^2\lambda^{1\Omega}}  \\
&\equiv & D^{\mathcal{R}} + D_{\Delta G^{1\mathbf{q}}}
\label{Dconstant3}
\end{eqnarray}
with
\begin{align}
\label{DR}
D^{\mathcal{R}}  \equiv &  \frac{{\cal{B}}^{2}}{q^{2}\omega\left(1+a\right)}
\int\limits_{\bf{k}}P_{ss,kl}({\bf k};{\bf q}) \\
 & \hspace{2em} \times\int\limits_{{\bf k}_{2}}\Phi_{kl,ij}({\bf k},{\bf k}_{2})P_{ij,tt}(\mathbf{q};{\bf k}_{2}) \, ,  \nonumber  \\
D_{\Delta G^{1\mathbf{q}}} \equiv & -     \frac{{\cal{B}}^{2}}{q^{2}\omega\left(1+a\right)}
\int\limits_{\bf{k}}P_{ss,kl}({\bf k};{\bf q}) \Delta G_{kl,tt}^{1\mathbf{q}}({\bf k})
\label{Dconstant2}
\end{align}
To obtain Eq. (\ref{Dconstant3}), in which the diffusion constant is written as the sum of two terms, we have substituted  the values for $\lambda^{2\mathbf{q}}$, $\lambda^{1\Omega}$ given by Eqs. (\ref{system2q}) and (\ref{firstomega}). The first one ensued from the form of $f^{1\mathbf{q}}({\bf k})$ ({See Appendix \ref{solu}}).
Thus, the expression for the diffusion constant in Eq. (\ref{Dconstant3}) is the sum of two contributions, as defined in (\ref{DR}) and (\ref{Dconstant2}). The computation, sketched in Appendix \ref{esti}  is laborious but a fairly straightforward generalization of a similar computation carried out in two dimension for elastic waves diffusing among many edge dislocations\cite{Churochkin2021}. The result is,  {the limit of small} $Im[K_{T,L}^{2}]$
\beq
\label{DCFintotleadexpl}
D^{lead}\approx
\frac{1}{(1+a)} \frac{\left( c_L^4 \frac{R_{2L}^{7/2}}{I_{2L}} + 2 c_T^4 \frac{R_{2T}^{7/2}}{I_{2T}}\right)}{3\om^3\left(2R_{2T}^{3/2}
+ R_{2L}^{3/2} \right) }
\eeq
with $a$ given by (\ref{defaaa}). In the limit of small frequencies this becomes
\beq\label{DCFintotleadexpllf}
D^{lead}_{\omega\rightarrow 0}\approx
\left(\frac{v_{T}^{3}c_{L}^{4}}{\left(2v_{L}^{3}+v_{T}^{3}\right)v_{L}^{4}}\frac{v_{L}l_{L}}{3}+\frac{2v_{L}^{3}c_{T}^{4}}{\left(2v_{L}^{3}+v_{T}^{3}\right)v_{T}^{4}}\frac{v_{T}l_{T}}{3}\right)
\eeq
where $v_{T,L}$ and $l_{T,L}$ are the effectve velocities and attenuation lengths introduced in Section \ref{sectionpreviousresults}, Eqs. (\ref{eq:effvel}) and (\ref{eq:attenlength}).

\section{Discussion}
\label{sec:relation}
The main result of this paper is expression (\ref{DCFintotleadexpl}) for the diffusion coefficient for elastic waves travelling in a continuum elastic medium populated with many, randomly placed and oriented, dislocation segments, and the simpler expression (\ref{DCFintotleadexpllf}), its value in the limit of low frequencies. It is valid (see below) for frequencies that are not too close to the fundamental string frequency $\om_F$. It is the sum of two terms, each one characterized by an attenuation length that appears because the imaginary part of the effective wave vector $Im[K_{T}^{2}]$ does not vanish. It has an overall factor $(1+a)$ with $a$ given by (\ref{aparaprox}). {Similar factors have been identified in the diffusion of sound in a layer with a rough interface\cite{Berman2000}, and of light waves in media with microstructure\cite{Livdan1996}}, in association with resonant scattering, as here. Indeed, if $\om = \om_F$, the diffusion coefficient vanishes. 
Having a frequency exactly equal to $\om_F$, however, takes us outside the domain of validity of the approximations employed in this work. {In any case}, it is allowed for the frequency to approach the resonant frequency, and the associated diffusion constant does get smaller.
This raises the question of looking mode closely at this regime (see below). The aforementioned models\cite{Livdan1996,Berman2000} also allow the possibility of an additional factor ``$(1+ \Delta)$'', associated with the extended nature of the scatterers present. Our formalism allows for the presence of this factor as well, it appears in Eq. (\ref{angularproof}). In our specific example, however, the analog of ``$\Delta$'' vanishes because we have taken scatterers that are effectively point-like.

\subsection{Restrictions placed by approximations employed}
\subsubsection{Long wavelength by comparison with dislocation segment length}
At the outset, in Section \ref{sectionpreviousresults}, we have formulated the wave-dislocation interaction problem in an approximation in which the whole interaction takes place at a single point, the dislocation center, although the specific interaction (\ref{potential}) does contain the information that the dislocation segment is a vibrating string of length $L$, with a specific eigenfrequency, at which a resonant interaction may occur.

\subsubsection{$| Im[K_{T,L}^{2}] | \ll | k^{2}-Re[K_{T,L}^{2}] |$}
\label{discone}
This approximation has been repeatedly used in the algebra, with $K_{T,L}$ the transverse ($T$) and longitudinal ($L$) effective wave vectors (\ref{poles}) characterizing the coherent propagation of waves. Using (\ref{poles}) for $B=0$, the case with no internal losses for which we have carried out the computations in the ISA, the inequality of this sub-section translates into
\beq
|\om^2 -\om_F^2| \gg \frac{1}{{\pi^2}} \frac{\om^3}{\om_F}
\eeq
so that the working frequency $\om$ can be close, but not equal to, the resonant string frequency $\om_F$.

\subsubsection{Independent scattering approximation (ISA)}
The ISA means that the random variables characterizing the dislocation segments, position and orientation, are statistically independent. It simplifies the computation of statistical averages, keeping only leading order terms in $n$, the number of dislocation segments per unit volume, in Eqs. (\ref{WTIisaexpressions}). In order to have a rough estimate of what this means in terms of dimensionless variables, consider the value of the $t$ matrix at low frequencies\cite{Churochkin2016}, and the following inequality results: $nL^3 \ll 1$. That is, the separation among dislocation segments must be larger that their length.

\subsection{Kubo representation for the diffusion constant}
\label{Kubo}
{We have obtained an explicit form for the diffusion constant of elastic wave energy when traveling through an elastic medium full of vibrating dislocation segments by use of a perturbation approach to the solution of the BS equation, regarded as an eigenvalue problem. In this subsection, we will show that the diffusion constant,
given by Eqn. (\ref{Dconstant}), admits a Kubo representation similar to that for diffusion of electromagnetic waves\cite{Barabanenkov1995}.

To achieve a Kubo representation for the diffusion constant we have to focus on the transformation of the $\Delta G^{1\mathbf{q}}_{kl,mm}({\bf k})$ from Eqn. (\ref{ApSoldeltaG}). According to \cite{Barabanenkov1995}, this implies, firstly, the construction of the equation similar to Eqn. (\ref{preWTI}), but for 
\begin{multline}\label{intensity--}
\mathbf{\Phi}^{--}({\bf k},{\bf k}^{\prime};{\bf q},\Omega) \equiv \Phi^{--}_{kl,mn}({\bf k},{\bf k}^{\prime};{\bf q},\Omega)  \\ \equiv <G^{-}_{km}(\mathbf{k}^{+},\mathbf{k}^{\prime +},\omega^{-})G^{-}_{nl}(\mathbf{k}^{\prime-},\mathbf{k}^{-},\omega^{-})>
\end{multline}
\begin{widetext}
It should be noted that the subtraction trick, briefly mentioned in \ref{sec:pre-WTI}, is rather general and  can be implemented without loss of generality to get the equation for $\mathbf{\Phi}^{--}({\bf k},{\bf k}^{\prime};{\bf q},\Omega)$. Passing through similar steps one can obtain
\begin{multline}\label{preWTI--}
{\int\limits_{\mathbf{k}}}\left((G^{0-})^{-1}_{ni}(\mb k^{-},\omega_{-})-(G^{0-})^{-1}_{ni}(\mb k^{+},\omega_{-})\right)\Phi^{--}_{ni,lj}({\bf k},{\bf k}^{\prime\prime\prime};{\bf q},\Omega) \\ +<G^{-}(\mb k^{\prime\prime\prime-},\mb k^{\prime\prime\prime-} ;\omega_{-})_{lj}>-<G^{-}(\mb k^{\prime\prime\prime+},\mb k^{\prime\prime\prime+} ;\omega_{-})_{lj}>\equiv 0
\end{multline}
or at $\Omega\rightarrow 0$
\begin{equation}\label{preWTI--II}
{\int\limits_{\mathbf{k}}}\frac{\partial L_{ni}({\bf k})}{\partial {\bf k}}\cdot\mathbf{q} Re[\Phi^{--}_{ni,lj}({\bf k},{\bf k}^{\prime\prime\prime};{\bf q},0)] \equiv Re[-<G^{-}(\mb k^{\prime\prime\prime-},\mb k^{\prime\prime\prime-} ;\omega)_{lj}>+<G^{-}(\mb k^{\prime\prime\prime+},\mb k^{\prime\prime\prime+} ;\omega)_{lj}>]
\end{equation}
and, to first order in ${\bf q}$, the identity reduces to
\beq\label{preWTIIorder}
\frac{\imath}{\rho}{\int\limits_{\mathbf{k}}}\frac{\partial L_{ni}({\bf k})}{2\partial {\bf k}}\cdot\mathbf{q} Re[\Phi^{--}_{ni,lj}({\bf k},{\bf k}^{\prime\prime\prime};{\bf 0},0)]\equiv\Delta G^{1\mathbf{q}}_{lj,mm}({\bf k}^{\prime\prime\prime})
\eeq
Hence,
we get for $\mathbf{f}^{1\mathbf{q}}_{kl}({\bf k}^{\prime\prime})$
\beq
\label{ApKR1q2}
\mathbf{f}^{1\mathbf{q}}_{kl}({\bf k}^{\prime\prime})=-\frac{\imath B}{\rho}\int\limits_{{\bf k}_{2}}\left(<G^{+}_{kk_{1}}({\bf k}^{\prime\prime},\mathbf{k}_{2};\omega)G^{-}_{l_{1}l}(\mathbf{k}_{2},{\bf k}^{\prime\prime};\omega)>-
Re[<G^{-}_{kk_{1}}({\bf k}^{\prime\prime},\mathbf{k}_{2};\omega)G^{-}_{l_{1}l}(\mathbf{k}_{2},{\bf k}^{\prime\prime};\omega)>]\right)\frac{\partial L_{k_{1}l_{1}}({\bf k}_{2})}{2\partial {\bf k}_{2}}\cdot\mathbf{q} \, .
\eeq
In Eqn. (\ref{ApKR1q2}) we deal with the difference of products of  complex numbers that may be symbolically presented  in the form
\beq
\label{ApKRcomp}
X^{*}_{kk_{1}}X_{l_{1}l}-Re[X_{kk_{1}}X_{l_{1}l}]=2Im[X]_{kk_{1}}Im[X]_{l_{1}l}+\imath\left(Re[X]_{kk_{1}}Im[X]_{l_{1}l}-Im[X]_{kk_{1}}Re[X]_{l_{1}l}\right)
\eeq
Using (\ref{ApKR1q2}), (\ref{ApKRcomp}) and (\ref{system2q}) we get
\beq
\label{ApKRlambda2q}
\lambda^{2\mathbf{q}} = \frac{2 B^{2}}{\rho^{2}}\int\limits_{\bf{k}}\int\limits_{{\bf k}_{2}} \mathbf{q}\cdot\frac{\partial L_{kl}({\bf k})}{2\partial {\bf k}}\\<Im[G^{-}_{kk_{1}}({\bf k},\mathbf{k}_{2};\omega)]Im[G^{-}_{l_{1}l}(\mathbf{k}_{2},{\bf k};\omega)]> \frac{\partial L_{k_{1}l_{1}}({\bf k}_{2})}{2\partial {\bf k}_{2}}\cdot\mathbf{q}
\eeq
so that. from (\ref{firstomega}), (\ref{Dconstant}) and (\ref{ApKRlambda2q}) the diffusion constant reads
\beq
\label{ApKRD}
D=\frac{-2 B^{2}}{\rho^{2}q^{2}\omega\left(1+a\right)}\\
\int\limits_{\bf{k}}\int\limits_{{\bf k}_{2}} \mathbf{q}\cdot\frac{\partial L_{kl}({\bf k})}{2\partial {\bf k}}<Im[G^{-}_{kk_{1}}({\bf k},\mathbf{k}_{2};\omega)]Im[G^{-}_{l_{1}l}(\mathbf{k}_{2},{\bf k};\omega)]> \frac{\partial L_{k_{1}l_{1}}({\bf k}_{2})}{2\partial {\bf k}_{2}}\cdot\mathbf{q}
\eeq
\end{widetext}
which is the desired Kubo representation.

{
\subsection{Transport equation approach and equipartition of energy}
\label{transport}
Ryzhyk et al.\cite{Ryzhik1996} have studied the transport of elastic energy density in a random medium. They showed that diffusive behavior occurs on long time and distance scales,  and  they have determined a diffusion coefficient. They, however, dealt with continuous random media and not, as in our case, with discrete scatterers that are randomly distributed in a medium. It is still of interest to compare our result (\ref{DCFintotleadexpllf}) with the value they give for the diffusion constant, which is their Eqn. (5.46) (in their notation):
\beq
D^{el} = \frac{1}{(2/v_S^3 + 1/v_P^3)} \left( \frac{l_P v_P}{3v_P^3} + \frac{2l_S v_S}{3v_S^3} \right) .
\eeq
Here ``$P$'' means ``primary'', or longitudinal ($L$) in our language, and ``$S$'' means ``secondary'', or transverse ($T$) in our case. The quantities $l_P$ and $l_S$ are longitudinal and transverse mean-free-paths that are determined by unspecified scattering cross sections.  We find there is a strong resemblance to (\ref{DCFintotleadexpllf}). One important difference, however, is that (\ref{DCFintotleadexpllf}), based as it is on a solution to the BS equation, involves not one phase velocity for each polarization, but two: the velocity in the absence of scatterers, and the velocity of coherent waves in the presence of scatterers. The latter quantity appears because of the relation between mass operator and irreducible kernel provided by the WTI. These considerations are absent in a transport equation approach. Both approaches coincide, however, in the limit of a very small density of dislocations, in which case $v_{L,T} \approx c_{L,T}$.
}

{Ryzhyk et al.\cite{Ryzhik1996} also noted that, in their diffusive limit, the energy of elastic waves is ``equipartitioned'', in the sense that, if ${\cal E}_L$ (resp. ${\cal E}_T$) is the longitudinal (resp. transverse) energy density so  that the total energy ${\cal E} = {\cal E}_T +{\cal E}_L$, then
\beq
\frac{{\cal E}_T}{{\cal E}_L} = 2 \gamma^3 \, .
\label{epartition}
\eeq
Earlier, Weaver\cite{Weaver1982} had obtained this result taking as the definition of the diffuse field a state in which energy is equipartitioned among all normal modes available to the elastic solid, and using the Debye density of states to compute the ratio between longitudinal and transverse modes. 

{In our formulation, the diffuse field energy tensor is defined by
\beq
\label{ed}
{\cal E} (\mathbf{q}, \Omega)_{ij,kl} = \lim\limits_{{\bf q}\rightarrow {\bf 0},\Omega\rightarrow 0}\int\limits_{\bf{k}}\int\limits_{\bf{k'}} \Phi_{ij,kl}({\bf k},{\bf k}^{\prime};{\bf q},\Omega) \, .
\eeq
It is a straightforward calculation, using the solution (\ref{solution}) to lowest order, Eqns. (\ref{Pseries2}) and (\ref{zeroorder}), to show that
\begin{align}
\label{diffener}
{\cal E} (\mathbf{0}, \Omega)_{ij,kl} &= \frac{i}{\Omega} \frac{\delta_{ij}\delta_{kl}}{36\pi\rho^{2}\omega^{3}} \frac{\left[2Re[K_{T}^{2}]^{3/2}+Re[K_{L}^{2}]^{3/2}\right]}{\left(1+a\right)}   \\
& \longrightarrow \frac{i}{\Omega} \frac{\delta_{ij}\delta_{kl}}{36\pi\rho^{2}} \left( \frac{2}{c_T^3} +\frac{1}{c_L^3} \right)
\end{align}
where the last limit is obtained when the density of dislocations is very small. Note that, in general, the diffuse energy density does not split into a sum of longitudinal and transverse terms, because of the $(1+a)$ denominator which, as we have discussed, is a consequence of the time scale introduced into the problem by the fundamental mode of the vibrating strings that are doing the scattering of the elastic waves.
}

{Additional insight into these results can be obtained noting that, using the result (\ref{green}) for the coherent Green's function, it is straightforward to verify that, in the limit $| Im[K_{T,L}^{2}] | \ll | k^{2}-Re[K_{T,L}^{2}] |$ already discussed in previous sections, 
\begin{align}
\label{APTrImgreenExp}
Tr[&Im[\langle \mathbf{G} \rangle^{+}({\bf k},\omega)]] =-\Delta G_{ii,mm}({\bf k},\omega)  \nonumber \\
&\approx
\frac{\pi k^2}{\rho \omega^{2}}\left(2\delta\left(k^2-Re[K_{T}^{2}]\right)+\delta\left(k^2-Re[K_{L}^{2}]\right)\right)
\end{align}
}

{Now, if we consider the diffusive energy as being carried by the coherent waves whose states are labelled by three polarizations and three real numbers, the components of a wave vector $\mb k$, we see that
\beq
\label{DS}
g_{T,L}(Re[K_{T,L}^{2}])=\sum \limits_{{\bf k}}\delta\left(k^2-Re[K_{T,L}^{2}]\right)
\eeq
counts the number of states that have the same $Re[K_{T,L}^{2}]$, and 
\begin{align}
\label{DSfriq}
g_{T,L}(\omega)=&\frac{1}{V} g_{T,L}(Re[K_{T,L}^{2}])\frac{\partial Re[K_{T,L}^{2}]}{\partial \omega} \nonumber\\
=&\frac{1}{6\pi^{2}}\frac{\partial (Re[K_{T,L}^{2}]^{3/2})}{\partial \omega}
\end{align}
is the density of states per unit frequency $\om$ and unit volume $V$. The second equality follows from (\ref{DS}). The ratio of transverse states to longitudinal states is then
\beq
\label{equipart}
\frac{2g_{T}(\omega)}{g_{L}(\omega)}=\frac{2\frac{\partial (Re[K_{T}^{2}]^{3/2})}{\partial \omega}}{\frac{\partial (Re[K_{L}^{2}]^{3/2})}{\partial \omega}} \longrightarrow 2 \gamma^3 \, .
\eeq
where the limiting behavior is obtained for a small density of dislocations. We see that, in general, diffuse energy density, given by Eqn. (\ref{diffener}), at a given frequency is not proportional to the density of states at that same frequency, given by Eqn. (\ref{DSfriq}). However, said proportionality (``equipartition'') is recovered in the limit of very few dislocations.
}

\section{Conclusions and outlook}
\label{sec:disc}
We have studied the diffusive behavior of elastic waves in a continuum that is populated by many edge-dislocation segments of length $L$, pinned at their ends. Their position is random, as well as the orientation of their tangent and Burgers vectors. The dislocations are modeled as elastic strings with internal losses, and are dynamical objects in their own right.
The study relies heavily on the existence of a regime where coherent wave behavior occurs, previously studied\cite{Churochkin2016}. The elastic waves are assumed to be monochromatic, with a frequency that is small compared to the first resonant frequency of the string-like pinned dislocations and computations are actually carried out in an independent scattering approximation, that is when the random variables, position and orientation, characterizing the dislocations, are statistically independent. In this case the coherent wave has an effective velocity and an attenuation that are, to leading order, proportional to the number $n$ of dislocation segments per unit volume, the small dimensionless parameter being $nL^3$.

The diffusion behavior is studied using a Bethe-Salpeter equation, supplemented by a Ward-Takahashi identity. Both equations hold in the presence of internal losses by the strings. However, in order to use the ISA, a necessary requirement for the actual computation of a diffusion coefficient, it is necessary to assume that these losses vanish. If this were not the case, the diffusive behavior would be influenced not only by the incoherent diffusion induced by the disordered dislocation segments, but also by a decay induced by the internal losses. It should be of interest to explore this regime, especially in view of the possible experimental measurements of the diffusion reported here.

Alternatively, one may ask about the origin of the internal losses. If they are due to inelastic scattering of the dislocation with phonons, a complete calculation of the phonon-dislocation interaction has been recently carried out \cite{Lund2019}, for phonons of arbitrary frequency. That is, without the requirement that their wavelength be long compared to dislocation length $L$. It should be of interest then to explore a BS equation, and attendant WTI, in this case, since the inelastic effects would be explicitly considered from the very beginning.

A study of the diffusion problem without the restriction of dislocation lengths small compared to wavelength would have the added benefit to clarify the role played by the vibrating string resonances. As it was indicated in the previous section, the diffusion constant that has been computed in the present work can, formally, vanish when the wave frequency coincides with the resonant frequency. {A similarly strong effect that resonances can have upon the diffusion of light has been considered by Lubatsch et al. \cite{Lubatsch2005}} This regime is outside the frame of approximations employed to carry out our computations however, and it would be of interest, in future, to explore in some detail the actual behavior of the diffusion coefficient for frequencies comparable to the resonant string frequency.

The continuum mechanics approach employed in the present work has the advantage of being applicable to any homogeneous solid material at all length scales down to several interatomic spacings. This is true even of the atomic structure does not have long range order, and it has been established\cite{Bianchi2020} that the coherent wave behavior already alluded to provides an adequate understanding of the behavior of amorphous materials in the THz range. Recently, Beltukov et al.\cite{Beltukov2018} have performed a numerical study of wave packet behavior in amorphous silicon, and have detected a transition from propagating to diffusive regimes, depending on the frequency of the waves. This phenomenology is relevant to the understanding of heat transport in amorphous solids, one of the significant unknowns in contemporary condensed-matter physics, and it looks tempting to apply the methods presented in this paper to try and elucidate the nature of heat propagation in glasses.

\acknowledgements
This work was supported in part by Fondecyt Grant 1191179.

\appendix
\section{Bethe-Salpeter Equation}
\label{BSE}
The key idea of the Bethe-Salpeter (BS) equation is the existence of an analogy of the Dyson equation for the intensity $<\mathbf{G^{+}}\otimes\mathbf{G^{-}}>$. In order to get it explicitly we use the representations
\begin{widetext}
\begin{eqnarray}\label{ApBSIntensity}
<\mathbf{G^{+}}\otimes\mathbf{G^{-}}> & = & <\mathbf{G^{+}}>\otimes<\mathbf{G^{-}}>+
\left(<\mathbf{G^{+}}\otimes\mathbf{G^{-}}>-<\mathbf{G^{+}}>\otimes<\mathbf{G^{-}}>\right)  \nonumber\\
 & = & <\mathbf{G^{+}}>\otimes<\mathbf{G^{-}}> \nonumber  \\
  & & +
<\mathbf{G^{+}}>\otimes<\mathbf{G^{-}}>:<\mathbf{G^{+}}>^{-1}\otimes<\mathbf{G^{-}}>^{-1} \nonumber \\
& & :
\left(<\mathbf{G^{+}}\otimes\mathbf{G^{-}}>-<\mathbf{G^{+}}>\otimes<\mathbf{G^{-}}>\right):
<\mathbf{G^{+}}\otimes\mathbf{G^{-}}>^{-1}:<\mathbf{G^{+}}\otimes\mathbf{G^{-}}> \nonumber\\
& = & <\mathbf{G^{+}}>\otimes<\mathbf{G^{-}}> \\
& &  +<\mathbf{G^{+}}>\otimes<\mathbf{G^{-}}>
:\left(<\mathbf{G^{+}}>^{-1}\otimes<\mathbf{G^{-}}>^{-1}-<\mathbf{G^{+}}\otimes\mathbf{G^{-}}>^{-1}\right)
:<\mathbf{G^{+}}\otimes\mathbf{G^{-}}> \, . \nonumber
\end{eqnarray}
\end{widetext}
From the last equality in (\ref{ApBSIntensity}) it is easy to introduce the pole structure for the intensity by defining the irreducible vertex $\mathbf{K}$ as
\beq
\label{ApBSIrred}
\mathbf{K}=<\mathbf{G^{+}}>^{-1}\otimes<\mathbf{G^{-}}>^{-1}-<\mathbf{G^{+}}\otimes\mathbf{G^{-}}>^{-1} \, .
\eeq
The BS equation in the form (\ref{ApBSIrred}) clearly corroborates the pole specificity of the intensity $<\mathbf{G^{+}}\otimes\mathbf{G^{-}}>$  in the sense that the $\mathbf{K}$ plays the same role as the self-energy $\mathbf{\Sigma}$ for both averaged $<\mathbf{G}>$ and free medium $\mathbf{G}_{0}$ Green's tensors in the Dyson equation
\begin{eqnarray}\label{ApBSDyson}
\mathbf{\Sigma}=\mathbf{G}_{0}^{-1}-<\mathbf{G}>^{-1} \, .
\end{eqnarray}
Replacing  (\ref{ApBSIrred}) into the last equality of (\ref{ApBSIntensity}) the BS equation takes the widely accepted form
\begin{eqnarray}\label{ApBSIntensityst}
<\mathbf{G^{+}}\otimes\mathbf{G^{-}}> & = &
<\mathbf{G^{+}}>\otimes<\mathbf{G^{-}}> \\
 & &  \hspace{-2em}+ <\mathbf{G^{+}}>\otimes<\mathbf{G^{-}}>:\mathbf{K}:<\mathbf{G^{+}}\otimes\mathbf{G^{-}}> \, . \nonumber
\end{eqnarray}

If we now define the Fourier transforms as\cite{Sheng2006}
\begin{eqnarray}
G_{i_{1}m_{1}}^{+}(\mathbf{x}_{1},\mathbf{x}^{\prime}_{1};\omega^{+}) & = &\int\limits_{\mathbf{k}_{1}} \int\limits_{\mathbf{k}^{\prime}_{1}}e^{\imath \mathbf{k}_{1}\mathbf{x}_{1}} G_{i_{1}m_{1}}^{+}(\mathbf{k}_{1},\mathbf{k}^{\prime}_{1};\omega^{+})e^{-\imath \mathbf{k}^{\prime}_{1}\mathbf{x}^{\prime}_{1}} \nonumber \\
G_{i_{2}m_{2}}^{-}(\mathbf{x}_{2},\mathbf{x}^{\prime}_{2};\omega^{-}) & = & \int\limits_{\mathbf{k}_{2}} \int\limits_{\mathbf{k}^{\prime}_{2}}e^{-\imath \mathbf{k}_{2}\mathbf{x}_{2}} G^{-}_{i_{2}m_{2}}(\mathbf{k}^{\prime}_{2},\mathbf{k}_{2};\omega^{-})e^{\imath \mathbf{k}^{\prime}_{2}\mathbf{x}^{\prime}_{2}} \, , \nonumber \\
 & &
\end{eqnarray}
then,
\beq
\label{ApBSft}
<\mathbf{G^{+}}\otimes\mathbf{G^{-}}>=\int\limits_{\mathbf{k}} \int\limits_{\mathbf{k}^{\prime}}\int\limits_{\mathbf{q}}\mathbf{\Phi}({\bf k},{\bf k}^{\prime};{\bf q},\Omega)
e^{\imath\left(\mathbf{k}\mathbf{r}-\mathbf{k}^{\prime}\mathbf{r}^{\prime}+\mathbf{q}\left(\mathbf{R}-\mathbf{R}^{\prime}\right)\right)}
\eeq
With space and momentum variables being specified as
\begin{align}\label{ApBSspace}
\mathbf{x}_{1}&=\mathbf{R}+\frac{\mathbf{r}}{2} \, ,  &
\mathbf{x}_{2}&=\mathbf{R}-\frac{\mathbf{r}}{2}\\
\mathbf{x}^{\prime}_{1}&=\mathbf{R}^{\prime}+\frac{\mathbf{r}^{\prime}}{2} \, ,& 
\mathbf{x}^{\prime}_{2}&=\mathbf{R}^{\prime}-\frac{\mathbf{r}^{\prime}}{2}\nonumber\\
\mathbf{k}_{1}=\mathbf{k}^{+}&=\mathbf{k}+\frac{\mathbf{q}}{2},   & 
\mathbf{k}^{\prime}_{1}=\mathbf{k}^{\prime+}&=\mathbf{k}^{\prime}+\frac{\mathbf{q}}{2}\nonumber\\
\mathbf{k}^{\prime}_{2}=\mathbf{k}^{\prime-}&=\mathbf{k}^{\prime}-\frac{\mathbf{q}}{2}, & \quad\mathbf{k}_{2}=\mathbf{k}^{-}&=\mathbf{k}-\frac{\mathbf{q}}{2} \, .\nonumber
\end{align}
And applying the inverse Fourier transform\cite{Stark1997}
\begin{eqnarray}\label{ApBSIft}
\int d\left(\mathbf{R}-\mathbf{R}^{\prime}\right)d\mathbf{r}d\mathbf{r}^{\prime}
e^{-\imath\left(\mathbf{k}\mathbf{r}-\mathbf{k}^{\prime}\mathbf{r}^{\prime}+\mathbf{q}\left(\mathbf{R}-\mathbf{R}^{\prime}\right)\right)} \end{eqnarray}
to Eqn. (\ref{ApBSIntensityst}) the BS equation in momentum space is obtained:
\begin{widetext}
\beq
\label{ApBSmom}
\mathbf{\Phi}({\bf k},{\bf k}^{\prime};{\bf q},\Omega)=
<\mathbf{G^{+}}>\otimes<\mathbf{G^{-}}>({\bf k};{\bf q},\Omega)\delta_{{\bf k},{\bf k}^{\prime}}+
<\mathbf{G^{+}}>\otimes<\mathbf{G^{-}}>({\bf k};{\bf q},\Omega):\mathbf{K}({\bf k},{\bf k}^{\prime\prime};{\bf q},\Omega):\mathbf{\Phi}({\bf k}^{\prime\prime},{\bf k}^{\prime};{\bf q},\Omega)
\eeq
where $\delta_{{\bf k},{\bf k}^{\prime}}=(2\pi)^{3}\delta({\bf k}-{\bf k}^{\prime})$ and the internal momentum variables, i.e. ${\bf k}^{\prime\prime}$, are integrated over.
To modify further Eqn. (\ref{ApBSmom}) to its kinetic form we use the following identity for the outer product of the averaged Green's tensors:
\beq
\label{ApBSidentity}
(<\mathbf{G^{+}}>^{-1}\otimes\mathbf{I}-\mathbf{I}\otimes<\mathbf{G^{-}}>^{-1}):<\mathbf{G^{+}}>\otimes<\mathbf{G^{-}}>=
\mathbf{I}\otimes<\mathbf{G^{-}}>-<\mathbf{G^{+}}>\otimes\mathbf{I}
\eeq
where $\mathbf{I}$ is a unit tensor. Acting from the left on Eqn. (\ref{ApBSmom}) with the tensor $(<\mathbf{G^{+}}>^{-1}\otimes\mathbf{I}-\mathbf{I}\otimes<\mathbf{G^{-}}>^{-1})$ and using the property (\ref{ApBSidentity}) the following relation is obtained:
\beq
\label{ApBSkinetic}
(<\mathbf{G^{+}}>^{-1}\otimes\mathbf{I}-\mathbf{I}\otimes<\mathbf{G^{-}}>^{-1}):\mathbf{\Phi}=
(\mathbf{I}\otimes<\mathbf{G^{-}}>-<\mathbf{G^{+}}>\otimes\mathbf{I}):
\left(
\mathbf{I}\otimes\mathbf{I}\delta_{{\bf k},{\bf k}^{\prime}}+
\mathbf{K}({\bf k},{\bf k}^{\prime\prime};{\bf q},\Omega):\mathbf{\Phi}({\bf k}^{\prime\prime},{\bf k}^{\prime};{\bf q},\Omega)\right)
\eeq
Finally, substituting (\ref{ApBSDyson}), (\ref{potBS}) and (\ref{delta}) into (\ref{ApBSkinetic}) as well as the explicit form of the Green's tensor for the  free medium \cite{Maurel2005b}
we obtain the BS equation in the form of  Eqn. (\ref{BS}).
\end{widetext}

\section{Integration over solid angles in 3D}
\label{Apaver}
The developed approach requires evaluation of the following integrals over a $n$-dimensional solid angle $\Omega^{n}$, comprised of the product of radial unit $n$-dimensional vectors $\hat{r}$ ($\hat{r}^{2}=1$)
\beq
I^{nk}=\int d\Omega^{(n)}_{\hat{r}}\hat{r}^{i_{1}}\cdots\hat{r}^{i_{k}}
\label{solangprodinttens}
\eeq
In a previous paper\cite{Churochkin2021} we were interested in the diffusion of waves in a two dimensional continuum. Now we have a problem in three dimensions and we are led to the evaluation of integrals
\bea
I^{32}=\int d\Omega^{(3)}_{\hat{\mb r}}\hat{r}^{i_{1}}\hat{r}^{i_{2}}  \nonumber \\
I^{34}=\int d\Omega^{(3)}_{\hat{\mb r}}\hat{r}^{i_{1}}\hat{r}^{i_{2}}\hat{r}^{i_{3}}\hat{r}^{i_{4}}
\label{solangprodinttenseval}
\eea
with $\Omega^{(3)}=4\pi$, $d\Omega^{(3)}_{\hat{r}}=\sin\theta d\theta d\phi$ and $\theta\in[0,\pi]$, $\phi\in[0,2\pi]$ are azimuthal and polar angles of a 3D spherical frame. In order to do this we use the results of \cite{Ee2017}, according to which the tensor integral of the product of $k$ radial unit $n$-dimensional vectors 
\beq
<\hat{r}^{i_{1}}\cdots\hat{r}^{i_{k}}>_{\hat{r}}=\frac{1}{\Omega^{(n)}}\int d\Omega^{(n)}_{\hat{\mb r}}\hat{r}^{i_{1}}\cdots\hat{r}^{i_{k}}=\frac{I^{nk}}{\Omega^{(n)}}
\label{prodint}
\eeq
vanishes when $k$ is odd, and is equal to a totally symmetric isotropic tensor when it is even
\beq
<\hat{r}^{i_{1}}\cdots\hat{r}^{i_{2k}}>_{\hat{r}}=\tilde{{\cal L}}^{i_{1}\cdots i_{2k}}_{(2k)}
\label{prodinttens}
\eeq
that is defined recursively,
\begin{multline}
\tilde{{\cal L}}^{i_{1}\cdots i_{2k}}_{(2k)}=\frac{1}{n+2k-2}
\left(\delta_{i_{1}i_{2}}\tilde{{\cal L}}^{i_{3}\cdots i_{2k}}_{(2k-2)} \right. \\ \left. +\delta_{i_{1}i_{3}}\tilde{{\cal L}}^{i_{2}i_{4}\cdots i_{2k}}_{(2k-2)}+\cdots+\delta_{i_{1}i_{2k}}\tilde{{\cal L}}^{i_{2}\cdots i_{2k-1}}_{(2k-2)}\right)
\label{recurrence}
\end{multline}
with initial condition condition $\tilde{{\cal L}}_{0}=1$.

These formulae provide us with the values we need for the integrals in (\ref{solangprodinttenseval}):

\bea
 <\hat{r}^{i}\hat{r}^{j}>&=&\frac{\delta_{ij}}{n}=\frac{I^{32}}{4\pi}\nonumber\\
<\hat{r}^{i}\hat{r}^{j}\hat{r}^{k}\hat{r}^{l}>&=&\frac{1}{n+2}\left(\delta_{ij}\tilde{{\cal L}}^{kl}_{(2)}+\delta_{ik}\tilde{{\cal L}}^{jl}_{(2)}+\delta_{il}\tilde{{\cal L}}^{jk}_{(2)}\right)\nonumber\\
&=&\frac{1}{n(n+2)}\left(\delta_{ij}\delta_{kl}+\delta_{ik}\delta_{jl}+\delta_{il}\delta_{jk}\right) \nonumber \\
&=&\frac{I^{34}}{4\pi} \, ,
\label{list}
\eea
where $n=3$ in Eq. (\ref{list}) for three dimensions, the case of interest here.
It should be noted that the meaning of an averaging symbol $<>$ is a bit different from the orientation averaging in the main text. The latter suggests averaging that includes integration over three Euler angles, whereas the former is just averaging over a solid angle defined by two angles of a spherical frame. At some limiting cases, the integration over Euler angles might be reduced to the integration over spherical angles only.
%
%

\section{Optical theorem}
\label{OT}
We need to show that Eq. (\ref{WTIeventzero})
\begin{multline}\label{exp}
\left(\Sigma^{\ast}_{ij}({\bf k})-\Sigma_{ij}({\bf k})\right)  \\ = {\int\limits_{\mathbf{k}_1}}\left(G^{0\ast}(\mb k_1)_{mn}-G^{0}(\mb k_1)_{mn}\right)
K_{mn,ij}({\bf k}_1,{\bf k})
\end{multline}
holds, in the ISA, { when $B=0$} . In this case the mass and irreducible vertex operators are related to the $t$ matrix by (\ref{WTIisaexpressions}), and the $t$ matrix itself is given by Eq. (29) from Ref. \onlinecite{Churochkin2016}. We have then, for the left-hand-side,
\bea
\left(\Sigma^{\ast}_{ij}({\bf k})-\Sigma_{ij}({\bf k})\right)  &=& 2inIm\left[\frac{{\mathcal A}}{1+{\mathcal A}I}\right]<\mM_{ik}\mM_{lj}>k_k k_l   \nonumber \\
&=& \frac{-2in{\mathcal A}^{2}Im[I]}{\left[1+{\mathcal A}I\right]\left[1+{\mathcal A}I\right]^{\ast}}<\mM_{ik}\mM_{lj}>k_k k_l
\label{lhsc14}   \nonumber \\
&&
\eea
And, for the right-hand-side,
\bwt
\bea
\label{proof}
{\int\limits_{\mathbf{k}_1}}\left(G^{0\ast}(\mb k_1)_{mn}-G^{0}(\mb k_1)_{mn}\right)
K_{mn,ij}({\bf k}_1,{\bf k})&  & \\
&& \hspace{-14em} = \frac{-in}{4\pi^{2}}\int\limits_{k_{1}}\int\limits_{\hat {\bf k}_{1}}k^{4}_{1}\left(\frac{\delta\left(k^{2}_{1}-k^{2}_{T}\right)}{\rho c^{2}_{T}}\delta_{mn}\hat k_{1s} \hat k_{1t}+\left(\frac{\delta\left(k^{2}_{1}-k^{2}_{L}\right)}{\rho c^{2}_{L}}-\frac{\delta\left(k^{2}_{1}-k^{2}_{T}\right)}{\rho c^{2}_{T}}\right)\hat k_{1m} \hat k_{1n}\hat k_{1s} \hat k_{1t}\right)\nonumber\\
&& \hspace{-7em}\times \frac{{\mathcal A}}{1+{\mathcal A}I}\left(\frac{{\mathcal A}}{1+{\mathcal A}I}\right)^{\ast} <\mM_{ms}\mM_{ki}\mM_{jl}\mM_{tn}>k_{k}k_{l} \nonumber\\
&& \hspace{-14em} \equiv \frac{-2in{\mathcal A}^{2}Im[I]}{\left[1+{\mathcal A}I\right]\left[1+{\mathcal A}I\right]^{\ast}}<\mM_{ki}\mM_{jl}>k_{k}k_{l}
\eea
\ewt
which coincides with the left-hand-side given by (\ref{lhsc14}). We have used results of Appendix \ref{Apaver}, properties of tensor $\mM$, as well as the explicit expressions
for tensors, which are included into Eq. (\ref{exp}). This calculation, being three-dimensional, differs from the analogous computation carried out in \cite{Churochkin2021} in two dimensions.
\section{Perturbation scheme for the spectral problem}
\label{pert}
To build up the system of equations for the determination of the diffusive pole structure we have to substitute the series from the Eqn. (\ref{Pseries}) 
into Eqn. (\ref{Hom}) and gather together all terms of the same order, either in $\Omega$ or in $\mathbf{q}$. Moreover, 
we assume that at every order of the perturbation scheme both WTI from Eqn. (\ref{WLWTI}) and symmetry constraints from (\ref{symmetry}) are valid 
This yields (omitting the $\Omega$ and $\mathbf{q}$ arguments, as well as indices for brevity)
\begin{widetext}
\begin{eqnarray}
\label{ApPSser}
\int\limits_{\bf{k}^{\prime\prime}}(\mathbf{H}({\bf k},{\bf k}^{\prime\prime})+\mathbf{H}^{1\Omega}({\bf k},{\bf k}^{\prime\prime})
+\mathbf{H}^{1\mathbf{q}}({\bf k},{\bf k}^{\prime\prime})+\mathbf{H}^{2\mathbf{q}}({\bf k},{\bf k}^{\prime\prime})+\dots)
(\mathbf{f}({\bf k}^{\prime\prime})+\mathbf{f}^{1\Omega}({\bf k}^{\prime\prime})
+\mathbf{f}^{1\mathbf{q}}({\bf k}^{\prime\prime})+\mathbf{f}^{2\mathbf{q}}({\bf k}^{\prime\prime})+\dots) &  & \\
\hspace{-7em}
 = (\lambda^{1\Omega}+\lambda^{1\mathbf{q}}+\lambda^{2\mathbf{q}} +\dots)(\mathbf{f}({\bf k})+\mathbf{f}^{1\Omega}({\bf k})
+\mathbf{f}^{1\mathbf{q}}({\bf k})+\mathbf{f}^{2\mathbf{q}}({\bf k}) +\dots) \nonumber \, .
\end{eqnarray}
\end{widetext}
At first order in $\Omega$ and zero order in $\mathbf{q}$, Eqn. (\ref{ApPSser}) easily leads to Eqns. (\ref{system0}) and (\ref{Omega}) in the text. In a similar manner, collecting the  first order in $\mathbf{q}$ terms from Eqn. (\ref{ApPSser}) we obtain the following equation for $\lambda^{1\mathbf{q}}$
\beq
\label{ApPSser1q}
\int\limits_{\bf{k}^{\prime\prime}}(\mathbf{H}({\bf k},{\bf k}^{\prime\prime})
\mathbf{f}^{1\mathbf{q}}({\bf k}^{\prime\prime})+\mathbf{H}^{1\mathbf{q}}({\bf k},{\bf k}^{\prime\prime})
\mathbf{f}({\bf k}^{\prime\prime}))=
\lambda^{1\mathbf{q}}\mathbf{f}({\bf k}) \, .
\eeq
Integrating (\ref{ApPSser1q}) over ${\bf k}$ and subsequently summing over the external indices cancels the contribution from the first term on its left hand side because of the WTI. So that, using (\ref{zeroorder}) we have
\beq
\label{ApPSser1q2}
\int\limits_{\bf{k}}\int\limits_{\bf{k}^{\prime\prime}}H^{1\mathbf{q}}_{ii,kl}({\bf k},{\bf k}^{\prime\prime})
\Delta G_{kl,mm}({\bf k}^{\prime\prime})=
\lambda^{1\mathbf{q}}\int\limits_{\bf{k}}\Delta G_{ii,mm}({\bf k})
\eeq
The left hand side of (\ref{ApPSser1q2}) is equal to zero because of the WTI written to first order in $\mathbf{q}$, as well as the odd in $\mathbf{k}$ character
of the tensor $P_{ii,kl}$ defined in (\ref{potBS2}).. Therefore, we obtain
\beq
\lambda^{1\mathbf{q}}=0.
\eeq

To complete the set of equations for the reconstruction of $\lambda_{0}\left(\mathbf{q},\Omega\right)$ we need $\lambda^{2\mathbf{q}}$. To second order in $\mathbf{q}$ Eqn. (\ref{ApPSser}) gives
\begin{eqnarray}\label{ApPSser2q}
\int\limits_{\bf{k}^{\prime\prime}}(\mathbf{H}({\bf k},{\bf k}^{\prime\prime})\mathbf{f}^{2\mathbf{q}}({\bf k}^{\prime\prime})
+\mathbf{H}^{1\mathbf{q}}({\bf k},{\bf k}^{\prime\prime})\mathbf{f}^{1\mathbf{q}}({\bf k}^{\prime\prime}) & & \\
& & \hspace{-9em}+\mathbf{H}^{2\mathbf{q}}({\bf k},{\bf k}^{\prime\prime})\mathbf{f}({\bf k}^{\prime\prime}))
=
\lambda^{2\mathbf{q}}\mathbf{f}({\bf k})\nonumber \, .
\end{eqnarray}
Then, Eqn. (\ref{system2q}) of the text is obtained integrating (\ref{ApPSser2q}) over ${\bf k}$, summing over the external indices and using the  explicit form of the WTI at corresponding orders.
%
%
\section{Solution  for $\mathbf{f}^{1\mathbf{q}}({\bf k})$}
\label{solu}
 $f^{1\mathbf{q}}({\bf k})$ is obtained by replacing (\ref{zeroorder}) into (\ref{system1q}), using the symmetry property from Eq. (\ref{symmetry}), applying the WTI to $H^{1\mathbf{q}}({\bf k},{\bf k}^{\prime\prime})$, and substituting $\delta_{\mathbf{k}^{\prime\prime},\mathbf{k}}\Delta G_{kl,mm}({\bf k})$ by its value given by (\ref{BS}) to get
 \bwt
\begin{eqnarray}\label{ApSolTran}
\int\limits_{\bf{k}^{\prime\prime}}H^{1\mathbf{q}}_{ii,kl}({\bf k},{\bf k}^{\prime\prime}){\cal{B}}\Delta G_{kl,mm}({\bf k}^{\prime\prime}) & = &
\int\limits_{\bf{k}^{\prime\prime}}{\cal{B}}\left(P_{ii,kl}(\mathbf{q};{\bf k}^{\prime\prime})\delta_{\mathbf{k}^{\prime\prime},\mathbf{k}}\Delta G_{kl,mm}({\bf k}) - H_{ii,kl}({\bf k},{\bf k}^{\prime\prime})\Delta G^{1\mathbf{q}}_{kl,mm}({\bf k}^{\prime\prime})\right) \nonumber \\
 & = & {\cal{B}}\int\limits_{\bf{k}^{\prime\prime}}H_{ii,kl}({\bf k},{\bf k}^{\prime\prime})
 [ \int\limits_{{\bf k}_{2}}\Phi_{kl,k_{1}l_{1}}({\bf k}^{\prime\prime},{\bf k}_{2})P_{k_{1}l_{1},ii}(\mathbf{q};{\bf k}_{2})- \Delta G^{1\mathbf{q}}_{kl,mm}({\bf k}^{\prime\prime}) ]\nonumber \, .
\end{eqnarray}
Hence,
\beq\label{ApSol1q}
f^{1\mathbf{q}}_{kl}({\bf k}^{\prime\prime})=
-{\cal{B}}\left(\int\limits_{{\bf k}_{2}}\Phi_{kl,k_{1}l_{1}}({\bf k}^{\prime\prime},{\bf k}_{2})P_{k_{1}l_{1},ii}(\mathbf{q};{\bf k}_{2})-\Delta G^{1\mathbf{q}}_{kl,mm}({\bf k}^{\prime\prime})\right)
\eeq
with
\bea\label{ApSoldeltaG}
\Delta G^{1\mathbf{q}}_{kl,mm}({\bf k})&=&{\bf q}\cdot\frac{\partial\Delta G_{kl,mm}({\bf k};{\bf q}^{\prime},0)}{\partial {\bf q}^{\prime}}|_{{\bf q}^{\prime}=0}\\ &=& -\frac{1}{2\imath\rho}{\bf q}\cdot\frac{\partial \left(Re[G_{kl}(\mathbf{k})]\right)}{\partial {\bf k}} \\
&=& -\frac{q_{t}Re[G_{L}-G_{T}]}{2\imath\rho}\frac{\partial P_{\textbf{\^{k}}}}{\partial k_{t}}
-\frac{q_{t}}{2\imath\rho}\left(\frac{\partial \left(Re[G_{T}]\right)}{\partial k_{t}}\left(\textbf{I}-P_{\textbf{\^{k}}}\right)+
\frac{\partial \left(Re[G_{L}]\right)}{\partial k_{t}}P_{\textbf{\^{k}}}\right)
\eea
and
\begin{eqnarray}\label{ApSolDerivative}
\frac{\partial P_{\textbf{\^{k}}}}{\partial k_{t}}&=&\frac{\partial \left(\frac{k_{k}k_{l}}{k^{2}}\right)}{\partial k_{t}}=\left(\frac{k_{l}\delta_{kt}+k_{k}\delta_{lt}}{k^{2}}\right)-\frac{2k_{k}k_{l}k_{t}}{k^{4}}\\
\label{ApSolDerivative2}
Re[G_{T,L}]&=&\frac{F_{T,L}(\omega,k)}{\rho\omega^{2}Im[K_{T,L}^{2}]}
\left(Re[K_{T,L}^{2}]\left(k^2-Re[K_{T,L}^{2}]\right)-Im[K_{T,L}^{2}]^{2}\right)\\
\label{ApSolDerivative3}
\frac{\partial \left(Re[G_{T,L}]\right)}{\partial k_{t}}&=&\frac{2 k_{t}F_{T,L}(\omega,k)}{\rho\omega^{2}Im[K_{T,L}^{2}]}
\left(2k^{2}F_{T,L}(\omega,k)Im[K_{T,L}^{2}]-Re[K_{T,L}^{2}]\right)\\
F_{T,L}(\omega,k)&=&\left(\frac{Im[K_{T,L}^{2}]}{\left(k^2-Re[K_{T,L}^{2}]\right)^{2}+Im[K_{T,L}^{2}]^{2}}\right)
\label{ApSolDerivative1}
\end{eqnarray}
%
%
\section{Calculation of  $D$}
\label{esti}
The calculation reported herein follows very closely an analogous computation in two dimensions\cite{Churochkin2021}, for which the reader is referred for a more detailed presentation. As we noted in Eq. (\ref{Dconstant3}) the diffusion constant is the sum of two  terms: $D = D^{\mathcal{R}} + D_{\Delta G^{1\mathbf{q}}}$ and we sketch how to compute each term.
%
\subsection{$D_{\Delta G^{1\mathbf{q}}}$}
Using Eqs. (\ref{parameters}) and (\ref{ApSoldeltaG}, \ref{ApSolDerivative}), Eq. (\ref{Dconstant2}) turns into
\begin{eqnarray}\label{Gq}
D_{\Delta G^{1\mathbf{q}}} & = &
\frac{-{\cal{B}}^{2}}{q^{2}\omega\left(1+a\right)}
\int\limits_{\bf{k}}P_{ii,kl}({\bf k};{\bf q})
\Delta G_{kl,mm}^{1\mathbf{q}}({\bf k})\\
& = &\frac{{\cal{B}}^{2}}{4\rho^{2}q^{2}\omega\left(1+a\right)}
\int\limits_{\bf{k}}q_{s}\frac{\partial L_{kl}({\bf k})}{\partial k_{s}}
\frac{\partial \left(Re[G^{-}_{kl}(\mathbf{k})]\right)}{\partial k_{t}}q_{t}\nonumber\\
& = &\frac{-{\cal{B}}^{2}q_{s}q_{t}\left(c_{L}^{2}-c_{T}^{2}\right)}{2\rho q^{2}\omega\left(1+a\right)}\int\limits_{\bf{k}}
\left(Re[G_{L}-G_{T}]\left(\delta_{st}-\frac{k_{s}k_{t}}{k^{2}}\right)\right)\nonumber\\
&  & \hspace{2em}+\frac{-B^{2}q_{s}q_{t}}{2\rho q^{2}\omega\left(1+a\right)}\int\limits_{\bf{k}}
\left(2c_{T}^{2}\frac{\partial \left(Re[G_{T}]\right)}{\partial k_{t}}+c_{L}^{2}\frac{\partial \left(Re[G_{L}]\right)}{\partial k_{t}}\right)k_{s}\nonumber
\end{eqnarray}

The following two types of integrals have to be considered in (\ref{Gq}):
\begin{eqnarray}\label{ApEstInt}
\mathbb{I}^{st}_{T,L}&=&\int\limits_{\bf{k}}
Re[G_{T,L}]\left(\delta_{st}-\frac{k_{s}k_{t}}{k^{2}} \right)
=\frac{\delta_{st}}{3\pi^{2}}\int\limits_{-\infty}^{\infty}k^{2}\Theta(k) Re[G_{T,L}]dk\\
\mathbb{J}^{st}_{T,L}&=&
\int\limits_{\bf{k}}
\left(\frac{\partial \left(Re[G_{T,L}]\right)}{\partial k_{t}}\right)k_{s}\nonumber
\end{eqnarray}
Using Eqs. (\ref{ApSolDerivative2},\ref{ApSolDerivative3}) we have
\begin{eqnarray}\label{ApEstInt2}
\mathbb{I}^{st}_{T,L}& = &\int\limits_{-\infty}^{\infty} \frac{\delta_{st}k^{2}\Theta(k)F_{T,L}(\omega,k)\left(Re[K_{T,L}^{2}]\left(k^2-Re[K_{T,L}^{2}]\right)-Im[K_{T,L}^{2}]^{2}\right)}{3\pi^{2}\rho\omega^{2}Im[K_{T,L}^{2}]}dk\\
\mathbb{J}^{st}_{T,L}& = & \int\limits_{\bf{k}}
\frac{2 k_{s}k_{t}\left(2k^{2}F_{T,L}^{2}(\omega,k)Im[K_{T,L}^{2}]-Re[K_{T,L}^{2}]F_{T,L}(\omega,k)\right)}{\rho\omega^{2}Im[K_{T,L}^{2}]}\nonumber\\
& = & \int\limits_{-\infty}^{\infty}\delta_{st}\Theta(k)k^{4}dk
\left(\frac{ 2k^{2}F_{T,L}^{2}(\omega,k)Im[K_{T,L}^{2}]-Re[K_{T,L}^{2}]F_{T,L}(\omega,k)}{3\pi^{2}\rho\omega^{2}Im[K_{T,L}^{2}]}
\right).\nonumber
\end{eqnarray}
\end{widetext}
 The integral $\mathbb{J}^{st}_{T,L}$ in Eq. (\ref{ApEstInt2}) includes an ill-defined term, proportional to $F_{T,L}^{2}(\omega,k)$, that can be regularized \cite{Mahan2000}   when $| Im[K_{T,L}^{2}] | \ll | k^{2}-Re[K_{T,L}^{2}] |$ to obtain
\begin{eqnarray}\label{FReg}
F_{T,L}(\omega,k) & = & \pi\delta\left(k^2-Re[K_{T,L}^{2}]\right),\\
F_{T,L}(\omega,k)^{2} & = & \frac{\pi\delta\left(k^2-Re[K_{T,L}^{2}]\right)}{2Im[K_{T,L}^{2}]}\nonumber
\end{eqnarray}

Consequently, Eqs. (\ref{Gq}), (\ref{ApEstInt}), (\ref{ApEstInt2}), and (\ref{FReg}) yield
\begin{eqnarray}\label{Dqeval}
\mathbb{I}^{st}_{T,L}& = & \frac{-\delta_{st}Re[K_{T,L}^{2}]^{\frac{1}{2}}Im[K_{T,L}^{2}]}{6\pi\rho\omega^{2}} \\
\mathbb{J}^{st}_{T,L}&=& 0
\eea
so that
\begin{multline}
D_{\Delta G^{1\mathbf{q}}} =
\frac{1}{3}\frac{-B^{2}\left(c_{L}^{2}-c_{T}^{2}\right)}{4\pi\rho^{2} \omega^{3}\left(1+a\right)}  \\ \times
\left(
Re[K_{T}^{2}]^{\frac{1}{2}}Im[K_{T}^{2}]-Re[K_{L}^{2}]^{\frac{1}{2}}Im[K_{L}^{2}]\right)\, .
\end{multline}

\subsection{$D^{\mathcal{R}}$}
In this calculation, which is similar to the analogous one carried out in two dimensions\cite{Churochkin2021}  we apply a method introduced in the treatment of light diffusion  \cite{Barabanenkov1992}, introducing an auxiliary tensor function $\mathbf{\Psi}_{,s}({\bf k})$ defined by
\begin{align}\label{DconstantAUX}
\mathbf{\Psi}_{,s}({\bf k})q_{s} \equiv& \Psi_{kl,s}({\bf k})q_{s} \\
\equiv &
\int\limits_{{\bf k}^{\prime}}\Phi_{kl,mn}({\bf k},{\bf k}^{\prime})P_{mn,tt}(\mathbf{q};{\bf k}^{\prime})  \\
=&
-\int\limits_{{\bf k}^{\prime}}\Phi_{kl,mn}({\bf k},{\bf k}^{\prime})\frac{1}{2i\rho}\frac{\partial L_{mn}({\bf k}^{\prime})}{\partial k^{\prime}_{s}}q_{s} \, .
\end{align}
Use of  (\ref{BS}) gives the following expression for $\mathbf{\Psi}_{,s}({\bf k})$:
\begin{align}\label{DconstantAUXEQ}
P_{ij,kl}({\bf p})\Psi_{kl,s}({\bf p})+\Delta\Sigma_{ij,k_1l_1}({\bf p})\Psi_{k_1l_1,s}({\bf p})
 \hspace{3em} &\\
-\int\limits_{\bf{p}^{\prime\prime}}
\Delta G_{ij,k_2l_2}({\bf p})K_{k_2l_2,k_1l_1}({\bf p},{\bf p}^{\prime\prime})\Psi_{k_1l_1,s}({\bf p}^{\prime\prime})&= \nonumber \\
-\Delta G_{ij,kl}({\bf p})\frac{1}{2i\rho}\frac{\partial L_{kl}({\bf p})}{\partial p_{s}} &  \nonumber
\end{align}
Using the explicit expression (\ref{g0ft}) for the free medium Green's function, as well as Eqs. (\ref{Dyson},\ref{parameters}) we get
\begin{multline}\label{greenexpl}
\Delta G_{ij,k_2l_2}({\bf p})=  \left(\Delta\Sigma_{ij,n_2 m_2}({\bf p}) \right. \\ \left. +P_{ij,n_2 m_2}({\bf p})\right)G_{n_2 k_2}({\bf p})G_{l_2 m_2}^{*}({\bf p})
\end{multline}
Next, we define an angular tensor $\mathbf{\Upsilon}$, in analogy  to the coefficient that relates transport mean free path and  extinction length in the diffusion of electromagnetic waves \cite{Sheng1990,Tiggelen2000}:
\beq\label{angular}
\Psi_{mn,s}({\bf p})q_{s}=G({\bf p})_{mi}G^{*}_{nj}({\bf p})\Upsilon_{ij}({\bf p},{\bf q}) \, .
\eeq
It obeys the following integral equation:
\begin{multline}
\label{angulareq}
P_{ij,tt}({\bf p};{\bf q}) =\Upsilon_{ij}({\bf p},{\bf q}) \\
-\int\limits_{\bf{p}^{\prime\prime}}
K_{ij,k_1 l_1}({\bf p},{\bf p}^{\prime\prime})G_{k_1 m_1}({\bf p}^{\prime\prime})G^{*}_{n_1 l_1}({\bf p}^{\prime\prime})\Upsilon_{m_1 n_1}({\bf p}^{\prime\prime},{\bf q})
\end{multline}
So that, using Eqs. (\ref{DconstantAUX}), (\ref{angular})
,   the following expression for $D^{\mathcal{R}}$  (defined by (\ref{DR})) results:
\beq\label{difften}
D^{\mathcal{R}} =
\frac{{\cal{B}}^{2}}{q^{2}\omega\left(1+a\right)}
\int\limits_{\bf{k}}P_{ss,ij}({\bf k};{\bf q})G_{im}({\bf k})
G^{*}_{nj}({\bf k})\Upsilon_{mn}({\bf k},{\bf q})
\eeq
Now, looking at Eq. (\ref{angular}) we make the ansatz that  $\mathbf{\Upsilon}({\bf p},{\bf q})$ is proportional to ${\bf q}$, and we look for a solution in the form
 \beq\label{angulsol}
\Upsilon_{mn}({\bf p},{\bf q})=\alpha P_{mn,kk}({\bf p};{\bf q})
\eeq
 Multiplying Eq. (\ref{angulareq}) on the left by  $P_{ss,k_1 l_1}({\bf p};{\bf q})G_{k_1 i}({\bf p})G_{j l_1}^{*}({\bf p})$  and integrating over  ${\bf p}$  we are left with
 \begin{widetext}
 \beq\label{angularproof}
\alpha^{-1}=1-
\frac{\int\limits_{\bf{p}}\int\limits_{\bf{p}^{\prime\prime}}P_{nn,t_1t_2}({\bf p};{\bf q})G_{t_1k_1}({\bf p})G_{k_2t_2}^{*}({\bf p})
K_{k_1k_2,m_1n_1}({\bf p},{\bf p}^{\prime\prime})G_{m_1k_3}({\bf p}^{\prime\prime})G^{*}_{k_4 n_1}({\bf p}^{\prime\prime})P_{k_3k_4,ll}({\bf p}^{\prime\prime},{\bf q})}{\int\limits_{\bf{k}}P_{ss,k_1 l_1}({\bf k};{\bf q})G_{k_1 i}({\bf k})G^{*}_{j l_1}({\bf k})P_{ij,tt}({\bf k};{\bf q})}  \, .
\eeq
\end{widetext}
{The second term on the right-hand-side is the analog of the $\langle \cos \theta \rangle$ term in the diffusion of electromagnetic waves \cite{Tiggelen2000}}. We are left with the following expression: 
\begin{multline}
\label{diffin}
D^{\mathcal{R}} =
\frac{{\cal{B}}^{2}}{q^{2}\omega\left(1+a\right)}  \\
\times \int\limits_{\bf{k}}\alpha P_{ii,kl}({\bf k};{\bf q})G_{km}({\bf k})G^{*}_{nl}({\bf k})P_{mn,tt}({\bf k};{\bf q}) \, .
\end{multline}

The coefficient $\alpha$ is now evaluated:  The
symmetry properties of the Green tensor, tensor $\mathbf{P}$, and the kernel $\mathbf{K}$ from Eqs. (\ref{green}), (\ref{parameters}),and (\ref{WTIisaexpressions}), respectively, yield
\begin{multline}
\label{WTIzeroproof}
\hspace{-1em} K_{ij,m_1 n_1}({\bf p},-{\bf p}^{\prime\prime})G_{m_1 t_1}(-{\bf p}^{\prime\prime})G^{*}_{s_1 n_1}(-{\bf p}^{\prime\prime})P_{t_1 s_1,ll}(-{\bf p}^{\prime\prime},{\bf q}) \\
=  -K_{ij,m_1 n_1}({\bf p},{\bf p}^{\prime\prime})G_{m_1 t_1}({\bf p}^{\prime\prime})G^{*}_{s_1 n_1}({\bf p}^{\prime\prime})P_{t_1 s_1,ll}({\bf p}^{\prime\prime},{\bf q})\nonumber
\end{multline}
Hence
\beq
 \int\limits_{{\bf p}^{\prime\prime}}K_{ij,m_1 n_1}({\bf p},{\bf p}^{\prime\prime})G_{m_1 t_1}({\bf p}^{\prime\prime})G^{*}_{s_1 n_1}({\bf p}^{\prime\prime})P_{t_1 s_1,ll}({\bf p}^{\prime\prime},{\bf q})=0
 \eeq
 and
$\alpha= 1$.

Therefore, $D^{\mathcal{R}}$ from the Eq.(\ref{diffin}) is given by
\beq\label{DCFin}
D^{\mathcal{R}} =
\frac{{\cal{B}}^{2}}{q^{2}\omega\left(1+a\right)}
\int\limits_{\bf{k}}P_{ll,mn}({\bf k};{\bf q})G_{mi}({\bf k})G_{jn}^{*}({\bf k})P_{ij,tt}({\bf k};{\bf q}) \, .
\eeq

Finally, using approximation from Eq. (\ref{FReg}) for Eq. (\ref{ApSolDerivative}) in Eq. (\ref{DCFin}) we can write
\bwt
\beq
\label{DCFinExpl0}
D^{\mathcal{R}} =\frac{-{\cal{B}}^{2}}{12\pi\rho^{2}\omega^{5}\left(1+a\right)}
 \times\left(c_{L}^{4}\left(\frac{Re[K_{L}^{2}]^{7/2}}{Im[K_{L}^{2}]}+Re[K_{L}^{2}]Im[K_{L}^{2}]^{3/2}\right)  +2c_{T}^{4}\left(\frac{Re[K_{T}^{2}]^{7/2}}{Im[K_{T}^{2}]}+Re[K_{T}^{2}]^{3/2}Im[K_{T}^{2}]\right)\right) \,
\eeq
Then, the total diffusion constant is
\bea
\label{DCFintot}
D&=&D^{\mathcal{R}}+D_{\Delta G^{1\mathbf{q}}} \\
&=&
\frac{-{\cal{B}}^{2}}{12\pi\rho^{2}\omega^{5}\left(1+a\right)} \left(c_{L}^{4}\left(\frac{Re[K_{L}^{2}]^{7/2}}{Im[K_{L}^{2}]}+Re[K_{L}^{2}]Im[K_{L}^{2}]^{3/2}\right) + 2c_{T}^{4}\left(\frac{Re[K_{T}^{2}]^{7/2}}{Im[K_{T}^{2}]}+Re[K_{T}^{2}]^{3/2}Im[K_{T}^{2}]\right)\right)\nonumber\\
&&\hspace{2em}+ \frac{{\cal{B}}^{2}\left(c_{L}^{2}-c_{T}^{2}\right)\left(Re[K_{L}^{2}]^{1/2}Im[K_{L}^{2}]-Re[K_{T}^{2}]^{1/2}Im[K_{T}^{2}]\right))}{12\pi\rho^{2} \omega^{3}\left(1+a\right)} \, , \nonumber
\eea
and the leading term in {the limit of small} $Im[K_{T,L}^{2}]$ {is}
\beq
\label{DCFintotlead}
D^{lead}\approx\frac{-{\cal{B}}^{2}}{12\pi\rho^{2}\omega^{5}\left(1+a\right)}
\left(c_{L}^{4}\left(\frac{Re[K_{L}^{2}]^{7/2}}{Im[K_{L}^{2}]}\right)+2c_{T}^{4}\left(\frac{Re[K_{T}^{2}]^{7/2}}{Im[K_{T}^{2}]}\right)\right)
\eeq
Explicitly, from Eq.(\ref{eq:B}), we have
\bea\label{Bexpl}
-{\cal{B}}^{2}=\frac{4\pi\rho^{2}\omega^{2}}{\left(2Re[K_{T}^{2}]^{3/2}+Re[K_{L}^{2}]^{3/2}\right)}
\eea
Then, using Eqs.(\ref{deltaGdelta}, \ref{eq:B}, \ref{aparaprox}, \ref{DCFintotlead})
\beq
\label{DCFintotleadexpla}
D^{lead}\approx
\left(1+\frac{2Re[K_{T}^{2}]^{3/2}\left(\frac{c_{T}^{2}}{\omega^{2}}Re[K_{T}^{2}]-1\right)+Re[K_{L}^{2}]^{3/2}\left(\frac{c_{L}^{2}}{\omega^{2}}Re[K_{L}^{2}]-1\right)}
{\left(\frac{\omega_{r1}^{2}}{\omega^{2}}-1\right)\left(2Re[K_{T}^{2}]^{3/2}+Re[K_{L}^{2}]^{3/2}\right)}\right)^{-1}
 \frac{\left(c_{L}^{4}\frac{Re[K_{L}^{2}]^{7/2}}{Im[K_{L}^{2}]}+2c_{T}^{4}\frac{Re[K_{T}^{2}]^{7/2}}{Im[K_{T}^{2}]}\right)}{3\omega^{3}\left(2Re[K_{T}^{2}]^{3/2}+Re[K_{L}^{2}]^{3/2}\right)}  \, ,
\eeq
which is Eq. (\ref{DCFintotleadexpl}). In the low-frequency limit it reads as
\beq\label{DCFintotleadexpllfa}
D^{lead}_{\omega\rightarrow 0}\approx
\left(\frac{v_{T}^{3}c_{L}^{4}}{\left(2v_{L}^{3}+v_{T}^{3}\right)v_{L}^{4}}\frac{v_{L}l_{L}}{3}+\frac{2v_{L}^{3}c_{T}^{4}}{\left(2v_{L}^{3}+v_{T}^{3}\right)v_{T}^{4}}\frac{v_{T}l_{T}}{3}\right)
\eeq
which is Eqn. (\ref{DCFintotleadexpllf}).
\ewt
%
%



%


\begin{thebibliography}{75}%
\makeatletter
\providecommand \@ifxundefined [1]{%
 \@ifx{#1\undefined}
}%
\providecommand \@ifnum [1]{%
 \ifnum #1\expandafter \@firstoftwo
 \else \expandafter \@secondoftwo
 \fi
}%
\providecommand \@ifx [1]{%
 \ifx #1\expandafter \@firstoftwo
 \else \expandafter \@secondoftwo
 \fi
}%
\providecommand \natexlab [1]{#1}%
\providecommand \enquote  [1]{``#1''}%
\providecommand \bibnamefont  [1]{#1}%
\providecommand \bibfnamefont [1]{#1}%
\providecommand \citenamefont [1]{#1}%
\providecommand \href@noop [0]{\@secondoftwo}%
\providecommand \href [0]{\begingroup \@sanitize@url \@href}%
\providecommand \@href[1]{\@@startlink{#1}\@@href}%
\providecommand \@@href[1]{\endgroup#1\@@endlink}%
\providecommand \@sanitize@url [0]{\catcode `\\12\catcode `\$12\catcode
  `\&12\catcode `\#12\catcode `\^12\catcode `\_12\catcode `\%12\relax}%
\providecommand \@@startlink[1]{}%
\providecommand \@@endlink[0]{}%
\providecommand \url  [0]{\begingroup\@sanitize@url \@url }%
\providecommand \@url [1]{\endgroup\@href {#1}{\urlprefix }}%
\providecommand \urlprefix  [0]{URL }%
\providecommand \Eprint [0]{\href }%
\providecommand \doibase [0]{http://dx.doi.org/}%
\providecommand \selectlanguage [0]{\@gobble}%
\providecommand \bibinfo  [0]{\@secondoftwo}%
\providecommand \bibfield  [0]{\@secondoftwo}%
\providecommand \translation [1]{[#1]}%
\providecommand \BibitemOpen [0]{}%
\providecommand \bibitemStop [0]{}%
\providecommand \bibitemNoStop [0]{.\EOS\space}%
\providecommand \EOS [0]{\spacefactor3000\relax}%
\providecommand \BibitemShut  [1]{\csname bibitem#1\endcsname}%
\let\auto@bib@innerbib\@empty
\bibitem [{\citenamefont {Nam}\ \emph {et~al.}(2012)\citenamefont {Nam},
  \citenamefont {Chung}, \citenamefont {Lo}, \citenamefont {Qi}, \citenamefont
  {Li}, \citenamefont {Lu}, \citenamefont {Johnson}, \citenamefont {Jung},
  \citenamefont {Nukala},\ and\ \citenamefont {Agarwal}}]{Nam2012}%
  \BibitemOpen
  \bibfield  {author} {\bibinfo {author} {\bibfnamefont {S.-W.}\ \bibnamefont
  {Nam}}, \bibinfo {author} {\bibfnamefont {H.-S.}\ \bibnamefont {Chung}},
  \bibinfo {author} {\bibfnamefont {Y.~C.}\ \bibnamefont {Lo}}, \bibinfo
  {author} {\bibfnamefont {L.}~\bibnamefont {Qi}}, \bibinfo {author}
  {\bibfnamefont {J.}~\bibnamefont {Li}}, \bibinfo {author} {\bibfnamefont
  {Y.}~\bibnamefont {Lu}}, \bibinfo {author} {\bibfnamefont {A.~C.}\
  \bibnamefont {Johnson}}, \bibinfo {author} {\bibfnamefont {Y.}~\bibnamefont
  {Jung}}, \bibinfo {author} {\bibfnamefont {P.}~\bibnamefont {Nukala}}, \ and\
  \bibinfo {author} {\bibfnamefont {R.}~\bibnamefont {Agarwal}},\ }\bibfield
  {title} {\enquote {\bibinfo {title} {Electrical wind force{\textendash}driven
  and dislocation-templated amorphization in phase-change nanowires},}\ }\href
  {\doibase 10.1126/science.1220119} {\bibfield  {journal} {\bibinfo  {journal}
  {Science}\ }\textbf {\bibinfo {volume} {336}},\ \bibinfo {pages} {1561}
  (\bibinfo {year} {2012})}\BibitemShut {NoStop}%
\bibitem [{\citenamefont {H{\"o}fling}\ \emph {et~al.}(2021)\citenamefont
  {H{\"o}fling}, \citenamefont {Zhou}, \citenamefont {Riemer}, \citenamefont
  {Bruder}, \citenamefont {Liu}, \citenamefont {Zhou}, \citenamefont
  {Groszewicz}, \citenamefont {Zhuo}, \citenamefont {Xu}, \citenamefont
  {Durst}, \citenamefont {Tan}, \citenamefont {Damjanovic}, \citenamefont
  {Koruza},\ and\ \citenamefont {R{\"o}del}}]{Hofling2021}%
  \BibitemOpen
  \bibfield  {author} {\bibinfo {author} {\bibfnamefont {M.}~\bibnamefont
  {H{\"o}fling}}, \bibinfo {author} {\bibfnamefont {X.}~\bibnamefont {Zhou}},
  \bibinfo {author} {\bibfnamefont {L.~M.}\ \bibnamefont {Riemer}}, \bibinfo
  {author} {\bibfnamefont {E.}~\bibnamefont {Bruder}}, \bibinfo {author}
  {\bibfnamefont {B.}~\bibnamefont {Liu}}, \bibinfo {author} {\bibfnamefont
  {L.}~\bibnamefont {Zhou}}, \bibinfo {author} {\bibfnamefont {P.~B.}\
  \bibnamefont {Groszewicz}}, \bibinfo {author} {\bibfnamefont
  {F.}~\bibnamefont {Zhuo}}, \bibinfo {author} {\bibfnamefont {B.-X.}\
  \bibnamefont {Xu}}, \bibinfo {author} {\bibfnamefont {K.}~\bibnamefont
  {Durst}}, \bibinfo {author} {\bibfnamefont {X.}~\bibnamefont {Tan}}, \bibinfo
  {author} {\bibfnamefont {D.}~\bibnamefont {Damjanovic}}, \bibinfo {author}
  {\bibfnamefont {J.}~\bibnamefont {Koruza}}, \ and\ \bibinfo {author}
  {\bibfnamefont {J.}~\bibnamefont {R{\"o}del}},\ }\bibfield  {title} {\enquote
  {\bibinfo {title} {Control of polarization in bulk ferroelectrics by
  mechanical dislocation imprint},}\ }\href {\doibase 10.1126/science.abe3810}
  {\bibfield  {journal} {\bibinfo  {journal} {Science}\ }\textbf {\bibinfo
  {volume} {372}},\ \bibinfo {pages} {961} (\bibinfo {year}
  {2021})}\BibitemShut {NoStop}%
\bibitem [{\citenamefont {Adepalli}\ \emph {et~al.}(2017)\citenamefont
  {Adepalli}, \citenamefont {Yang}, \citenamefont {Maier}, \citenamefont
  {Tuller},\ and\ \citenamefont {Yildiz}}]{Adepalli2017}%
  \BibitemOpen
  \bibfield  {author} {\bibinfo {author} {\bibfnamefont {K.~K.}\ \bibnamefont
  {Adepalli}}, \bibinfo {author} {\bibfnamefont {J.}~\bibnamefont {Yang}},
  \bibinfo {author} {\bibfnamefont {J.}~\bibnamefont {Maier}}, \bibinfo
  {author} {\bibfnamefont {H.~L.}\ \bibnamefont {Tuller}}, \ and\ \bibinfo
  {author} {\bibfnamefont {B.}~\bibnamefont {Yildiz}},\ }\bibfield  {title}
  {\enquote {\bibinfo {title} {Tunable oxygen diffusion and electronic
  conduction in {S}r{T}i{O}3 by dislocation-induced space charge fields},}\
  }\href {\doibase https://doi.org/10.1002/adfm.201700243} {\bibfield
  {journal} {\bibinfo  {journal} {Advanced Functional Materials}\ }\textbf
  {\bibinfo {volume} {27}},\ \bibinfo {pages} {1700243} (\bibinfo {year}
  {2017})}\BibitemShut {NoStop}%
\bibitem [{\citenamefont {Porz}\ \emph {et~al.}(2021)\citenamefont {Porz},
  \citenamefont {Fr{\"o}mling}, \citenamefont {Nakamura}, \citenamefont {Li},
  \citenamefont {Maruyama}, \citenamefont {Matsunaga}, \citenamefont {Gao},
  \citenamefont {Simons}, \citenamefont {Dietz}, \citenamefont {Rohnke},
  \citenamefont {Janek},\ and\ \citenamefont {R{\"o}del}}]{Porz2021}%
  \BibitemOpen
  \bibfield  {author} {\bibinfo {author} {\bibfnamefont {L.}~\bibnamefont
  {Porz}}, \bibinfo {author} {\bibfnamefont {T.}~\bibnamefont {Fr{\"o}mling}},
  \bibinfo {author} {\bibfnamefont {A.}~\bibnamefont {Nakamura}}, \bibinfo
  {author} {\bibfnamefont {N.}~\bibnamefont {Li}}, \bibinfo {author}
  {\bibfnamefont {R.}~\bibnamefont {Maruyama}}, \bibinfo {author}
  {\bibfnamefont {K.}~\bibnamefont {Matsunaga}}, \bibinfo {author}
  {\bibfnamefont {P.}~\bibnamefont {Gao}}, \bibinfo {author} {\bibfnamefont
  {H.}~\bibnamefont {Simons}}, \bibinfo {author} {\bibfnamefont
  {C.}~\bibnamefont {Dietz}}, \bibinfo {author} {\bibfnamefont
  {M.}~\bibnamefont {Rohnke}}, \bibinfo {author} {\bibfnamefont
  {J.}~\bibnamefont {Janek}}, \ and\ \bibinfo {author} {\bibfnamefont
  {J.}~\bibnamefont {R{\"o}del}},\ }\bibfield  {title} {\enquote {\bibinfo
  {title} {Conceptual framework for dislocation-modified conductivity in oxide
  ceramics deconvoluting mesoscopic structure, core, and space charge
  exemplified for {S}r{T}i{O}3},}\ }\bibfield  {booktitle} {\emph {\bibinfo
  {booktitle} {ACS Nano}},\ }\href {\doibase 10.1021/acsnano.0c04491}
  {\bibfield  {journal} {\bibinfo  {journal} {ACS Nano}\ }\textbf {\bibinfo
  {volume} {15}},\ \bibinfo {pages} {9355} (\bibinfo {year}
  {2021})}\BibitemShut {NoStop}%
\bibitem [{\citenamefont {Massabuau}\ \emph {et~al.}(2017)\citenamefont
  {Massabuau}, \citenamefont {Chen}, \citenamefont {Horton}, \citenamefont
  {Rhode}, \citenamefont {Ren}, \citenamefont {O'Hanlon}, \citenamefont
  {Kov\'{a}cs}, \citenamefont {Kappers}, \citenamefont {Humphreys},
  \citenamefont {Dunin-Borkowski},\ and\ \citenamefont
  {Oliver}}]{Massabau2017}%
  \BibitemOpen
  \bibfield  {author} {\bibinfo {author} {\bibfnamefont {F.~C.-P.}\
  \bibnamefont {Massabuau}}, \bibinfo {author} {\bibfnamefont {P.}~\bibnamefont
  {Chen}}, \bibinfo {author} {\bibfnamefont {M.~K.}\ \bibnamefont {Horton}},
  \bibinfo {author} {\bibfnamefont {S.~L.}\ \bibnamefont {Rhode}}, \bibinfo
  {author} {\bibfnamefont {C.~X.}\ \bibnamefont {Ren}}, \bibinfo {author}
  {\bibfnamefont {T.~J.}\ \bibnamefont {O'Hanlon}}, \bibinfo {author}
  {\bibfnamefont {A.}~\bibnamefont {Kov\'{a}cs}}, \bibinfo {author}
  {\bibfnamefont {M.~J.}\ \bibnamefont {Kappers}}, \bibinfo {author}
  {\bibfnamefont {C.~J.}\ \bibnamefont {Humphreys}}, \bibinfo {author}
  {\bibfnamefont {R.~E.}\ \bibnamefont {Dunin-Borkowski}}, \ and\ \bibinfo
  {author} {\bibfnamefont {R.~A.}\ \bibnamefont {Oliver}},\ }\bibfield  {title}
  {\enquote {\bibinfo {title} {Carrier localization in the vicinity of
  dislocations in {I}n{G}a{N}},}\ }\href {\doibase 10.1063/1.4973278}
  {\bibfield  {journal} {\bibinfo  {journal} {Journal of Applied Physics}\
  }\textbf {\bibinfo {volume} {121}},\ \bibinfo {pages} {013104} (\bibinfo
  {year} {2017})}\BibitemShut {NoStop}%
\bibitem [{\citenamefont {Kotchetkov}\ \emph {et~al.}(2001)\citenamefont
  {Kotchetkov}, \citenamefont {Zou}, \citenamefont {Balandin}, \citenamefont
  {Florescu},\ and\ \citenamefont {Pollak}}]{Kotchetkov2001}%
  \BibitemOpen
  \bibfield  {author} {\bibinfo {author} {\bibfnamefont {D.}~\bibnamefont
  {Kotchetkov}}, \bibinfo {author} {\bibfnamefont {J.}~\bibnamefont {Zou}},
  \bibinfo {author} {\bibfnamefont {A.~A.}\ \bibnamefont {Balandin}}, \bibinfo
  {author} {\bibfnamefont {D.~I.}\ \bibnamefont {Florescu}}, \ and\ \bibinfo
  {author} {\bibfnamefont {F.~H.}\ \bibnamefont {Pollak}},\ }\bibfield  {title}
  {\enquote {\bibinfo {title} {Effect of dislocations on thermal conductivity
  of {G}a{N} layers},}\ }\href {\doibase 10.1063/1.1427153} {\bibfield
  {journal} {\bibinfo  {journal} {Applied Physics Letters}\ }\textbf {\bibinfo
  {volume} {79}},\ \bibinfo {pages} {4316} (\bibinfo {year}
  {2001})}\BibitemShut {NoStop}%
\bibitem [{\citenamefont {Kamatagi}\ \emph {et~al.}(2007)\citenamefont
  {Kamatagi}, \citenamefont {Sankeshwar},\ and\ \citenamefont
  {Mulimani}}]{Kamatagi2007}%
  \BibitemOpen
  \bibfield  {author} {\bibinfo {author} {\bibfnamefont {M.}~\bibnamefont
  {Kamatagi}}, \bibinfo {author} {\bibfnamefont {N.}~\bibnamefont
  {Sankeshwar}}, \ and\ \bibinfo {author} {\bibfnamefont {B.}~\bibnamefont
  {Mulimani}},\ }\bibfield  {title} {\enquote {\bibinfo {title} {Thermal
  conductivity of {G}a{N}},}\ }\href {\doibase
  https://doi.org/10.1016/j.diamond.2006.04.004} {\bibfield  {journal}
  {\bibinfo  {journal} {Diamond and Related Materials}\ }\textbf {\bibinfo
  {volume} {16}},\ \bibinfo {pages} {98} (\bibinfo {year} {2007})}\BibitemShut
  {NoStop}%
\bibitem [{\citenamefont {Ma}\ \emph {et~al.}(2013)\citenamefont {Ma},
  \citenamefont {Wang}, \citenamefont {Huang},\ and\ \citenamefont
  {Luo}}]{Ma2013}%
  \BibitemOpen
  \bibfield  {author} {\bibinfo {author} {\bibfnamefont {J.}~\bibnamefont
  {Ma}}, \bibinfo {author} {\bibfnamefont {X.}~\bibnamefont {Wang}}, \bibinfo
  {author} {\bibfnamefont {B.}~\bibnamefont {Huang}}, \ and\ \bibinfo {author}
  {\bibfnamefont {X.}~\bibnamefont {Luo}},\ }\bibfield  {title} {\enquote
  {\bibinfo {title} {Effects of point defects and dislocations on spectral
  phonon transport properties of wurtzite {G}a{N}},}\ }\href {\doibase
  10.1063/1.4817083} {\bibfield  {journal} {\bibinfo  {journal} {Journal of
  Applied Physics}\ }\textbf {\bibinfo {volume} {114}},\ \bibinfo {pages}
  {074311} (\bibinfo {year} {2013})}\BibitemShut {NoStop}%
\bibitem [{\citenamefont {Kamatagi}\ \emph {et~al.}(2009)\citenamefont
  {Kamatagi}, \citenamefont {Vaidya}, \citenamefont {Sankeshwar},\ and\
  \citenamefont {Mulimani}}]{Kamatagi2009}%
  \BibitemOpen
  \bibfield  {author} {\bibinfo {author} {\bibfnamefont {M.}~\bibnamefont
  {Kamatagi}}, \bibinfo {author} {\bibfnamefont {R.}~\bibnamefont {Vaidya}},
  \bibinfo {author} {\bibfnamefont {N.}~\bibnamefont {Sankeshwar}}, \ and\
  \bibinfo {author} {\bibfnamefont {B.}~\bibnamefont {Mulimani}},\ }\bibfield
  {title} {\enquote {\bibinfo {title} {Low-temperature lattice thermal
  conductivity in free-standing {G}a{N} thin films},}\ }\href {\doibase
  https://doi.org/10.1016/j.ijheatmasstransfer.2008.10.032} {\bibfield
  {journal} {\bibinfo  {journal} {International Journal of Heat and Mass
  Transfer}\ }\textbf {\bibinfo {volume} {52}},\ \bibinfo {pages} {2885}
  (\bibinfo {year} {2009})}\BibitemShut {NoStop}%
\bibitem [{\citenamefont {Singh}\ \emph {et~al.}(2006)\citenamefont {Singh},
  \citenamefont {Menon},\ and\ \citenamefont {Sood}}]{Singh2006}%
  \BibitemOpen
  \bibfield  {author} {\bibinfo {author} {\bibfnamefont {B.~K.}\ \bibnamefont
  {Singh}}, \bibinfo {author} {\bibfnamefont {V.~J.}\ \bibnamefont {Menon}}, \
  and\ \bibinfo {author} {\bibfnamefont {K.~C.}\ \bibnamefont {Sood}},\
  }\bibfield  {title} {\enquote {\bibinfo {title} {Phonon conductivity of
  plastically deformed crystals: Role of stacking faults and dislocations},}\
  }\href {\doibase 10.1103/PhysRevB.74.184302} {\bibfield  {journal} {\bibinfo
  {journal} {Phys. Rev. B}\ }\textbf {\bibinfo {volume} {74}},\ \bibinfo
  {pages} {184302} (\bibinfo {year} {2006})}\BibitemShut {NoStop}%
\bibitem [{\citenamefont {Shuai}\ \emph {et~al.}(2016)\citenamefont {Shuai},
  \citenamefont {Geng}, \citenamefont {Lan}, \citenamefont {Zhu}, \citenamefont
  {Wang}, \citenamefont {Liu}, \citenamefont {Bao}, \citenamefont {Chu},
  \citenamefont {Sui},\ and\ \citenamefont {Ren}}]{Shuai2016}%
  \BibitemOpen
  \bibfield  {author} {\bibinfo {author} {\bibfnamefont {J.}~\bibnamefont
  {Shuai}}, \bibinfo {author} {\bibfnamefont {H.}~\bibnamefont {Geng}},
  \bibinfo {author} {\bibfnamefont {Y.}~\bibnamefont {Lan}}, \bibinfo {author}
  {\bibfnamefont {Z.}~\bibnamefont {Zhu}}, \bibinfo {author} {\bibfnamefont
  {C.}~\bibnamefont {Wang}}, \bibinfo {author} {\bibfnamefont {Z.}~\bibnamefont
  {Liu}}, \bibinfo {author} {\bibfnamefont {J.}~\bibnamefont {Bao}}, \bibinfo
  {author} {\bibfnamefont {C.-W.}\ \bibnamefont {Chu}}, \bibinfo {author}
  {\bibfnamefont {J.}~\bibnamefont {Sui}}, \ and\ \bibinfo {author}
  {\bibfnamefont {Z.}~\bibnamefont {Ren}},\ }\bibfield  {title} {\enquote
  {\bibinfo {title} {Higher thermoelectric performance of {Z}intl phases
  ({E}u$_{0.5}${Y}b$_{0.5}$)$_{1-x}${C}a$_x${M}g$_2${B}i$_2$ by band
  engineering and strain fluctuation},}\ }\href {\doibase
  10.1073/pnas.1608794113} {\bibfield  {journal} {\bibinfo  {journal}
  {Proceedings of the National Academy of Sciences}\ }\textbf {\bibinfo
  {volume} {113}},\ \bibinfo {pages} {E4125} (\bibinfo {year}
  {2016})}\BibitemShut {NoStop}%
\bibitem [{\citenamefont {Wu}\ \emph {et~al.}(2019)\citenamefont {Wu},
  \citenamefont {Chen}, \citenamefont {Nan}, \citenamefont {Xiong},
  \citenamefont {Lin}, \citenamefont {Zhang}, \citenamefont {Chen},
  \citenamefont {Chen}, \citenamefont {Ge},\ and\ \citenamefont
  {Pei}}]{Wu2019}%
  \BibitemOpen
  \bibfield  {author} {\bibinfo {author} {\bibfnamefont {Y.}~\bibnamefont
  {Wu}}, \bibinfo {author} {\bibfnamefont {Z.}~\bibnamefont {Chen}}, \bibinfo
  {author} {\bibfnamefont {P.}~\bibnamefont {Nan}}, \bibinfo {author}
  {\bibfnamefont {F.}~\bibnamefont {Xiong}}, \bibinfo {author} {\bibfnamefont
  {S.}~\bibnamefont {Lin}}, \bibinfo {author} {\bibfnamefont {X.}~\bibnamefont
  {Zhang}}, \bibinfo {author} {\bibfnamefont {Y.}~\bibnamefont {Chen}},
  \bibinfo {author} {\bibfnamefont {L.}~\bibnamefont {Chen}}, \bibinfo {author}
  {\bibfnamefont {B.}~\bibnamefont {Ge}}, \ and\ \bibinfo {author}
  {\bibfnamefont {Y.}~\bibnamefont {Pei}},\ }\bibfield  {title} {\enquote
  {\bibinfo {title} {Lattice strain advances thermoelectrics},}\ }\href
  {\doibase https://doi.org/10.1016/j.joule.2019.02.008} {\bibfield  {journal}
  {\bibinfo  {journal} {Joule}\ }\textbf {\bibinfo {volume} {3}},\ \bibinfo
  {pages} {1276 } (\bibinfo {year} {2019})}\BibitemShut {NoStop}%
\bibitem [{\citenamefont {You}\ \emph {et~al.}(2018)\citenamefont {You},
  \citenamefont {Liu}, \citenamefont {Li}, \citenamefont {Nan}, \citenamefont
  {Ge}, \citenamefont {Jiang}, \citenamefont {Luo}, \citenamefont {Pan},
  \citenamefont {Pei}, \citenamefont {Zhang}, \citenamefont {Snyder},
  \citenamefont {Yang}, \citenamefont {Zhang},\ and\ \citenamefont
  {Luo}}]{You2018}%
  \BibitemOpen
  \bibfield  {author} {\bibinfo {author} {\bibfnamefont {L.}~\bibnamefont
  {You}}, \bibinfo {author} {\bibfnamefont {Y.}~\bibnamefont {Liu}}, \bibinfo
  {author} {\bibfnamefont {X.}~\bibnamefont {Li}}, \bibinfo {author}
  {\bibfnamefont {P.}~\bibnamefont {Nan}}, \bibinfo {author} {\bibfnamefont
  {B.}~\bibnamefont {Ge}}, \bibinfo {author} {\bibfnamefont {Y.}~\bibnamefont
  {Jiang}}, \bibinfo {author} {\bibfnamefont {P.}~\bibnamefont {Luo}}, \bibinfo
  {author} {\bibfnamefont {S.}~\bibnamefont {Pan}}, \bibinfo {author}
  {\bibfnamefont {Y.}~\bibnamefont {Pei}}, \bibinfo {author} {\bibfnamefont
  {W.}~\bibnamefont {Zhang}}, \bibinfo {author} {\bibfnamefont {G.~J.}\
  \bibnamefont {Snyder}}, \bibinfo {author} {\bibfnamefont {J.}~\bibnamefont
  {Yang}}, \bibinfo {author} {\bibfnamefont {J.}~\bibnamefont {Zhang}}, \ and\
  \bibinfo {author} {\bibfnamefont {J.}~\bibnamefont {Luo}},\ }\bibfield
  {title} {\enquote {\bibinfo {title} {Boosting the thermoelectric performance
  of {P}b{S}e through dynamic doping and hierarchical phonon scattering},}\
  }\href {\doibase 10.1039/C8EE00418H} {\bibfield  {journal} {\bibinfo
  {journal} {Energy Environ. Sci.}\ }\textbf {\bibinfo {volume} {11}},\
  \bibinfo {pages} {1848} (\bibinfo {year} {2018})}\BibitemShut {NoStop}%
\bibitem [{\citenamefont {Xin}\ \emph {et~al.}(2017)\citenamefont {Xin},
  \citenamefont {Wu}, \citenamefont {Liu}, \citenamefont {Zhu}, \citenamefont
  {Yu},\ and\ \citenamefont {Zhao}}]{Xin2017}%
  \BibitemOpen
  \bibfield  {author} {\bibinfo {author} {\bibfnamefont {J.}~\bibnamefont
  {Xin}}, \bibinfo {author} {\bibfnamefont {H.}~\bibnamefont {Wu}}, \bibinfo
  {author} {\bibfnamefont {X.}~\bibnamefont {Liu}}, \bibinfo {author}
  {\bibfnamefont {T.}~\bibnamefont {Zhu}}, \bibinfo {author} {\bibfnamefont
  {G.}~\bibnamefont {Yu}}, \ and\ \bibinfo {author} {\bibfnamefont
  {X.}~\bibnamefont {Zhao}},\ }\bibfield  {title} {\enquote {\bibinfo {title}
  {Mg vacancy and dislocation strains as strong phonon scatterers in
  {M}g$_2${S}i$_{1-x}${S}b$_x$ thermoelectric materials},}\ }\href {\doibase
  https://doi.org/10.1016/j.nanoen.2017.03.012} {\bibfield  {journal} {\bibinfo
   {journal} {Nano Energy}\ }\textbf {\bibinfo {volume} {34}},\ \bibinfo
  {pages} {428 } (\bibinfo {year} {2017})}\BibitemShut {NoStop}%
\bibitem [{\citenamefont {Zhou}\ \emph {et~al.}(2018)\citenamefont {Zhou},
  \citenamefont {Lee}, \citenamefont {Cha}, \citenamefont {Yoo}, \citenamefont
  {Cho}, \citenamefont {Hyeon},\ and\ \citenamefont {Chung}}]{Zhou2018}%
  \BibitemOpen
  \bibfield  {author} {\bibinfo {author} {\bibfnamefont {C.}~\bibnamefont
  {Zhou}}, \bibinfo {author} {\bibfnamefont {Y.~K.}\ \bibnamefont {Lee}},
  \bibinfo {author} {\bibfnamefont {J.}~\bibnamefont {Cha}}, \bibinfo {author}
  {\bibfnamefont {B.}~\bibnamefont {Yoo}}, \bibinfo {author} {\bibfnamefont
  {S.-P.}\ \bibnamefont {Cho}}, \bibinfo {author} {\bibfnamefont
  {T.}~\bibnamefont {Hyeon}}, \ and\ \bibinfo {author} {\bibfnamefont
  {I.}~\bibnamefont {Chung}},\ }\bibfield  {title} {\enquote {\bibinfo {title}
  {Defect engineering for high-performance n-{T}ype {P}b{S}e
  thermoelectrics},}\ }\href {\doibase 10.1021/jacs.8b05741} {\bibfield
  {journal} {\bibinfo  {journal} {Journal of the American Chemical Society}\
  }\textbf {\bibinfo {volume} {140}},\ \bibinfo {pages} {9282} (\bibinfo {year}
  {2018})}\BibitemShut {NoStop}%
\bibitem [{\citenamefont {Yu}\ \emph {et~al.}(2018)\citenamefont {Yu},
  \citenamefont {Zhang}, \citenamefont {Mio}, \citenamefont {Gault},
  \citenamefont {Sheskin}, \citenamefont {Scheu}, \citenamefont {Raabe},
  \citenamefont {Zu}, \citenamefont {Wuttig}, \citenamefont {Amouyal},\ and\
  \citenamefont {Cojocaru-Mir\'edin}}]{Yu2018}%
  \BibitemOpen
  \bibfield  {author} {\bibinfo {author} {\bibfnamefont {Y.}~\bibnamefont
  {Yu}}, \bibinfo {author} {\bibfnamefont {S.}~\bibnamefont {Zhang}}, \bibinfo
  {author} {\bibfnamefont {A.~M.}\ \bibnamefont {Mio}}, \bibinfo {author}
  {\bibfnamefont {B.}~\bibnamefont {Gault}}, \bibinfo {author} {\bibfnamefont
  {A.}~\bibnamefont {Sheskin}}, \bibinfo {author} {\bibfnamefont
  {C.}~\bibnamefont {Scheu}}, \bibinfo {author} {\bibfnamefont
  {D.}~\bibnamefont {Raabe}}, \bibinfo {author} {\bibfnamefont
  {F.}~\bibnamefont {Zu}}, \bibinfo {author} {\bibfnamefont {M.}~\bibnamefont
  {Wuttig}}, \bibinfo {author} {\bibfnamefont {Y.}~\bibnamefont {Amouyal}}, \
  and\ \bibinfo {author} {\bibfnamefont {O.}~\bibnamefont
  {Cojocaru-Mir\'edin}},\ }\bibfield  {title} {\enquote {\bibinfo {title}
  {Ag-segregation to dislocations in {P}b{T}e-based thermoelectric
  materials},}\ }\href {\doibase 10.1021/acsami.7b17142} {\bibfield  {journal}
  {\bibinfo  {journal} {ACS Applied Materials \& Interfaces}\ }\textbf
  {\bibinfo {volume} {10}},\ \bibinfo {pages} {3609} (\bibinfo {year}
  {2018})}\BibitemShut {NoStop}%
\bibitem [{\citenamefont {Giaremis}\ \emph {et~al.}(2020)\citenamefont
  {Giaremis}, \citenamefont {Kioseoglou}, \citenamefont {Desmarchelier},
  \citenamefont {Tanguy}, \citenamefont {Isaiev}, \citenamefont {Belabbas},
  \citenamefont {Komninou},\ and\ \citenamefont {Termentzidis}}]{Giaremis2020}%
  \BibitemOpen
  \bibfield  {author} {\bibinfo {author} {\bibfnamefont {S.}~\bibnamefont
  {Giaremis}}, \bibinfo {author} {\bibfnamefont {J.}~\bibnamefont
  {Kioseoglou}}, \bibinfo {author} {\bibfnamefont {P.}~\bibnamefont
  {Desmarchelier}}, \bibinfo {author} {\bibfnamefont {A.}~\bibnamefont
  {Tanguy}}, \bibinfo {author} {\bibfnamefont {M.}~\bibnamefont {Isaiev}},
  \bibinfo {author} {\bibfnamefont {I.}~\bibnamefont {Belabbas}}, \bibinfo
  {author} {\bibfnamefont {P.}~\bibnamefont {Komninou}}, \ and\ \bibinfo
  {author} {\bibfnamefont {K.}~\bibnamefont {Termentzidis}},\ }\bibfield
  {title} {\enquote {\bibinfo {title} {Decorated dislocations against phonon
  propagation for thermal management},}\ }\href {\doibase
  10.1021/acsaem.9b02368} {\bibfield  {journal} {\bibinfo  {journal} {ACS
  Applied Energy Materials}\ }\textbf {\bibinfo {volume} {3}},\ \bibinfo
  {pages} {2682} (\bibinfo {year} {2020})}\BibitemShut {NoStop}%
\bibitem [{\citenamefont {Minnich}\ \emph {et~al.}(2009)\citenamefont
  {Minnich}, \citenamefont {Dresselhaus}, \citenamefont {Ren},\ and\
  \citenamefont {Chen}}]{Minnich2009}%
  \BibitemOpen
  \bibfield  {author} {\bibinfo {author} {\bibfnamefont {A.~J.}\ \bibnamefont
  {Minnich}}, \bibinfo {author} {\bibfnamefont {M.~S.}\ \bibnamefont
  {Dresselhaus}}, \bibinfo {author} {\bibfnamefont {Z.~F.}\ \bibnamefont
  {Ren}}, \ and\ \bibinfo {author} {\bibfnamefont {G.}~\bibnamefont {Chen}},\
  }\bibfield  {title} {\enquote {\bibinfo {title} {Bulk nanostructured
  thermoelectric materials: current research and future prospects},}\ }\href
  {\doibase 10.1039/B822664B} {\bibfield  {journal} {\bibinfo  {journal}
  {Energy Environ. Sci.}\ }\textbf {\bibinfo {volume} {2}},\ \bibinfo {pages}
  {466} (\bibinfo {year} {2009})}\BibitemShut {NoStop}%
\bibitem [{\citenamefont {Lindsay}\ \emph {et~al.}(2013)\citenamefont
  {Lindsay}, \citenamefont {Broido},\ and\ \citenamefont
  {Reinecke}}]{Lindsay2013}%
  \BibitemOpen
  \bibfield  {author} {\bibinfo {author} {\bibfnamefont {L.}~\bibnamefont
  {Lindsay}}, \bibinfo {author} {\bibfnamefont {D.~A.}\ \bibnamefont {Broido}},
  \ and\ \bibinfo {author} {\bibfnamefont {T.~L.}\ \bibnamefont {Reinecke}},\
  }\bibfield  {title} {\enquote {\bibinfo {title} {Ab initio thermal transport
  in compound semiconductors},}\ }\href {\doibase 10.1103/PhysRevB.87.165201}
  {\bibfield  {journal} {\bibinfo  {journal} {Phys. Rev. B}\ }\textbf {\bibinfo
  {volume} {87}},\ \bibinfo {pages} {165201} (\bibinfo {year}
  {2013})}\BibitemShut {NoStop}%
\bibitem [{\citenamefont {Lee}\ \emph {et~al.}(2014)\citenamefont {Lee},
  \citenamefont {Esfarjani}, \citenamefont {Mendoza}, \citenamefont
  {Dresselhaus},\ and\ \citenamefont {Chen}}]{Lee2014}%
  \BibitemOpen
  \bibfield  {author} {\bibinfo {author} {\bibfnamefont {S.}~\bibnamefont
  {Lee}}, \bibinfo {author} {\bibfnamefont {K.}~\bibnamefont {Esfarjani}},
  \bibinfo {author} {\bibfnamefont {J.}~\bibnamefont {Mendoza}}, \bibinfo
  {author} {\bibfnamefont {M.~S.}\ \bibnamefont {Dresselhaus}}, \ and\ \bibinfo
  {author} {\bibfnamefont {G.}~\bibnamefont {Chen}},\ }\bibfield  {title}
  {\enquote {\bibinfo {title} {Lattice thermal conductivity of {B}i, {S}b, and
  {B}i-{S}b alloy from first principles},}\ }\href {\doibase
  10.1103/PhysRevB.89.085206} {\bibfield  {journal} {\bibinfo  {journal} {Phys.
  Rev. B}\ }\textbf {\bibinfo {volume} {89}},\ \bibinfo {pages} {085206}
  (\bibinfo {year} {2014})}\BibitemShut {NoStop}%
\bibitem [{\citenamefont {Tian}\ \emph {et~al.}(2014)\citenamefont {Tian},
  \citenamefont {Lee},\ and\ \citenamefont {Chen}}]{Tian2014}%
  \BibitemOpen
  \bibfield  {author} {\bibinfo {author} {\bibfnamefont {Z.}~\bibnamefont
  {Tian}}, \bibinfo {author} {\bibfnamefont {S.}~\bibnamefont {Lee}}, \ and\
  \bibinfo {author} {\bibfnamefont {G.}~\bibnamefont {Chen}},\ }\bibfield
  {title} {\enquote {\bibinfo {title} {Comprehensive review of heat transfer in
  thermoelectric materials and devices},}\ }\href@noop {} {\bibfield  {journal}
  {\bibinfo  {journal} {Annual review of heat transfer}\ }\textbf {\bibinfo
  {volume} {17}} (\bibinfo {year} {2014})}\BibitemShut {NoStop}%
\bibitem [{\citenamefont {Jund}\ and\ \citenamefont
  {Jullien}(1999)}]{Jund1999}%
  \BibitemOpen
  \bibfield  {author} {\bibinfo {author} {\bibfnamefont {P.}~\bibnamefont
  {Jund}}\ and\ \bibinfo {author} {\bibfnamefont {R.}~\bibnamefont {Jullien}},\
  }\bibfield  {title} {\enquote {\bibinfo {title} {Molecular-dynamics
  calculation of the thermal conductivity of vitreous silica},}\ }\href
  {\doibase 10.1103/PhysRevB.59.13707} {\bibfield  {journal} {\bibinfo
  {journal} {Phys. Rev. B}\ }\textbf {\bibinfo {volume} {59}},\ \bibinfo
  {pages} {13707} (\bibinfo {year} {1999})}\BibitemShut {NoStop}%
\bibitem [{\citenamefont {M{\"{u}}ller-Plathe}(1997)}]{Muller1997}%
  \BibitemOpen
  \bibfield  {author} {\bibinfo {author} {\bibfnamefont {F.}~\bibnamefont
  {M{\"{u}}ller-Plathe}},\ }\bibfield  {title} {\enquote {\bibinfo {title} {A
  simple nonequilibrium molecular dynamics method for calculating the thermal
  conductivity},}\ }\href {\doibase 10.1063/1.473271} {\bibfield  {journal}
  {\bibinfo  {journal} {The Journal of Chemical Physics}\ }\textbf {\bibinfo
  {volume} {106}},\ \bibinfo {pages} {6082} (\bibinfo {year}
  {1997})}\BibitemShut {NoStop}%
\bibitem [{\citenamefont {Bedoya-Mart\'{\i}nez}\ \emph
  {et~al.}(2014)\citenamefont {Bedoya-Mart\'{\i}nez}, \citenamefont {Barrat},\
  and\ \citenamefont {Rodney}}]{Bedoya2014}%
  \BibitemOpen
  \bibfield  {author} {\bibinfo {author} {\bibfnamefont {O.~N.}\ \bibnamefont
  {Bedoya-Mart\'{\i}nez}}, \bibinfo {author} {\bibfnamefont {J.-L.}\
  \bibnamefont {Barrat}}, \ and\ \bibinfo {author} {\bibfnamefont
  {D.}~\bibnamefont {Rodney}},\ }\bibfield  {title} {\enquote {\bibinfo {title}
  {Computation of the thermal conductivity using methods based on classical and
  quantum molecular dynamics},}\ }\href {\doibase 10.1103/PhysRevB.89.014303}
  {\bibfield  {journal} {\bibinfo  {journal} {Phys. Rev. B}\ }\textbf {\bibinfo
  {volume} {89}},\ \bibinfo {pages} {014303} (\bibinfo {year}
  {2014})}\BibitemShut {NoStop}%
\bibitem [{\citenamefont {Klemens}(1955)}]{Klemens1955}%
  \BibitemOpen
  \bibfield  {author} {\bibinfo {author} {\bibfnamefont {P.~G.}\ \bibnamefont
  {Klemens}},\ }\bibfield  {title} {\enquote {\bibinfo {title} {The scattering
  of low-frequency lattice waves by static imperfections},}\ }\href
  {http://stacks.iop.org/0370-1298/68/i=12/a=303} {\bibfield  {journal}
  {\bibinfo  {journal} {Proceedings of the Physical Society. Section A}\
  }\textbf {\bibinfo {volume} {68}},\ \bibinfo {pages} {1113} (\bibinfo {year}
  {1955})}\BibitemShut {NoStop}%
\bibitem [{\citenamefont {Carruthers}(1959)}]{Carruthers1959}%
  \BibitemOpen
  \bibfield  {author} {\bibinfo {author} {\bibfnamefont {P.}~\bibnamefont
  {Carruthers}},\ }\bibfield  {title} {\enquote {\bibinfo {title} {Scattering
  of phonons by elastic strain fields and the thermal resistance of
  dislocations},}\ }\href {\doibase 10.1103/PhysRev.114.995} {\bibfield
  {journal} {\bibinfo  {journal} {Phys. Rev.}\ }\textbf {\bibinfo {volume}
  {114}},\ \bibinfo {pages} {995} (\bibinfo {year} {1959})}\BibitemShut
  {NoStop}%
\bibitem [{\citenamefont {Lund}\ and\ \citenamefont
  {Scheihing~H.}(2019)}]{Lund2019}%
  \BibitemOpen
  \bibfield  {author} {\bibinfo {author} {\bibfnamefont {F.}~\bibnamefont
  {Lund}}\ and\ \bibinfo {author} {\bibfnamefont {B.}~\bibnamefont
  {Scheihing~H.}},\ }\bibfield  {title} {\enquote {\bibinfo {title} {Scattering
  of phonons by quantum-dislocation segments in an elastic continuum},}\ }\href
  {\doibase 10.1103/PhysRevB.99.214102} {\bibfield  {journal} {\bibinfo
  {journal} {Phys. Rev. B}\ }\textbf {\bibinfo {volume} {99}},\ \bibinfo
  {pages} {214102} (\bibinfo {year} {2019})}\BibitemShut {NoStop}%
\bibitem [{\citenamefont {Lund}\ and\ \citenamefont
  {Scheihing-Hitschfeld}(2020)}]{Lund2020}%
  \BibitemOpen
  \bibfield  {author} {\bibinfo {author} {\bibfnamefont {F.}~\bibnamefont
  {Lund}}\ and\ \bibinfo {author} {\bibfnamefont {B.}~\bibnamefont
  {Scheihing-Hitschfeld}},\ }\bibfield  {title} {\enquote {\bibinfo {title}
  {The scattering of phonons by infinitely long quantum dislocations segments
  and the generation of thermal transport anisotropy in a solid threaded by
  many parallel dislocations},}\ }\href {\doibase 10.3390/nano10091711}
  {\bibfield  {journal} {\bibinfo  {journal} {Nanomaterials}\ }\textbf
  {\bibinfo {volume} {10}} (\bibinfo {year} {2020}),\
  10.3390/nano10091711}\BibitemShut {NoStop}%
\bibitem [{\citenamefont {Granato}\ and\ \citenamefont
  {L\"ucke}(1956{\natexlab{a}})}]{Granato1956a}%
  \BibitemOpen
  \bibfield  {author} {\bibinfo {author} {\bibfnamefont {A.}~\bibnamefont
  {Granato}}\ and\ \bibinfo {author} {\bibfnamefont {K.}~\bibnamefont
  {L\"ucke}},\ }\bibfield  {title} {\enquote {\bibinfo {title} {Theory of
  mechanical damping due to dislocations},}\ }\href {\doibase
  10.1063/1.1722436} {\bibfield  {journal} {\bibinfo  {journal} {Journal of
  Applied Physics}\ }\textbf {\bibinfo {volume} {27}},\ \bibinfo {pages} {583}
  (\bibinfo {year} {1956}{\natexlab{a}})}\BibitemShut {NoStop}%
\bibitem [{\citenamefont {Granato}\ and\ \citenamefont
  {L\"ucke}(1956{\natexlab{b}})}]{Granato1956b}%
  \BibitemOpen
  \bibfield  {author} {\bibinfo {author} {\bibfnamefont {A.}~\bibnamefont
  {Granato}}\ and\ \bibinfo {author} {\bibfnamefont {K.}~\bibnamefont
  {L\"ucke}},\ }\bibfield  {title} {\enquote {\bibinfo {title} {Application of
  dislocation theory to internal friction phenomena at high frequencies},}\
  }\href {\doibase 10.1063/1.1722485} {\bibfield  {journal} {\bibinfo
  {journal} {Journal of Applied Physics}\ }\textbf {\bibinfo {volume} {27}},\
  \bibinfo {pages} {789} (\bibinfo {year} {1956}{\natexlab{b}})}\BibitemShut
  {NoStop}%
\bibitem [{\citenamefont {L\"ucke}\ and\ \citenamefont
  {Granato}(1981)}]{Lucke1981}%
  \BibitemOpen
  \bibfield  {author} {\bibinfo {author} {\bibfnamefont {K.}~\bibnamefont
  {L\"ucke}}\ and\ \bibinfo {author} {\bibfnamefont {A.~V.}\ \bibnamefont
  {Granato}},\ }\bibfield  {title} {\enquote {\bibinfo {title} {Simplified
  theory of dislocation damping including point-defect drag. i. theory of drag
  by equidistant point defects},}\ }\href {\doibase 10.1103/PhysRevB.24.6991}
  {\bibfield  {journal} {\bibinfo  {journal} {Phys. Rev. B}\ }\textbf {\bibinfo
  {volume} {24}},\ \bibinfo {pages} {6991} (\bibinfo {year}
  {1981})}\BibitemShut {NoStop}%
\bibitem [{\citenamefont {Kneezel}\ and\ \citenamefont
  {Granato}(1982)}]{Kneezel1982}%
  \BibitemOpen
  \bibfield  {author} {\bibinfo {author} {\bibfnamefont {G.~A.}\ \bibnamefont
  {Kneezel}}\ and\ \bibinfo {author} {\bibfnamefont {A.~V.}\ \bibnamefont
  {Granato}},\ }\bibfield  {title} {\enquote {\bibinfo {title} {Effect of
  independent and coupled vibrations of dislocations on low-temperature thermal
  conductivity in alkali halides},}\ }\href {\doibase 10.1103/PhysRevB.25.2851}
  {\bibfield  {journal} {\bibinfo  {journal} {Phys. Rev. B}\ }\textbf {\bibinfo
  {volume} {25}},\ \bibinfo {pages} {2851} (\bibinfo {year}
  {1982})}\BibitemShut {NoStop}%
\bibitem [{\citenamefont {Shilo}\ and\ \citenamefont
  {Zolotoyabko}(2007)}]{Shilo2007}%
  \BibitemOpen
  \bibfield  {author} {\bibinfo {author} {\bibfnamefont {D.}~\bibnamefont
  {Shilo}}\ and\ \bibinfo {author} {\bibfnamefont {E.}~\bibnamefont
  {Zolotoyabko}},\ }\bibfield  {title} {\enquote {\bibinfo {title} {X-ray
  imaging of phonon interaction with dislocations},}\ \ }(\bibinfo  {publisher}
  {Elsevier},\ \bibinfo {year} {2007})\ Chap.~\bibinfo {chapter} {80}, pp.\
  \bibinfo {pages} {603--639}\BibitemShut {NoStop}%
\bibitem [{\citenamefont {Maurel}\ \emph
  {et~al.}(2005{\natexlab{a}})\citenamefont {Maurel}, \citenamefont {Pagneux},
  \citenamefont {Barra},\ and\ \citenamefont {Lund}}]{Maurel2005a}%
  \BibitemOpen
  \bibfield  {author} {\bibinfo {author} {\bibfnamefont {A.}~\bibnamefont
  {Maurel}}, \bibinfo {author} {\bibfnamefont {V.}~\bibnamefont {Pagneux}},
  \bibinfo {author} {\bibfnamefont {F.}~\bibnamefont {Barra}}, \ and\ \bibinfo
  {author} {\bibfnamefont {F.}~\bibnamefont {Lund}},\ }\bibfield  {title}
  {\enquote {\bibinfo {title} {Interaction between an elastic wave and a single
  pinned dislocation},}\ }\href {\doibase 10.1103/PhysRevB.72.174110}
  {\bibfield  {journal} {\bibinfo  {journal} {Phys. Rev. B}\ }\textbf {\bibinfo
  {volume} {72}},\ \bibinfo {pages} {174110} (\bibinfo {year}
  {2005}{\natexlab{a}})}\BibitemShut {NoStop}%
\bibitem [{\citenamefont {Mujica}\ \emph {et~al.}(2012)\citenamefont {Mujica},
  \citenamefont {Cerda}, \citenamefont {Espinoza}, \citenamefont {Lisoni},\
  and\ \citenamefont {Lund}}]{Mujica2012}%
  \BibitemOpen
  \bibfield  {author} {\bibinfo {author} {\bibfnamefont {N.}~\bibnamefont
  {Mujica}}, \bibinfo {author} {\bibfnamefont {M.~T.}\ \bibnamefont {Cerda}},
  \bibinfo {author} {\bibfnamefont {R.}~\bibnamefont {Espinoza}}, \bibinfo
  {author} {\bibfnamefont {J.}~\bibnamefont {Lisoni}}, \ and\ \bibinfo {author}
  {\bibfnamefont {F.}~\bibnamefont {Lund}},\ }\bibfield  {title} {\enquote
  {\bibinfo {title} {Ultrasound as a probe of dislocation density in
  aluminum},}\ }\href {\doibase https://doi.org/10.1016/j.actamat.2012.07.023}
  {\bibfield  {journal} {\bibinfo  {journal} {Acta Materialia}\ }\textbf
  {\bibinfo {volume} {60}},\ \bibinfo {pages} {5828 } (\bibinfo {year}
  {2012})}\BibitemShut {NoStop}%
\bibitem [{\citenamefont {Barra}\ \emph {et~al.}(2015)\citenamefont {Barra},
  \citenamefont {Espinoza-Gonz{\'a}lez}, \citenamefont {Fern{\'a}ndez},
  \citenamefont {Lund}, \citenamefont {Maurel},\ and\ \citenamefont
  {Pagneux}}]{Barra2015}%
  \BibitemOpen
  \bibfield  {author} {\bibinfo {author} {\bibfnamefont {F.}~\bibnamefont
  {Barra}}, \bibinfo {author} {\bibfnamefont {R.}~\bibnamefont
  {Espinoza-Gonz{\'a}lez}}, \bibinfo {author} {\bibfnamefont {H.}~\bibnamefont
  {Fern{\'a}ndez}}, \bibinfo {author} {\bibfnamefont {F.}~\bibnamefont {Lund}},
  \bibinfo {author} {\bibfnamefont {A.}~\bibnamefont {Maurel}}, \ and\ \bibinfo
  {author} {\bibfnamefont {V.}~\bibnamefont {Pagneux}},\ }\bibfield  {title}
  {\enquote {\bibinfo {title} {The use of ultrasound to measure dislocation
  density},}\ }\href {\doibase 10.1007/s11837-015-1458-9} {\bibfield  {journal}
  {\bibinfo  {journal} {JOM}\ }\textbf {\bibinfo {volume} {67}},\ \bibinfo
  {pages} {1856} (\bibinfo {year} {2015})}\BibitemShut {NoStop}%
\bibitem [{\citenamefont {Salinas}\ \emph {et~al.}(2017)\citenamefont
  {Salinas}, \citenamefont {Aguilar}, \citenamefont {Espinoza-Gonz{\'a}lez},
  \citenamefont {Lund},\ and\ \citenamefont {Mujica}}]{Salinas2017}%
  \BibitemOpen
  \bibfield  {author} {\bibinfo {author} {\bibfnamefont {V.}~\bibnamefont
  {Salinas}}, \bibinfo {author} {\bibfnamefont {C.}~\bibnamefont {Aguilar}},
  \bibinfo {author} {\bibfnamefont {R.}~\bibnamefont {Espinoza-Gonz{\'a}lez}},
  \bibinfo {author} {\bibfnamefont {F.}~\bibnamefont {Lund}}, \ and\ \bibinfo
  {author} {\bibfnamefont {N.}~\bibnamefont {Mujica}},\ }\bibfield  {title}
  {\enquote {\bibinfo {title} {In situ monitoring of dislocation proliferation
  during plastic deformation using ultrasound},}\ }\href {\doibase
  https://doi.org/10.1016/j.ijplas.2017.06.001} {\bibfield  {journal} {\bibinfo
   {journal} {International Journal of Plasticity}\ }\textbf {\bibinfo {volume}
  {97}},\ \bibinfo {pages} {178 } (\bibinfo {year} {2017})}\BibitemShut
  {NoStop}%
\bibitem [{\citenamefont {Espinoza}\ \emph {et~al.}(2018)\citenamefont
  {Espinoza}, \citenamefont {Feli\'u}, \citenamefont {Aguilar}, \citenamefont
  {Espinoza-Gonz\'alez}, \citenamefont {Lund}, \citenamefont {Salinas},\ and\
  \citenamefont {Mujica}}]{Espinoza2018}%
  \BibitemOpen
  \bibfield  {author} {\bibinfo {author} {\bibfnamefont {C.}~\bibnamefont
  {Espinoza}}, \bibinfo {author} {\bibfnamefont {D.}~\bibnamefont {Feli\'u}},
  \bibinfo {author} {\bibfnamefont {C.}~\bibnamefont {Aguilar}}, \bibinfo
  {author} {\bibfnamefont {R.}~\bibnamefont {Espinoza-Gonz\'alez}}, \bibinfo
  {author} {\bibfnamefont {F.}~\bibnamefont {Lund}}, \bibinfo {author}
  {\bibfnamefont {V.}~\bibnamefont {Salinas}}, \ and\ \bibinfo {author}
  {\bibfnamefont {N.}~\bibnamefont {Mujica}},\ }\bibfield  {title} {\enquote
  {\bibinfo {title} {Linear versus nonlinear acoustic probing of plasticity in
  metals: A quantitative assessment},}\ }\href {\doibase 10.3390/ma11112217}
  {\bibfield  {journal} {\bibinfo  {journal} {Materials}\ }\textbf {\bibinfo
  {volume} {11}} (\bibinfo {year} {2018}),\ 10.3390/ma11112217}\BibitemShut
  {NoStop}%
\bibitem [{\citenamefont {Maurel}\ \emph
  {et~al.}(2005{\natexlab{b}})\citenamefont {Maurel}, \citenamefont {Pagneux},
  \citenamefont {Barra},\ and\ \citenamefont {Lund}}]{Maurel2005b}%
  \BibitemOpen
  \bibfield  {author} {\bibinfo {author} {\bibfnamefont {A.}~\bibnamefont
  {Maurel}}, \bibinfo {author} {\bibfnamefont {V.}~\bibnamefont {Pagneux}},
  \bibinfo {author} {\bibfnamefont {F.}~\bibnamefont {Barra}}, \ and\ \bibinfo
  {author} {\bibfnamefont {F.}~\bibnamefont {Lund}},\ }\bibfield  {title}
  {\enquote {\bibinfo {title} {Wave propagation through a random array of
  pinned dislocations: Velocity change and attenuation in a generalized
  {G}ranato and {L}\"ucke theory},}\ }\href {\doibase
  10.1103/PhysRevB.72.174111} {\bibfield  {journal} {\bibinfo  {journal} {Phys.
  Rev. B}\ }\textbf {\bibinfo {volume} {72}},\ \bibinfo {pages} {174111}
  (\bibinfo {year} {2005}{\natexlab{b}})}\BibitemShut {NoStop}%
\bibitem [{\citenamefont {Churochkin}\ \emph {et~al.}(2016)\citenamefont
  {Churochkin}, \citenamefont {Barra}, \citenamefont {Lund}, \citenamefont
  {Maurel},\ and\ \citenamefont {Pagneux}}]{Churochkin2016}%
  \BibitemOpen
  \bibfield  {author} {\bibinfo {author} {\bibfnamefont {D.}~\bibnamefont
  {Churochkin}}, \bibinfo {author} {\bibfnamefont {F.}~\bibnamefont {Barra}},
  \bibinfo {author} {\bibfnamefont {F.}~\bibnamefont {Lund}}, \bibinfo {author}
  {\bibfnamefont {A.}~\bibnamefont {Maurel}}, \ and\ \bibinfo {author}
  {\bibfnamefont {V.}~\bibnamefont {Pagneux}},\ }\bibfield  {title} {\enquote
  {\bibinfo {title} {Multiple scattering of elastic waves by pinned dislocation
  segments in a continuum},}\ }\href {\doibase
  https://doi.org/10.1016/j.wavemoti.2015.10.005} {\bibfield  {journal}
  {\bibinfo  {journal} {Wave Motion}\ }\textbf {\bibinfo {volume} {60}},\
  \bibinfo {pages} {220 } (\bibinfo {year} {2016})}\BibitemShut {NoStop}%
\bibitem [{\citenamefont {Sheng}(2006)}]{Sheng2006}%
  \BibitemOpen
  \bibfield  {author} {\bibinfo {author} {\bibfnamefont {P.}~\bibnamefont
  {Sheng}},\ }\href {\doibase 10.1007/3-540-29156-3} {\emph {\bibinfo {title}
  {Introduction to Wave Scattering, Localization and Mesoscopic Phenomena}}}\
  (\bibinfo  {publisher} {Academic, New York},\ \bibinfo {year}
  {2006})\BibitemShut {NoStop}%
\bibitem [{\citenamefont {Vollhardt}\ and\ \citenamefont
  {W\"olfle}(1980)}]{Vollhardt1980}%
  \BibitemOpen
  \bibfield  {author} {\bibinfo {author} {\bibfnamefont {D.}~\bibnamefont
  {Vollhardt}}\ and\ \bibinfo {author} {\bibfnamefont {P.}~\bibnamefont
  {W\"olfle}},\ }\bibfield  {title} {\enquote {\bibinfo {title} {Diagrammatic,
  self-consistent treatment of the anderson localization problem in
  $d\ensuremath{\le}2$ dimensions},}\ }\href {\doibase
  10.1103/PhysRevB.22.4666} {\bibfield  {journal} {\bibinfo  {journal} {Phys.
  Rev. B}\ }\textbf {\bibinfo {volume} {22}},\ \bibinfo {pages} {4666}
  (\bibinfo {year} {1980})}\BibitemShut {NoStop}%
\bibitem [{\citenamefont {W\"olfle}\ and\ \citenamefont
  {Bhatt}(1984)}]{Wolfle1984}%
  \BibitemOpen
  \bibfield  {author} {\bibinfo {author} {\bibfnamefont {P.}~\bibnamefont
  {W\"olfle}}\ and\ \bibinfo {author} {\bibfnamefont {R.~N.}\ \bibnamefont
  {Bhatt}},\ }\bibfield  {title} {\enquote {\bibinfo {title} {Electron
  localization in anisotropic systems},}\ }\href {\doibase
  10.1103/PhysRevB.30.3542} {\bibfield  {journal} {\bibinfo  {journal} {Phys.
  Rev. B}\ }\textbf {\bibinfo {volume} {30}},\ \bibinfo {pages} {3542}
  (\bibinfo {year} {1984})}\BibitemShut {NoStop}%
\bibitem [{\citenamefont {Bhatt}\ \emph {et~al.}(1985)\citenamefont {Bhatt},
  \citenamefont {W\"olfle},\ and\ \citenamefont {Ramakrishnan}}]{Bhatt1985}%
  \BibitemOpen
  \bibfield  {author} {\bibinfo {author} {\bibfnamefont {R.~N.}\ \bibnamefont
  {Bhatt}}, \bibinfo {author} {\bibfnamefont {P.}~\bibnamefont {W\"olfle}}, \
  and\ \bibinfo {author} {\bibfnamefont {T.~V.}\ \bibnamefont {Ramakrishnan}},\
  }\bibfield  {title} {\enquote {\bibinfo {title} {Localization and interaction
  effects in anisotropic disordered electronic systems},}\ }\href {\doibase
  10.1103/PhysRevB.32.569} {\bibfield  {journal} {\bibinfo  {journal} {Phys.
  Rev. B}\ }\textbf {\bibinfo {volume} {32}},\ \bibinfo {pages} {569} (\bibinfo
  {year} {1985})}\BibitemShut {NoStop}%
\bibitem [{\citenamefont {Kirkpatrick}(1985)}]{Kirkpatrick1985}%
  \BibitemOpen
  \bibfield  {author} {\bibinfo {author} {\bibfnamefont {T.~R.}\ \bibnamefont
  {Kirkpatrick}},\ }\bibfield  {title} {\enquote {\bibinfo {title}
  {Localization of acoustic waves},}\ }\href {\doibase
  10.1103/PhysRevB.31.5746} {\bibfield  {journal} {\bibinfo  {journal} {Phys.
  Rev. B}\ }\textbf {\bibinfo {volume} {31}},\ \bibinfo {pages} {5746}
  (\bibinfo {year} {1985})}\BibitemShut {NoStop}%
\bibitem [{\citenamefont {Barabanenkov}\ and\ \citenamefont
  {Ozrin}(1991)}]{Barabanenkov1991}%
  \BibitemOpen
  \bibfield  {author} {\bibinfo {author} {\bibfnamefont {Y.}~\bibnamefont
  {Barabanenkov}}\ and\ \bibinfo {author} {\bibfnamefont {V.}~\bibnamefont
  {Ozrin}},\ }\bibfield  {title} {\enquote {\bibinfo {title} {Asymptotic
  solution of the {B}ethe-{S}alpeter equation and the {G}reen-{K}ubo formula
  for the diffusion constant for wave propagation in random media},}\ }\href
  {\doibase https://doi.org/10.1016/0375-9601(91)90425-8} {\bibfield  {journal}
  {\bibinfo  {journal} {Physics Letters A}\ }\textbf {\bibinfo {volume}
  {154}},\ \bibinfo {pages} {38 } (\bibinfo {year} {1991})}\BibitemShut
  {NoStop}%
\bibitem [{\citenamefont {Barabanenkov}\ and\ \citenamefont
  {Ozrin}(1995)}]{Barabanenkov1995}%
  \BibitemOpen
  \bibfield  {author} {\bibinfo {author} {\bibfnamefont {Y.}~\bibnamefont
  {Barabanenkov}}\ and\ \bibinfo {author} {\bibfnamefont {V.}~\bibnamefont
  {Ozrin}},\ }\bibfield  {title} {\enquote {\bibinfo {title} {Diffusion
  asymptotics of the {B}ethe-{S}alpeter equation for electromagnetic waves in
  discrete random media},}\ }\href {\doibase
  https://doi.org/10.1016/0375-9601(95)00576-O} {\bibfield  {journal} {\bibinfo
   {journal} {Physics Letters A}\ }\textbf {\bibinfo {volume} {206}},\ \bibinfo
  {pages} {116 } (\bibinfo {year} {1995})}\BibitemShut {NoStop}%
\bibitem [{\citenamefont {Stark}\ and\ \citenamefont
  {Lubensky}(1997)}]{Stark1997}%
  \BibitemOpen
  \bibfield  {author} {\bibinfo {author} {\bibfnamefont {H.}~\bibnamefont
  {Stark}}\ and\ \bibinfo {author} {\bibfnamefont {T.~C.}\ \bibnamefont
  {Lubensky}},\ }\bibfield  {title} {\enquote {\bibinfo {title} {Multiple light
  scattering in anisotropic random media},}\ }\href {\doibase
  10.1103/PhysRevE.55.514} {\bibfield  {journal} {\bibinfo  {journal} {Phys.
  Rev. E}\ }\textbf {\bibinfo {volume} {55}},\ \bibinfo {pages} {514} (\bibinfo
  {year} {1997})}\BibitemShut {NoStop}%
\bibitem [{\citenamefont {Weaver}(1990)}]{Weaver1990}%
  \BibitemOpen
  \bibfield  {author} {\bibinfo {author} {\bibfnamefont {R.}~\bibnamefont
  {Weaver}},\ }\bibfield  {title} {\enquote {\bibinfo {title} {Diffusivity of
  ultrasound in polycrystals},}\ }\href {\doibase
  https://doi.org/10.1016/0022-5096(90)90021-U} {\bibfield  {journal} {\bibinfo
   {journal} {Journal of the Mechanics and Physics of Solids}\ }\textbf
  {\bibinfo {volume} {38}},\ \bibinfo {pages} {55 } (\bibinfo {year}
  {1990})}\BibitemShut {NoStop}%
\bibitem [{\citenamefont {van Tiggelen}\ \emph {et~al.}(2001)\citenamefont {van
  Tiggelen}, \citenamefont {Margerin},\ and\ \citenamefont
  {Campillo}}]{Tiggelen2001}%
  \BibitemOpen
  \bibfield  {author} {\bibinfo {author} {\bibfnamefont {B.~A.}\ \bibnamefont
  {van Tiggelen}}, \bibinfo {author} {\bibfnamefont {L.}~\bibnamefont
  {Margerin}}, \ and\ \bibinfo {author} {\bibfnamefont {M.}~\bibnamefont
  {Campillo}},\ }\bibfield  {title} {\enquote {\bibinfo {title} {Coherent
  backscattering of elastic waves: Specific role of source, polarization, and
  near field},}\ }\href {\doibase 10.1121/1.1388017} {\bibfield  {journal}
  {\bibinfo  {journal} {The Journal of the Acoustical Society of America}\
  }\textbf {\bibinfo {volume} {110}},\ \bibinfo {pages} {1291} (\bibinfo {year}
  {2001})}\BibitemShut {NoStop}%
\bibitem [{\citenamefont {Tr{\'{e}}gour{\`{e}}s}\ and\ \citenamefont {van
  Tiggelen}(2002)}]{Tregoures2002a}%
  \BibitemOpen
  \bibfield  {author} {\bibinfo {author} {\bibfnamefont {N.~P.}\ \bibnamefont
  {Tr{\'{e}}gour{\`{e}}s}}\ and\ \bibinfo {author} {\bibfnamefont {B.~A.}\
  \bibnamefont {van Tiggelen}},\ }\bibfield  {title} {\enquote {\bibinfo
  {title} {Generalized diffusion equation for multiple scattered elastic
  waves},}\ }\href {\doibase 10.1088/0959-7174/12/1/302} {\bibfield  {journal}
  {\bibinfo  {journal} {Waves in Random Media}\ }\textbf {\bibinfo {volume}
  {12}},\ \bibinfo {pages} {21} (\bibinfo {year} {2002})}\BibitemShut {NoStop}%
\bibitem [{\citenamefont {Tr\'egour\`es}\ and\ \citenamefont {van
  Tiggelen}(2002)}]{Tregoures2002b}%
  \BibitemOpen
  \bibfield  {author} {\bibinfo {author} {\bibfnamefont {N.~P.}\ \bibnamefont
  {Tr\'egour\`es}}\ and\ \bibinfo {author} {\bibfnamefont {B.~A.}\ \bibnamefont
  {van Tiggelen}},\ }\bibfield  {title} {\enquote {\bibinfo {title}
  {Quasi-two-dimensional transfer of elastic waves},}\ }\href {\doibase
  10.1103/PhysRevE.66.036601} {\bibfield  {journal} {\bibinfo  {journal} {Phys.
  Rev. E}\ }\textbf {\bibinfo {volume} {66}},\ \bibinfo {pages} {036601}
  (\bibinfo {year} {2002})}\BibitemShut {NoStop}%
\bibitem [{\citenamefont {Trujillo}\ \emph {et~al.}(2010)\citenamefont
  {Trujillo}, \citenamefont {Peniche},\ and\ \citenamefont
  {Sigalotti}}]{Trujillo2010}%
  \BibitemOpen
  \bibfield  {author} {\bibinfo {author} {\bibfnamefont {L.}~\bibnamefont
  {Trujillo}}, \bibinfo {author} {\bibfnamefont {F.}~\bibnamefont {Peniche}}, \
  and\ \bibinfo {author} {\bibfnamefont {L.~D.~G.}\ \bibnamefont {Sigalotti}},\
  }\bibfield  {title} {\enquote {\bibinfo {title} {Derivation of a
  schr{\"o}dinger-like equation for elastic waves in granular media},}\ }\href
  {\doibase 10.1007/s10035-010-0190-y} {\bibfield  {journal} {\bibinfo
  {journal} {Granular Matter}\ }\textbf {\bibinfo {volume} {12}},\ \bibinfo
  {pages} {417} (\bibinfo {year} {2010})}\BibitemShut {NoStop}%
\bibitem [{\citenamefont {Cobus}\ \emph {et~al.}(2018)\citenamefont {Cobus},
  \citenamefont {Hildebrand}, \citenamefont {Skipetrov}, \citenamefont {van
  Tiggelen},\ and\ \citenamefont {Page}}]{Cobus2018}%
  \BibitemOpen
  \bibfield  {author} {\bibinfo {author} {\bibfnamefont {L.~A.}\ \bibnamefont
  {Cobus}}, \bibinfo {author} {\bibfnamefont {W.~K.}\ \bibnamefont
  {Hildebrand}}, \bibinfo {author} {\bibfnamefont {S.~E.}\ \bibnamefont
  {Skipetrov}}, \bibinfo {author} {\bibfnamefont {B.~A.}\ \bibnamefont {van
  Tiggelen}}, \ and\ \bibinfo {author} {\bibfnamefont {J.~H.}\ \bibnamefont
  {Page}},\ }\bibfield  {title} {\enquote {\bibinfo {title} {Transverse
  confinement of ultrasound through the anderson transition in
  three-dimensional mesoglasses},}\ }\href {\doibase
  10.1103/PhysRevB.98.214201} {\bibfield  {journal} {\bibinfo  {journal} {Phys.
  Rev. B}\ }\textbf {\bibinfo {volume} {98}},\ \bibinfo {pages} {214201}
  (\bibinfo {year} {2018})}\BibitemShut {NoStop}%
\bibitem [{\citenamefont {Go\"{\i}coechea}\ \emph {et~al.}(2020)\citenamefont
  {Go\"{\i}coechea}, \citenamefont {Skipetrov},\ and\ \citenamefont
  {Page}}]{Goicoechea2020}%
  \BibitemOpen
  \bibfield  {author} {\bibinfo {author} {\bibfnamefont {A.}~\bibnamefont
  {Go\"{\i}coechea}}, \bibinfo {author} {\bibfnamefont {S.~E.}\ \bibnamefont
  {Skipetrov}}, \ and\ \bibinfo {author} {\bibfnamefont {J.~H.}\ \bibnamefont
  {Page}},\ }\bibfield  {title} {\enquote {\bibinfo {title} {Suppression of
  transport anisotropy at the anderson localization transition in
  three-dimensional anisotropic media},}\ }\href {\doibase
  10.1103/PhysRevB.102.220201} {\bibfield  {journal} {\bibinfo  {journal}
  {Phys. Rev. B}\ }\textbf {\bibinfo {volume} {102}},\ \bibinfo {pages}
  {220201} (\bibinfo {year} {2020})}\BibitemShut {NoStop}%
\bibitem [{\citenamefont {Lund}(2015)}]{Lund2015}%
  \BibitemOpen
  \bibfield  {author} {\bibinfo {author} {\bibfnamefont {F.}~\bibnamefont
  {Lund}},\ }\bibfield  {title} {\enquote {\bibinfo {title} {Normal modes and
  acoustic properties of an elastic solid with line defects},}\ }\href
  {\doibase 10.1103/PhysRevB.91.094102} {\bibfield  {journal} {\bibinfo
  {journal} {Phys. Rev. B}\ }\textbf {\bibinfo {volume} {91}},\ \bibinfo
  {pages} {094102} (\bibinfo {year} {2015})}\BibitemShut {NoStop}%
\bibitem [{\citenamefont {Bianchi}\ \emph {et~al.}(2020)\citenamefont
  {Bianchi}, \citenamefont {Giordano},\ and\ \citenamefont
  {Lund}}]{Bianchi2020}%
  \BibitemOpen
  \bibfield  {author} {\bibinfo {author} {\bibfnamefont {E.}~\bibnamefont
  {Bianchi}}, \bibinfo {author} {\bibfnamefont {V.~M.}\ \bibnamefont
  {Giordano}}, \ and\ \bibinfo {author} {\bibfnamefont {F.}~\bibnamefont
  {Lund}},\ }\bibfield  {title} {\enquote {\bibinfo {title} {Elastic anomalies
  in glasses: Elastic string theory understanding of the cases of glycerol and
  silica},}\ }\href {\doibase 10.1103/PhysRevB.101.174311} {\bibfield
  {journal} {\bibinfo  {journal} {Phys. Rev. B}\ }\textbf {\bibinfo {volume}
  {101}},\ \bibinfo {pages} {174311} (\bibinfo {year} {2020})}\BibitemShut
  {NoStop}%
\bibitem [{\citenamefont {Beltukov}\ \emph {et~al.}(2018)\citenamefont
  {Beltukov}, \citenamefont {Parshin}, \citenamefont {Giordano},\ and\
  \citenamefont {Tanguy}}]{Beltukov2018}%
  \BibitemOpen
  \bibfield  {author} {\bibinfo {author} {\bibfnamefont {Y.~M.}\ \bibnamefont
  {Beltukov}}, \bibinfo {author} {\bibfnamefont {D.~A.}\ \bibnamefont
  {Parshin}}, \bibinfo {author} {\bibfnamefont {V.~M.}\ \bibnamefont
  {Giordano}}, \ and\ \bibinfo {author} {\bibfnamefont {A.}~\bibnamefont
  {Tanguy}},\ }\bibfield  {title} {\enquote {\bibinfo {title} {Propagative and
  diffusive regimes of acoustic damping in bulk amorphous material},}\ }\href
  {\doibase 10.1103/PhysRevE.98.023005} {\bibfield  {journal} {\bibinfo
  {journal} {Phys. Rev. E}\ }\textbf {\bibinfo {volume} {98}},\ \bibinfo
  {pages} {023005} (\bibinfo {year} {2018})}\BibitemShut {NoStop}%
\bibitem [{\citenamefont {Ryzhik}\ \emph {et~al.}(1996)\citenamefont {Ryzhik},
  \citenamefont {Papanicolaou},\ and\ \citenamefont {Keller}}]{Ryzhik1996}%
  \BibitemOpen
  \bibfield  {author} {\bibinfo {author} {\bibfnamefont {L.}~\bibnamefont
  {Ryzhik}}, \bibinfo {author} {\bibfnamefont {G.}~\bibnamefont
  {Papanicolaou}}, \ and\ \bibinfo {author} {\bibfnamefont {J.~B.}\
  \bibnamefont {Keller}},\ }\bibfield  {title} {\enquote {\bibinfo {title}
  {Transport equations for elastic and other waves in random media},}\ }\href
  {\doibase https://doi.org/10.1016/S0165-2125(96)00021-2} {\bibfield
  {journal} {\bibinfo  {journal} {Wave Motion}\ }\textbf {\bibinfo {volume}
  {24}},\ \bibinfo {pages} {327 } (\bibinfo {year} {1996})}\BibitemShut
  {NoStop}%
\bibitem [{\citenamefont {Lund}(1988)}]{Lund1988}%
  \BibitemOpen
  \bibfield  {author} {\bibinfo {author} {\bibfnamefont {F.}~\bibnamefont
  {Lund}},\ }\bibfield  {title} {\enquote {\bibinfo {title} {Response of a
  stringlike dislocation loop to an external stress},}\ }\href {\doibase
  10.1557/JMR.1988.0280} {\bibfield  {journal} {\bibinfo  {journal} {Journal of
  Materials Research}\ }\textbf {\bibinfo {volume} {3}},\ \bibinfo {pages}
  {280} (\bibinfo {year} {1988})}\BibitemShut {NoStop}%
\bibitem [{\citenamefont {Churochkin}\ and\ \citenamefont
  {Lund}(2017)}]{Churochkin2017}%
  \BibitemOpen
  \bibfield  {author} {\bibinfo {author} {\bibfnamefont {D.}~\bibnamefont
  {Churochkin}}\ and\ \bibinfo {author} {\bibfnamefont {F.}~\bibnamefont
  {Lund}},\ }\bibfield  {title} {\enquote {\bibinfo {title} {Diffusion of
  elastic waves in a two dimensional continuum with a random distribution of
  screw dislocations},}\ }\href {\doibase
  https://doi.org/10.1016/j.wavemoti.2016.11.007} {\bibfield  {journal}
  {\bibinfo  {journal} {Wave Motion}\ }\textbf {\bibinfo {volume} {69}},\
  \bibinfo {pages} {16 } (\bibinfo {year} {2017})}\BibitemShut {NoStop}%
\bibitem [{\citenamefont {Churochkin}\ and\ \citenamefont
  {Lund}(2021)}]{Churochkin2021}%
  \BibitemOpen
  \bibfield  {author} {\bibinfo {author} {\bibfnamefont {D.}~\bibnamefont
  {Churochkin}}\ and\ \bibinfo {author} {\bibfnamefont {F.}~\bibnamefont
  {Lund}},\ }\bibfield  {title} {\enquote {\bibinfo {title} {Coherent
  propagation and incoherent diffusion of elastic waves in a two dimensional
  continuum with a random distribution of edge dislocations},}\ }\href
  {\doibase https://doi.org/10.1016/j.wavemoti.2021.102768} {\bibfield
  {journal} {\bibinfo  {journal} {Wave Motion}\ }\textbf {\bibinfo {volume}
  {105}},\ \bibinfo {pages} {102768} (\bibinfo {year} {2021})}\BibitemShut
  {NoStop}%
\bibitem [{\citenamefont {Nieh}\ \emph {et~al.}(1998)\citenamefont {Nieh},
  \citenamefont {Chen},\ and\ \citenamefont {Sheng}}]{Nieh1998}%
  \BibitemOpen
  \bibfield  {author} {\bibinfo {author} {\bibfnamefont {H.~T.}\ \bibnamefont
  {Nieh}}, \bibinfo {author} {\bibfnamefont {L.}~\bibnamefont {Chen}}, \ and\
  \bibinfo {author} {\bibfnamefont {P.}~\bibnamefont {Sheng}},\ }\bibfield
  {title} {\enquote {\bibinfo {title} {Ward identities for transport of
  classical waves in disordered media},}\ }\href {\doibase
  10.1103/PhysRevE.57.1145} {\bibfield  {journal} {\bibinfo  {journal} {Phys.
  Rev. E}\ }\textbf {\bibinfo {volume} {57}},\ \bibinfo {pages} {1145}
  (\bibinfo {year} {1998})}\BibitemShut {NoStop}%
\bibitem [{\citenamefont {Barabanenkov}\ and\ \citenamefont
  {Ozrin}(2001)}]{Barabanenkov2001}%
  \BibitemOpen
  \bibfield  {author} {\bibinfo {author} {\bibfnamefont {Y.~N.}\ \bibnamefont
  {Barabanenkov}}\ and\ \bibinfo {author} {\bibfnamefont {V.~D.}\ \bibnamefont
  {Ozrin}},\ }\bibfield  {title} {\enquote {\bibinfo {title} {Comment on
  ``{W}ard identities for transport of classical waves in disordered
  media''},}\ }\href {\doibase 10.1103/PhysRevE.64.018601} {\bibfield
  {journal} {\bibinfo  {journal} {Phys. Rev. E}\ }\textbf {\bibinfo {volume}
  {64}},\ \bibinfo {pages} {018601} (\bibinfo {year} {2001})}\BibitemShut
  {NoStop}%
\bibitem [{\citenamefont {Nieh}\ \emph {et~al.}(2001)\citenamefont {Nieh},
  \citenamefont {Chen},\ and\ \citenamefont {Sheng}}]{Nieh2001}%
  \BibitemOpen
  \bibfield  {author} {\bibinfo {author} {\bibfnamefont {H.~T.}\ \bibnamefont
  {Nieh}}, \bibinfo {author} {\bibfnamefont {L.}~\bibnamefont {Chen}}, \ and\
  \bibinfo {author} {\bibfnamefont {P.}~\bibnamefont {Sheng}},\ }\bibfield
  {title} {\enquote {\bibinfo {title} {Reply to `{C}omment on {W}ard identities
  for transport of classical waves in disordered media'},}\ }\href {\doibase
  10.1103/PhysRevE.64.018602} {\bibfield  {journal} {\bibinfo  {journal} {Phys.
  Rev. E}\ }\textbf {\bibinfo {volume} {64}},\ \bibinfo {pages} {018602}
  (\bibinfo {year} {2001})}\BibitemShut {NoStop}%
\bibitem [{\citenamefont {Berman}(2000)}]{Berman2000}%
  \BibitemOpen
  \bibfield  {author} {\bibinfo {author} {\bibfnamefont {D.~H.}\ \bibnamefont
  {Berman}},\ }\bibfield  {title} {\enquote {\bibinfo {title} {Diffusion of
  waves in a layer with a rough interface},}\ }\href {\doibase
  10.1103/PhysRevE.62.7365} {\bibfield  {journal} {\bibinfo  {journal} {Phys.
  Rev. E}\ }\textbf {\bibinfo {volume} {62}},\ \bibinfo {pages} {7365}
  (\bibinfo {year} {2000})}\BibitemShut {NoStop}%
\bibitem [{\citenamefont {van Tiggelen}\ and\ \citenamefont
  {Lagendijk}(1993)}]{Tiggelen1993}%
  \BibitemOpen
  \bibfield  {author} {\bibinfo {author} {\bibfnamefont {B.~A.}\ \bibnamefont
  {van Tiggelen}}\ and\ \bibinfo {author} {\bibfnamefont {A.}~\bibnamefont
  {Lagendijk}},\ }\bibfield  {title} {\enquote {\bibinfo {title} {Rigorous
  treatment of the speed of diffusing classical waves},}\ }\href {\doibase
  10.1209/0295-5075/23/5/002} {\bibfield  {journal} {\bibinfo  {journal}
  {Europhysics Letters ({EPL})}\ }\textbf {\bibinfo {volume} {23}},\ \bibinfo
  {pages} {311} (\bibinfo {year} {1993})}\BibitemShut {NoStop}%
\bibitem [{\citenamefont {Livdan}\ and\ \citenamefont
  {Lisyansky}(1996)}]{Livdan1996}%
  \BibitemOpen
  \bibfield  {author} {\bibinfo {author} {\bibfnamefont {D.}~\bibnamefont
  {Livdan}}\ and\ \bibinfo {author} {\bibfnamefont {A.~A.}\ \bibnamefont
  {Lisyansky}},\ }\bibfield  {title} {\enquote {\bibinfo {title} {Transport
  properties of waves in absorbing random media with microstructure},}\ }\href
  {\doibase 10.1103/PhysRevB.53.14843} {\bibfield  {journal} {\bibinfo
  {journal} {Phys. Rev. B}\ }\textbf {\bibinfo {volume} {53}},\ \bibinfo
  {pages} {14843} (\bibinfo {year} {1996})}\BibitemShut {NoStop}%
\bibitem [{\citenamefont {Weaver}(1982)}]{Weaver1982}%
  \BibitemOpen
  \bibfield  {author} {\bibinfo {author} {\bibfnamefont {R.~L.}\ \bibnamefont
  {Weaver}},\ }\bibfield  {title} {\enquote {\bibinfo {title} {On diffuse waves
  in solid media},}\ }\href {\doibase 10.1121/1.387816} {\bibfield  {journal}
  {\bibinfo  {journal} {The Journal of the Acoustical Society of America}\
  }\textbf {\bibinfo {volume} {71}},\ \bibinfo {pages} {1608} (\bibinfo {year}
  {1982})}\BibitemShut {NoStop}%
\bibitem [{\citenamefont {Lubatsch}\ \emph {et~al.}(2005)\citenamefont
  {Lubatsch}, \citenamefont {Kroha},\ and\ \citenamefont
  {Busch}}]{Lubatsch2005}%
  \BibitemOpen
  \bibfield  {author} {\bibinfo {author} {\bibfnamefont {A.}~\bibnamefont
  {Lubatsch}}, \bibinfo {author} {\bibfnamefont {J.}~\bibnamefont {Kroha}}, \
  and\ \bibinfo {author} {\bibfnamefont {K.}~\bibnamefont {Busch}},\ }\bibfield
   {title} {\enquote {\bibinfo {title} {Theory of light diffusion in disordered
  media with linear absorption or gain},}\ }\href {\doibase
  10.1103/PhysRevB.71.184201} {\bibfield  {journal} {\bibinfo  {journal} {Phys.
  Rev. B}\ }\textbf {\bibinfo {volume} {71}},\ \bibinfo {pages} {184201}
  (\bibinfo {year} {2005})}\BibitemShut {NoStop}%
\bibitem [{\citenamefont {Ee}\ \emph {et~al.}(2017)\citenamefont {Ee},
  \citenamefont {Jung}, \citenamefont {Kim},\ and\ \citenamefont
  {Lee}}]{Ee2017}%
  \BibitemOpen
  \bibfield  {author} {\bibinfo {author} {\bibfnamefont {J.-H.}\ \bibnamefont
  {Ee}}, \bibinfo {author} {\bibfnamefont {D.-W.}\ \bibnamefont {Jung}},
  \bibinfo {author} {\bibfnamefont {U.-R.}\ \bibnamefont {Kim}}, \ and\
  \bibinfo {author} {\bibfnamefont {J.}~\bibnamefont {Lee}},\ }\bibfield
  {title} {\enquote {\bibinfo {title} {Combinatorics in tensor-integral
  reduction},}\ }\href {\doibase 10.1088/1361-6404/aa54ce} {\bibfield
  {journal} {\bibinfo  {journal} {European Journal of Physics}\ }\textbf
  {\bibinfo {volume} {38}},\ \bibinfo {pages} {025801} (\bibinfo {year}
  {2017})}\BibitemShut {NoStop}%
\bibitem [{\citenamefont {Mahan}(2000)}]{Mahan2000}%
  \BibitemOpen
  \bibfield  {author} {\bibinfo {author} {\bibfnamefont {G.~D.}\ \bibnamefont
  {Mahan}},\ }\href@noop {} {\emph {\bibinfo {title} {Many Particle Physics,
  Third Edition}}}\ (\bibinfo  {publisher} {Plenum},\ \bibinfo {address} {New
  York},\ \bibinfo {year} {2000})\BibitemShut {NoStop}%
\bibitem [{\citenamefont {Barabanenkov}\ and\ \citenamefont
  {Ozrin}(1992)}]{Barabanenkov1992}%
  \BibitemOpen
  \bibfield  {author} {\bibinfo {author} {\bibfnamefont {Y.~N.}\ \bibnamefont
  {Barabanenkov}}\ and\ \bibinfo {author} {\bibfnamefont {V.~D.}\ \bibnamefont
  {Ozrin}},\ }\bibfield  {title} {\enquote {\bibinfo {title} {Problem of light
  diffusion in strongly scattering media},}\ }\href {\doibase
  10.1103/PhysRevLett.69.1364} {\bibfield  {journal} {\bibinfo  {journal}
  {Phys. Rev. Lett.}\ }\textbf {\bibinfo {volume} {69}},\ \bibinfo {pages}
  {1364} (\bibinfo {year} {1992})}\BibitemShut {NoStop}%
\bibitem [{\citenamefont {{Sheng}}(1990)}]{Sheng1990}%
  \BibitemOpen
  \bibfield  {author} {\bibinfo {author} {\bibfnamefont {P.}~\bibnamefont
  {{Sheng}}},\ }\href {\doibase 10.1142/0565} {\emph {\bibinfo {title}
  {{Scattering and Localization of Classical Waves in Random Media}}}}\
  (\bibinfo  {publisher} {World Scientific, Singapore},\ \bibinfo {year}
  {1990})\BibitemShut {NoStop}%
\bibitem [{\citenamefont {van Tiggelen}\ and\ \citenamefont
  {Stark}(2000)}]{Tiggelen2000}%
  \BibitemOpen
  \bibfield  {author} {\bibinfo {author} {\bibfnamefont {B.}~\bibnamefont {van
  Tiggelen}}\ and\ \bibinfo {author} {\bibfnamefont {H.}~\bibnamefont
  {Stark}},\ }\bibfield  {title} {\enquote {\bibinfo {title} {Nematic liquid
  crystals as a new challenge for radiative transfer},}\ }\href {\doibase
  10.1103/RevModPhys.72.1017} {\bibfield  {journal} {\bibinfo  {journal} {Rev.
  Mod. Phys.}\ }\textbf {\bibinfo {volume} {72}},\ \bibinfo {pages} {1017}
  (\bibinfo {year} {2000})}\BibitemShut {NoStop}%
\end{thebibliography}
\end{document}